# Evolution and sudden change of steady interactions of low enthalpy hypersonic double wedge flows with fore angle

Yihui Weng[a], Yi Duan[b], Qin Li[a*], Yunchuan Wu[a], Mengyu Wang[a], Pan Yan[a], Siyi Li[b]

a. School of Aerospace Engineering, Xiamen University, China, 361102

b. Science and Technology on Space Physics Laboratory, China Academy of Launch Vehicle Technology, Beijing 100076, China

**Abstract:** This study conducts numerical simulations and analyses to investigate the evolution and sudden change of *steady* interaction structures with the fore wedge angle $\theta_1$ in a low enthalpy hypersonic double wedge configuration. It particularly focuses on the conditions of Swantek and Austin's experiments [AIAA 2012-284] where $Ma = 7$, and $h_0 = 2\ MJ/kg$ but with a reduced Reynolds number ($Re$). The sudden structural change indicates that when $\theta_1$ reaches a critical value, minor angular variations can trigger a discontinuous transformation in flow structures. The analysis is based on the laminar Navier-Stokes equations, using ideal gas and non-equilibrium gas models to capture differences in the evolution of interaction patterns. A third-order scheme WENO3-PRM$^2_{1,1}$ [Li et al. J. Sci. Comput., 88 (2021)75-130] is employed to achieve high accuracy. A grid convergence study determines the required grids based on results from the ideal gas model at $\theta_1 = 32°$ and the non-equilibrium gas model at $\theta_1 = 35°$. Then, under the condition of $Re = 1 \times 10^5\ /m$, detailed numerical simulations are conducted to investigate the evolution and sudden change of *steady* flow interactions as $\theta_1$ varies over $0° - 40°$. The simulations identify the boundary of $\theta_1$ that distinguishes steady from unsteady flows and analyzes the flow evolutions, including wave systems, vortex structures, and parameter distributions. This study yields the following findings: (a) The upper and lower boundaries of $\theta_1$ for the onset of unsteady flow are identified. When $\theta_1$ lies outside these boundaries, the flow remains steady. (b) As $\theta_1$ increases, the interaction patterns evolve sequentially, progressing from Type VI through Type VI → V, Type III, Type IV$_\mathrm{r}$, and ultimately to a flow dominated solely by a bow shock. This evolution defines the boundaries between different interaction patterns and provides a comprehensive understanding of their progression with $\theta_1$. Sudden structural changes occur during the transitions from Type III to Type IV$_\mathrm{r}$ and from Type IV$_\mathrm{r}$ to a bow shock-dominated flow. In addition, a comparative study is performed through shock polar analysis to compare its predictions with computational results. (c) An unconventional reflection pattern of the transmitted shock over the separation zone, called Type III$_\mathrm{r}$, is observed in non-equilibrium gas flows, which differs from classical interaction patterns. (d) The aerodynamic distribution of wall properties under various interactions is obtained, indicating distinct features before and after the sudden structural change. (e) A comparison of the characteristics and evolutionary trends of interactions for different gas models demonstrates that the evolution rate of interaction patterns is faster in the case of the ideal gas model.

**Keywords**: Hypersonic; Double wedge; Low enthalpy; Steady interaction; Sudden structural change

## 1 Introduction

As one of the common geometries, the double wedge configuration is widely applied in hypersonic vehicles, scramjet engines, and supersonic inlets. In the case of hypersonic flows, typical phenomena such as shock/shock and shock wave/boundary layer interactions are prevalent. Such interactions can induce flow separations over the configuration, increase localized heating and pressure, and cause potential structural damage to the vehicle. Therefore, conducting research on the double wedge is crucial for the design of hypersonic vehicles.

[*] Corresponding author, email: qin-li@vip.tom.com

In computational studies of hypersonic double wedge flows, the configuration and corresponding experimental results by Swantek and Austin [1, 2] have been selected by NATO AVT-205 as the reference for evaluating the capability of CFD in predicting hypersonic laminar boundary layer interaction. For this configuration, numerical simulation studies have been conducted, including those by Komives et al. [3], Badr and Knight [4], Knight et al. [5], and Reinert et al. [6]. In the experiments [1, 2], the double wedge has fore and aft angles of $\theta_1 = 30°$ and $\theta_2 = 55°$, respectively, with corresponding surface lengths of $L_1 = 50.8\ mm$ and $L_2 = 25.4\ mm$, and a width of $L_Z = 101.6\ mm$. In addition, a horizontal extension was connected to the aft wedge with $L_3 = 10.82\ mm$. The computations considered a range of Mach numbers ($Ma$) under hypersonic conditions ($Ma \approx 7$), featuring both low enthalpy ($h_0 \approx 2\ MJ/kg$) and high enthalpy ($h_0 \approx 8\ MJ/kg$).

Under hypersonic high enthalpy conditions (M7_8), the flow field exhibits significant thermochemical non-equilibrium, resulting in aerodynamic features such as temperature and species distributions, shock/vortex structures, and peak heat fluxes that differ considerably from those in ideal gas cases. Therefore, chemical and thermochemical non-equilibrium models are required for CFD studies. Knight et al. [5] reviewed works in this regard, and then research, including [7-10], explored this topic. However, detailed discussions are not included.

The hypersonic low enthalpy condition (M7_2) is another key area of research and is the primary focus of this study. Although the complexity of heat fluxes and chemical reactions in the flow field under low enthalpy conditions is less pronounced than in high enthalpy conditions, ensuring consistency between CFD results and experimental data remains challenging. For example, as shown in [5], Celik et al., Badr et al., Komives et al., and Lani et al. conducted numerical simulations on the configuration. They identified spatial oscillation patterns in the flow field that were not observed in experiments. Although the results demonstrated qualitative consistency, quantitative predictions (heat fluxes) exhibited discrepancies with experimental data, such as mismatches in reattachment locations, peak heat transfer values, and triple-point positions. Recent computational studies [11-14] have employed approaches to comparing time-averaged heat fluxes over finite periods with experimental data. However, these methods are inconsistent with the experimental measurements, which have a temporal resolution of 1 $\mu s$. Beyond prediction accuracy, researchers have also been concerned with the evolution of interactions under varying parameters.

Early research efforts focused on the evolution patterns induced by geometric variations of the double wedge, primarily examining the aft wedge angle $\theta_2$ and the length ratio $L_2/L_1$. Olejniczak et al. [15] conducted pioneering studies on inviscid flows at different Mach numbers, specifically $Ma = 2.8, 9$ with a fore wedge angle $\theta_1 = 15°$ and $L_2/L_1 = 1$. The evolution of shock interaction types was analyzed by increasing $\theta_2$. The results revealed that under $Ma = 9$, when $\theta_2 = 35°$, the primary interaction in the double wedge flow field was Type VI. As $\theta_2$ increased to 45°, the Type VI interaction transitioned to Type V, exhibiting an intermediate structure. When $\theta_2 = 45°$, the transformation to Type V was completed, and at $\theta_2 = 50°$, the interaction further evolved into Type IV. When $\theta_2$ increased to 60°, the Type IV interaction developed into a structure termed Type $IV_r$ in [15], as shown in Fig. 1, where the transmitted shock reflected on the wall and gradually formed a subsequent jet. Under $Ma = 2.8$, the evolution trend of the interaction structures was similar to that observed at $Ma = 9$. Based on these findings, Olejniczak et al. [15] summarized the characteristics of various shock interaction types and the transition criteria between them, providing critical conditions for these transitions. Then, Hashimoto [16] conducted experimental studies on double wedge configurations under high enthalpy conditions with $\theta_1 = 25°$ and $\theta_2 = 40°$, $50°$, and $68°$. The study found that when $\theta_2 = 40°$, the interaction was Type VI, and the flow remained steady. When $\theta_2 = 50°$, the interaction transitioned from Type VI to Type V due to the formation of a bow shock on the aft wedge surface. When $\theta_2 = 68°$, the flow separation region expanded, the separation point moved upstream, and the transmitted shock extending from the triple-point nearly perpendicularly impacted the wedge surface, leading to unsteady flow. Durna and Celik [17] conducted simulations for low enthalpy flows at $Ma = 7$, focusing on a three-dimensional double wedge configuration with $\theta_1 = 30°$, $\theta_2 = 45° - 60°$. They found that when $\theta_2 < 50°$, the flow exhibited two-dimensional characteristics. However, as $\theta_2$ increased, three-dimensional effects became pronounced, with streamwise vortices appearing within the shear layer. These vortices intensified with increasing $\theta_2$, significantly influencing heat flux distributions. Kumar and De [14] performed simulations for double wedge configurations with $\theta_1 = 30°$, $\theta_2 = 45°$, $50°$, $55°$, and $60°$, and $L_1/L_2 = 0.5$, 1, 2, and 2.5. They observed steady flow patterns and three unsteady flow modes: vibration, oscillation, and pulsation. They reported that in the absence of incident shock interactions with the separation zone, the flow remained steady for the selected $L_1/L_2$. Increasing $\theta_2$ expanded the separation zone and raised a transition to unsteady flow. In addition, increasing $L_2$ caused the flow to transition from a steady state to oscillation mode, then to pulsation mode, and finally back to a steady state. Hong et al. [18] investigated interactions and their evolution under extremely high enthalpy conditions ($21.77\ MJ/kg$) for varying aft wedge angles. They considered $\theta_1 = 15°$, $\theta_2 = 25° - 55°$, and $L_1 = L_2 = 100\ mm$, employing frozen, thermal non-equilibrium, and thermochemical non-equilibrium gas models, respectively. The results showed that as $\theta_2$ increased, the interaction transitioned from Type VI to Type V. Thermochemical non-equilibrium effects reduced the extent of curved shocks and the size of separation bubbles, delaying flow pulsations. In addition, global stability analysis was utilized to study the flow's stability characteristics, revealing that the flow became unstable when the wedge angle exceeded a critical value.

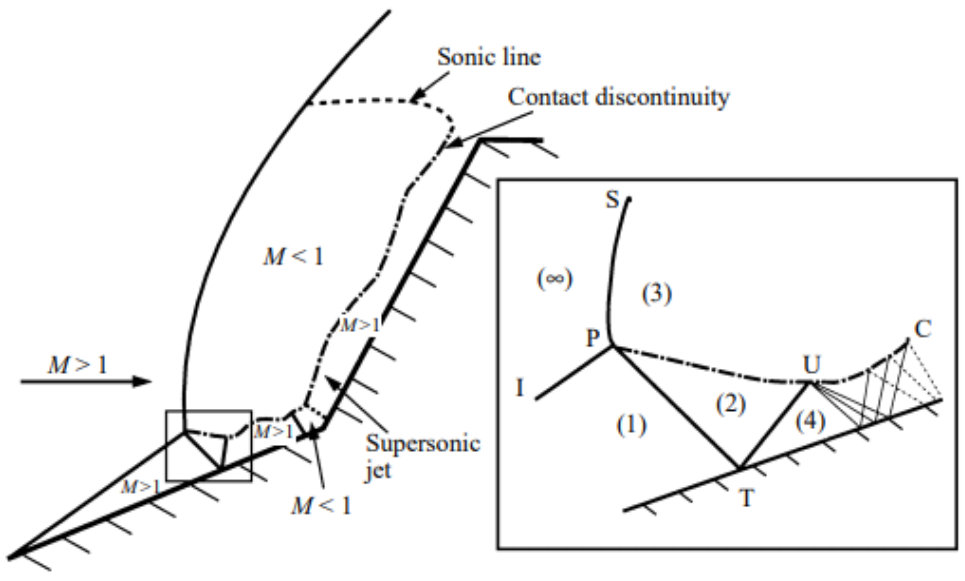

Fig. 1 Diagram of Type $IV_r$ interaction presented by Olejniczak et al. with the courtesy from [15]

Based on the above studies on the evolution of interactions with geometric parameters, research under low enthalpy conditions still has several shortcomings: (a) Previous studies have primarily focused on changes in the double wedge flow induced by variations in $\theta_2$, while systematic investigations into the effects of modifying the fore-wedge angle $\theta_1$ remain lacking. There is insufficient understanding of the significant impact of $\theta_1$ on flow separation characteristics and shock structures, as demonstrated in this study. (b) When modifying the geometry, there is a lack of research on the influence of different gas models on the interaction flow field. A recent study [19] investigated the evolution of steady interaction flow fields in the same double wedge using different gas models for $Re = 1{\times}10^4 - 1{\times}10^5/m$. The study revealed that different gas models significantly affect the interaction structures and the steady/unsteady characteristics of the flow field, even under low enthalpy M7_2 conditions. However, the study primarily focused on the evolution of steady interactions with flow parameters, leaving a gap in understanding the evolution of interactions with geometric parameters, particularly $\theta_1$, as emphasized in this work. (c) As previously mentioned, long-time computations under M7_2 conditions yield results (heat flux distributions and interaction structures) that are entirely inconsistent with experimental observations, where the transmitted shock impinges on the aft wedge surface. Comparisons based on time-averaged results from short-time computations do not align with the temporal resolution of experimental measurements, which is 1 $ns$. There is a lack of deterministic studies on the steady evolution of interaction structures that closely resemble experimental observations.

Therefore, further research and analysis are essential to address the above issues. The $30 - 55°$ double-wedge configuration from Swantek and Austin's experiments and the low enthalpy condition with a relatively low $Re$ ($1 \times 10^5\ /m$) are referenced. Numerical simulations and analyses of the evolution and sudden changes in steady interaction structures in two-dimensional double wedge flow with varying fore wedge angles are performed using ideal and non-equilibrium gas models. The focus on steady interactions stems from their typically higher computational reliability. The sudden change observed in this study refers to rapid transformations in the interaction structures as the fore wedge angle approaches a critical value, where minor angular variations can trigger discontinuous changes in the steady interaction structures, further termed a sudden structural change. Given the low $Re$, the laminar Navier-Stokes equations are employed, and the WENO3-PRM$^2_{1,1}$ scheme [20] is adopted to enhance the resolution and accuracy of the interaction structures. The paper is organized as follows: Section 2 introduces the governing equations, the WENO3-PRM$^2_{1,1}$ scheme, and computational validations; Section 3 presents the case setup and grid convergence study, followed by analyses of the overall characteristics of interactions with varying fore wedge angles, interaction structures within different fore wedge angle ranges, and wall aerodynamic properties. The numerical simulations are conducted under conditions referring to M7_2 but with $\theta_1 = 0° - 40°$. Finally, Section 4 provides the conclusions.

## 2 Governing equations, gas models, and numerical methods

### 2.1 Governing equations and gas models

The numerical simulations are based on the laminar Navier-Stokes equations, which employ ideal gas and non-equilibrium gas models. Below, the governing equations for chemically non-equilibrium flows are provided, from which the equations for an ideal gas can be derived. For a gas mixture consisting of $ns$ species, the equations are as follows:

$$\frac{\partial Q}{\partial t} + \frac{\partial E}{\partial x} + \frac{\partial F}{\partial y} + \frac{\partial G}{\partial z} - \left(\frac{\partial E_v}{\partial x} + \frac{\partial F_v}{\partial y} + \frac{\partial G_v}{\partial z}\right) = S \tag{1}$$

where $Q=\begin{bmatrix}\rho_1\\ \vdots\\ \rho_{ns}\\ \rho u\\ \rho v\\ \rho w\\ \rho E\end{bmatrix}$, $E=\begin{bmatrix}\rho_1 u\\ \vdots\\ \rho_{ns} u\\ \rho u^2+p\\ \rho uv\\ \rho uw\\ \rho u h_0\end{bmatrix}$, $F=\begin{bmatrix}\rho_1 v\\ \vdots\\ \rho_{ns} v\\ \rho vu\\ \rho v^2+p\\ \rho vw\\ \rho v h_0\end{bmatrix}$, $G=\begin{bmatrix}\rho_1 w\\ \vdots\\ \rho_{ns} w\\ \rho wu\\ \rho wv\\ \rho w^2+p\\ \rho w h_0\end{bmatrix}$, $S=\begin{bmatrix}\omega_1\\ \vdots\\ \omega_{ns}\\ 0\\ 0\\ 0\\ 0\end{bmatrix}$, $E_v=\begin{bmatrix}q_{x1}\\ \vdots\\ q_{xns}\\ \tau_{xx}\\ \tau_{xy}\\ \tau_{xz}\\ u\tau_{xx}+v\tau_{xy}+w\tau_{xz}+\\ q_x+\rho\sum_{i=1}^{ns}D_i h_i\frac{\partial Y_i}{\partial x}\end{bmatrix}$, $F_v=\begin{bmatrix}q_{y1}\\ \vdots\\ q_{yns}\\ \tau_{yx}\\ \tau_{yy}\\ \tau_{yz}\\ u\tau_{yx}+v\tau_{yy}+w\tau_{yz}+\\ q_y+\rho\sum_{i=1}^{ns}D_i h_i\frac{\partial Y_i}{\partial y}\end{bmatrix}$, $G_v=\begin{bmatrix}q_{z1}\\ \vdots\\ q_{zns}\\ \tau_{zx}\\ \tau_{zy}\\ \tau_{zz}\\ u\tau_{zx}+v\tau_{zy}+w\tau_{zz}+\\ q_z+\rho\sum_{i=1}^{ns}D_i h_i\frac{\partial Y_i}{\partial z}\end{bmatrix}$,

$\rho_i$ is the density of the *i*-th species, $Y_i$ is its mass fraction ($\rho_i/\rho$), $\omega_i$ is the generation source term, and the mass diffusion is $q_{x_j i}=\rho D_i \partial Y_i/\partial x_j$, with $D_i$ as the diffusion coefficient and $h_i$ representing the specific enthalpy. In addition, the specific total enthalpy is $h_0=\sum_{i=1}^{ns}Y_i h_i+(\sum_{j=1}^{3}u_j^2)/2$. The viscous stress is $\tau_{x_i x_j}=-\frac{2}{3}\mu\nabla\cdot\vec{V}+\mu(\frac{\partial u_i}{\partial x_j}+\frac{\partial u_j}{\partial x_i})$, where $\mu$ is the viscosity coefficient. $p=\sum_{i=1}^{ns}p_i=\sum_{i=1}^{ns}\rho_i R_i T$, where $R_i$ is the gas constant for species *i*. In addition, $q_{x_j}=k\,\partial T/\partial x_j$, where $k$ is the heat conductivity. When the multispecies component is not considered in the case of an ideal gas, *ns* can be reduced to 1, $\rho_i$ can be replaced by $\rho$, and the quantities $\omega_i$, $q_{x_j i}$, and $\partial Y_i/\partial x_j$ disappear. The relationships between transport coefficients and thermodynamic properties vary across different gas models. Therefore, the relationships between transport coefficients and thermodynamic properties adopted in this study are introduced.

First, the viscosity coefficient $\mu$ is determined for an ideal gas using the Sutherland formula:

$$\mu/\mu_0=(T/T_0)^{\frac{3}{2}}\frac{1+T_S/T_0}{T/T_0+T_S/T_0} \quad (2)$$

where $T_0=273.16\,K$, $\mu_0=1.7161\times10^{-5}\,/(Pa\cdot s)$, $T_s=124\,K$. The thermal conductivity is $k=\mu C_p/Pr$, where $Pr=0.72$. The relationship between thermodynamic properties is described by the state equation $p=\rho RT$ for air.

Finite rate chemical reaction processes are considered in chemical non-equilibrium gases. If there are *ns* species and *nr* reversible reactions, the reactions can be expressed as follows:

$$\sum_{i=1}^{ns}\alpha_{ji}A_i\rightleftharpoons\sum_{i=1}^{ns}\beta_{ji}A_i\ (j=1,\cdots,nr) \quad (3)$$

where $\alpha_{ji}$ and $\beta_{ji}$ are the stoichiometric coefficients of the reactant; the stoichiometric coefficient for colliders is 0, and $A_i$ is the specie *i*. Based on Eq. (3), the source generation of the *i*-th species in Eq. (1) is defined as follows: $\omega_i=M_i\sum_{j=1}^{nr}(d(\rho_i/M_i)/dt)_j$, where $M_i$ is the molecular weight. The solution process can be found in the literature [19] and is not explained here.

Similar to [19], this study adopts the 5 species and 6 reaction air model proposed by Gupta et al. [21], as listed in Table 1. The colliders are defined as: $M_1=M_3=\{O,N,O_2,N_2,NO\}$ and $M_2=\{O,O_2,N_2,NO\}$. Detailed information on other coefficients can be found in Gupta's study [21]. In addition, the non-catalytic condition is employed at the wall, considering the metal material of the double wedge.

Table 1. Reactions of the gas model with 5 species and 6 reactions [21]

| Index | Reaction equation |
|---|---|
| 1 | $O_2+M_1\rightleftharpoons 2O+M_1$ |
| 2 | $N_2+M_2\rightleftharpoons 2N+M_2$ |
| 3 | $N_2+N\rightleftharpoons 2N+N$ |
| 4 | $NO+M_3\rightleftharpoons N+O+M_3$ |
| 5 | $NO + O \rightleftharpoons O_2 + N$ |
| 6 | $N_2 + O \rightleftharpoons NO + N$ |

## 2.2 Numerical methods

Li et al. [20] proposed the WENO3-PRM$^2_{1,1}$ scheme by developing a new piecewise rational polynomial mapping method (PRM) with good performance, based on the third-order improvement of WENO schemes by Jiang and Shu [22] (WENO3-JS). The WENO3-PRM$^2_{1,1}$ scheme provides enhanced accuracy while preserving as much stability as possible. A detailed introduction and application of the scheme were provided in our previous study [23]. Only the specific form of the new mapping is presented:

$$\begin{cases} PRM^{L,n+1}_{n,m;m_1;c_1,c_2} = d_k + \frac{(\omega-d_k)^{n+1}}{(\omega-d_k)^n+(-1)^{1+n}c_2^L(\omega-d_k)\omega^{m_1}+(-1)c_1^L\omega^{m+1}}, when\ \omega < d_k \\ PRM^{R,n+1}_{n,m;m_1;c_1,c_2} = d_k + \frac{(\omega-d_k)^{n+1}}{(\omega-d_k)^n+c_2^R(\omega-d_k)(1-\omega)^{m_1}+c_1^R(1-\omega)^{m+1}}, when\ \omega \geq d_k \end{cases} \tag{4}$$

where $d_k$ is the linear weight, $\omega$ is the nonlinear weight. Superscript $L$ indicates that the function is for $\omega < d_k$, and the superscript $R$ indicates that it is for $\omega \geq d_k$. In addition, $n$ corresponds to the order of the first non-zero term in the Taylor series expansion of the mapping function at $d_k$, and $m$ controls the behavior of the mapping function as it approaches the endpoints $\{0, 1\}$. The mapping function exhibits good flatness at $d_k$, and optimized convergence properties at the endpoints $\{0,1\}$ by adjusting the parameters $c_1$, $c_2$, and $m_1$ in Eq. (4). Therefore, the mapped WENO schemes achieve high resolution and robustness. In WENO3-PRM$^2_{1,1}$, the parameters in Eq. (4) are set to $n$=1 and $m$=1, while the values of other parameters are listed in Table 2. More detailed information can be found in the literature [20].

Table 2. Definitions of parameters ($c_1$, $c_2$, $m_1$) in the third-order WENO3-PRM$^2_{1,1}$ scheme

| | | $c_1$ | $c_2$ | $m_1$ |
|---|---|---|---|---|
| $d_k$=1/3 | $L$ | 1 | $7 \times 10^7$ | 5 |
| | $R$ | 1 | $3 \times 10^6$ | 5 |
| $d_k$=2/3 | $L$ | 1 | $1 \times 10^5$ | 4 |
| | $R$ | 1 | $3 \times 10^6$ | 4 |

In addition, given that flow steadiness is the main focus of this study, the temporal discretization uses the canonical LU-SGS method with a local time step to efficiently determine flow steadiness [24].

## 2.3 Validation tests

Weng et al. [23] validated the accuracy and precision of the WENO3-PRM$^2_{1,1}$ scheme using the Hollow Cylinder Extended Flare case at $Ma = 11.35$. This study presents the following cases for computational validation to further demonstrate numerical performance.

(1) Double cone flow at $Ma = 11.3$ (RUN35) [25]

The calculations employ the ideal gas model with the following inflow conditions: $Ma = 11.3$, $Re = 1.333\times10^5\ /m$, $T_\infty = 138.69\ K$, and $T_w = 296.11\ K$.

Fig. 2 compares the heat flux prediction by WENO3-PRM$^2_{1,1}$ to the experimental results, indicating a good agreement. Fig. 3 depicts the numerical schlieren based on density gradients, where the green solid lines represent streamlines. The results demonstrate that the computation resolves the triple separation vortices and the series of wave reflections formed by the interaction in the jet channel, showing the high resolution of the WENO3-PRM$^2_{1,1}$ scheme.

---

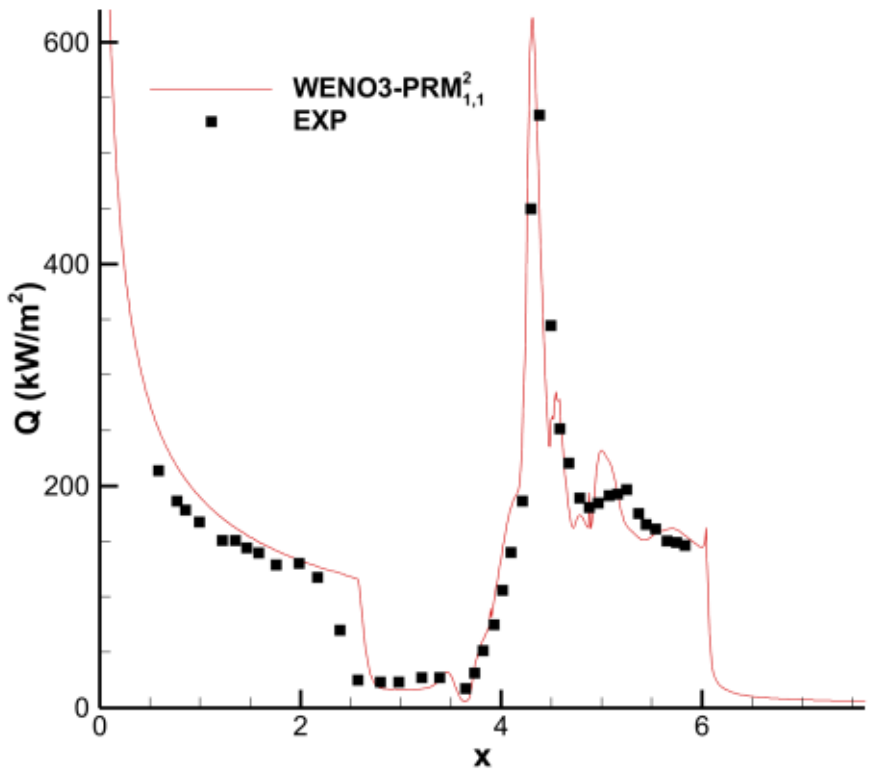

Fig. 2 Heat flux predictions of the double cone in the case of Run35 compared to experimental results [25]

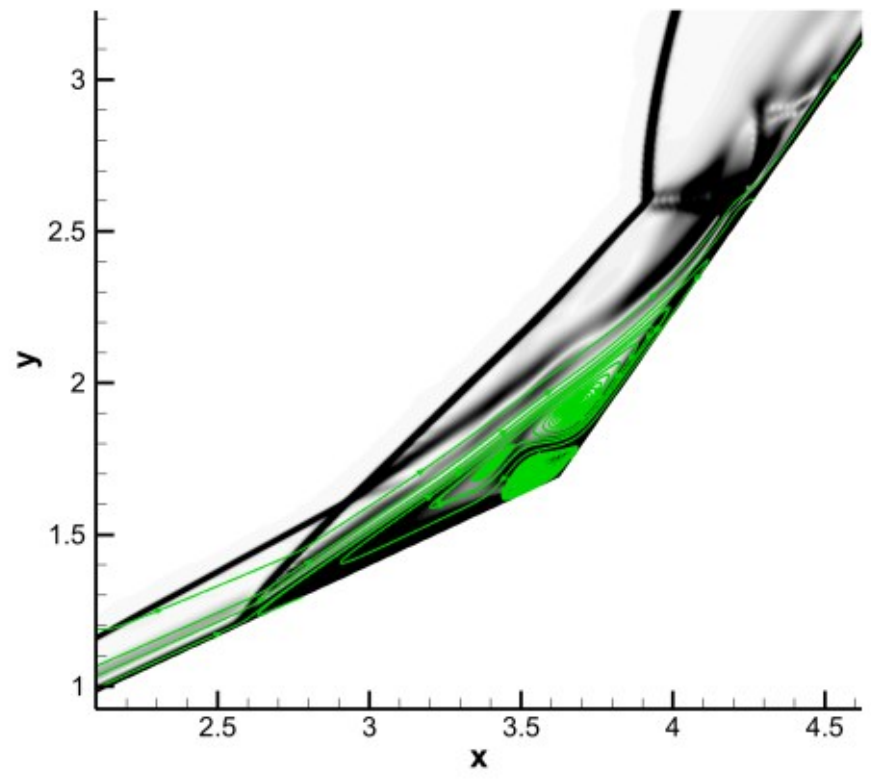

Fig. 3 Numerical schlieren of the double cone in case Run35 by WENO3-PRM$^2_{1,1}$ (Green solid lines denote streamlines)

(2) High enthalpy cylinder flow at $Ma = 8.76$ [26]

The calculations employ the non-equilibrium gas model listed in Table 1, with the following inflow conditions: $Ma = 8.76$, $Re = 4.7\times10^5\ /m$, $T_\infty = 694\ K$, and $T_w = 300\ K$.

Fig. 4 compares the computed wall pressure distribution with experimental data. It indicates good overall agreement between the results and the experimental values. Fig. 5 presents the distribution of mass fractions of species along the stagnation line, with reference results compared to those from HYFLOW [27]. The consistency between the two sets of results validates the correctness and accuracy of the calculations combining the non-equilibrium gas model and the WENO3-PRM$^2_{1,1}$ scheme.

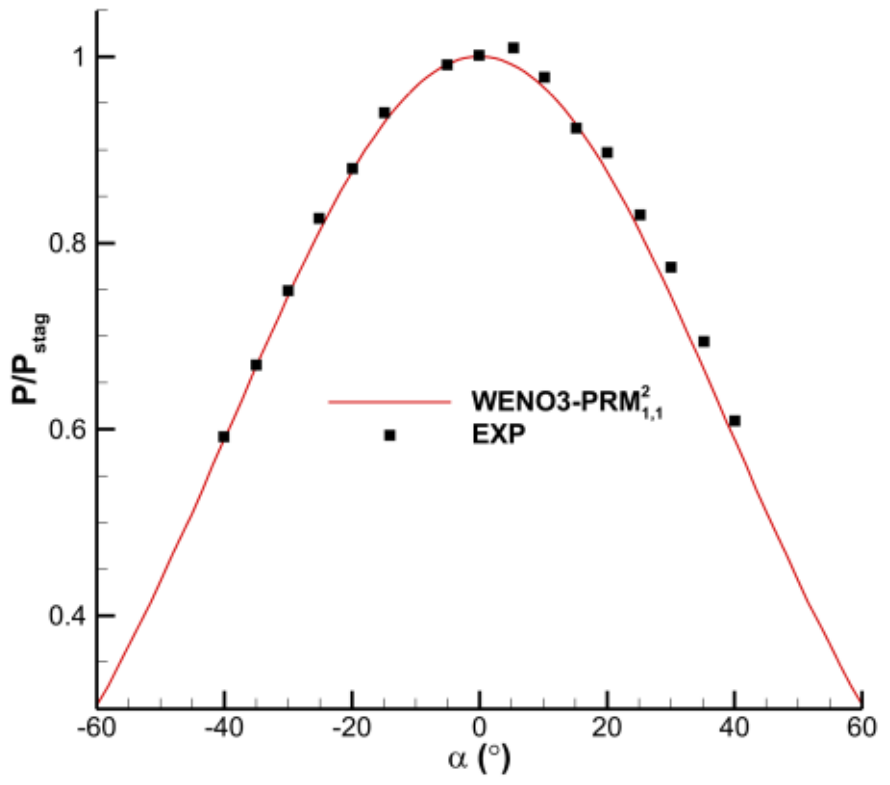

Fig. 4 Wall pressure of the high enthalpy cylinder flow at $Ma = 8.76$ compared to that of the experiment [26]

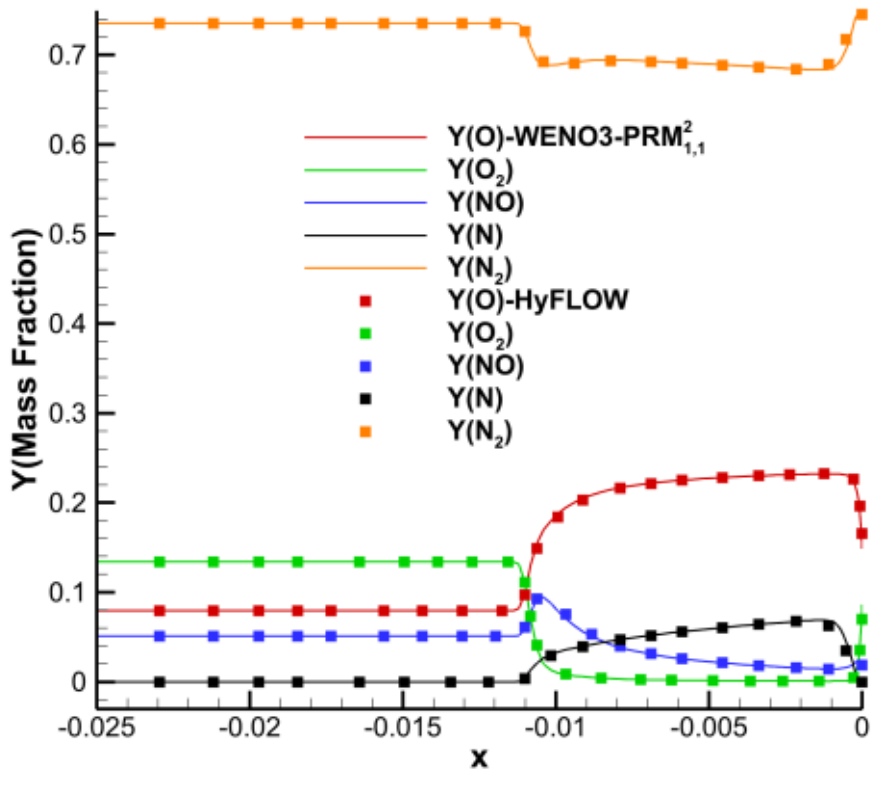

Fig. 5 Distributions of mass fractions of species along the stagnation line compared to the results of HYFLOW [27]

## 3 Numerical simulation and analysis of the evolution of complex interactions in double wedge flows with different $\theta_1$

As stated in the introduction, current studies on double wedge primarily focus on adjusting inflow conditions and the magnitude of $\theta_2$ to investigate the evolution of flow characteristics with varying parameters while studies on various $\theta_1$ are lacking. This study indicates that the flow field exhibits multiple interaction patterns within different $\theta_1$. The sudden structural change occurs during specific transformation processes, where a discontinuous structural change in the interaction occurs immediately after $\theta_1$ is altered by 1°. This study references the experimental conditions of Swantek and Austin [1, 2], selects the standard $30 - 55°$ double wedge configuration as a reference, and conducts numerical simulations of two-dimensional double wedge flow with varying $\theta_1$ under low enthalpy, laminar conditions. Although varying $\theta_1$, the wedge surface lengths and $\theta_2$ remain unchanged. The $Re$ in the calculations is smaller than in the experiments to obtain more deterministic steady interaction results, particularly interactions that resemble the structure observed in [1, 2] with the transmitted shock impinges on the aft wedge. The range of $\theta_1$ is from 0° to 40°.

### 3.1 Grid convergence study

The numerical simulations reference the low enthalpy experimental conditions: $Ma = 7.11$, $h_0 = 2.1\ MJ/kg$, $Re = 1.1 \times 10^6\ /m$, $T_\infty = 191\ K$, and $T_w = 298\ K$. This study adopts the $Re = 1 \times 10^5\ /m$ to obtain steady results predominantly during the evolution of interactions with varying $\theta_1$. The origin of the coordinate system is placed at the tip of the double wedge in the computational setup. The fore wedge angle $\theta_1$ varies when rotating the fore wedge surface counterclockwise around the coordinate origin. In contrast, the angle between the aft wedge surface and the horizontal line remains unchanged, and the third wedge surface remains horizontal. A schematic of the double wedge geometry is shown in Fig. 6. The relevant parameters are as follows: $\theta_1 = 0° - 40°$, $\theta_2 = 55°$, $L_1 = 50.8\ mm$, $L_2 = 25.4\ mm$, and $L_3 = 21.64\ mm$.

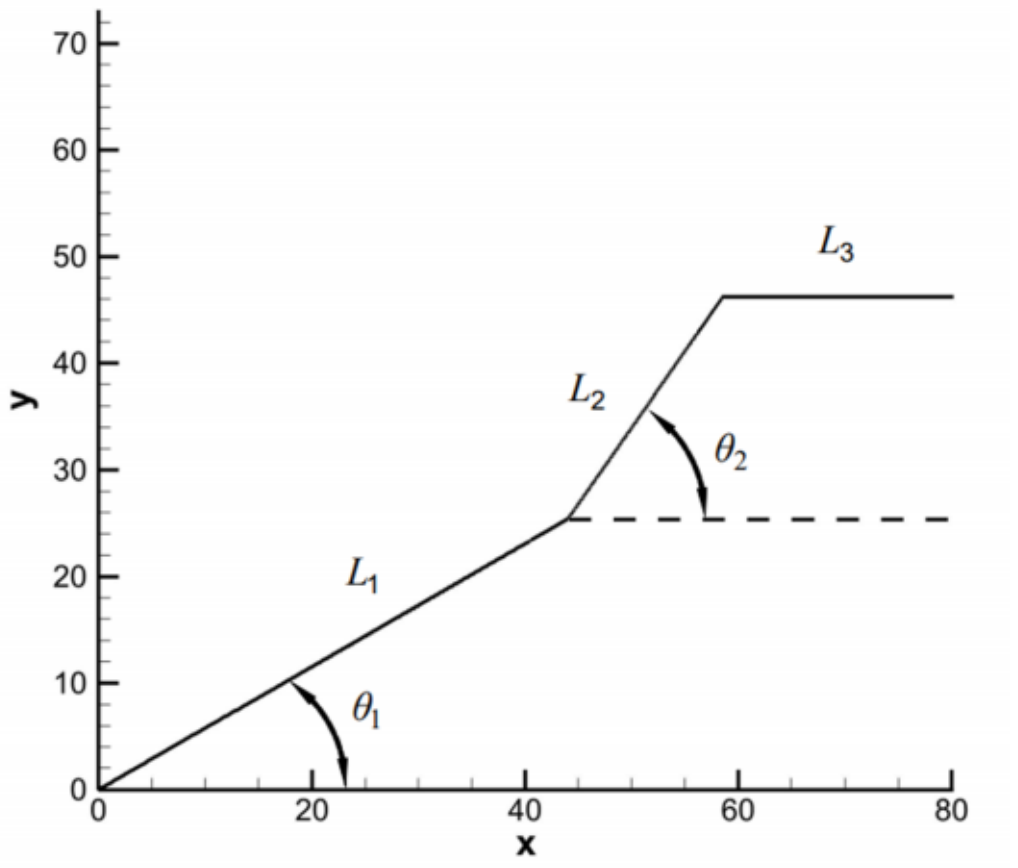

Fig. 6 Two-dimensional double wedge with geometric characteristics indicated

The grid convergence study was conducted to determine the appropriate computational grids. Initially, the coarse grids of $541 \times 256$ were employed to quickly obtain flow fields of investigating cases. Then, two representative cases with complex structures were selected from both gas models for further grid convergence analysis. The case with $\theta_1 = 32°$ was selected for the ideal gas (IG) model, while the case with $\theta_1 = 35°$ was chosen for the non-equilibrium gas (NEG) model. In addition, three grid resolutions were defined for study: coarse grids of $541 \times 256$, medium grids of $812 \times 382$, and fine grids of $1093 \times 512$.

Figs. 7 (a) and (b) show the two gas models' wall pressure coefficients and skin friction distributions on three different types of grids. The results for $\theta_1 = 32°$ (IG) and $\theta_1 = 35°$ (NEG) in the pressure coefficient distributions show qualitatively consistent trends, including the initial rise after the separation shock, the pressure plateau caused by vortex separation, the secondary rise during reattachment, and the drop after the expansion corner. Some characteristic locations in the skin friction distributions correspond to those in the pressure distributions. The skin friction gradually decreases to negative values after the separation, remains constant within the most separation zone, reaches its minimum at $x \approx 50\ mm$, and finally peaks near the expansion corner. The results for the three grids indicate minimal differences in pressure and skin friction distributions.

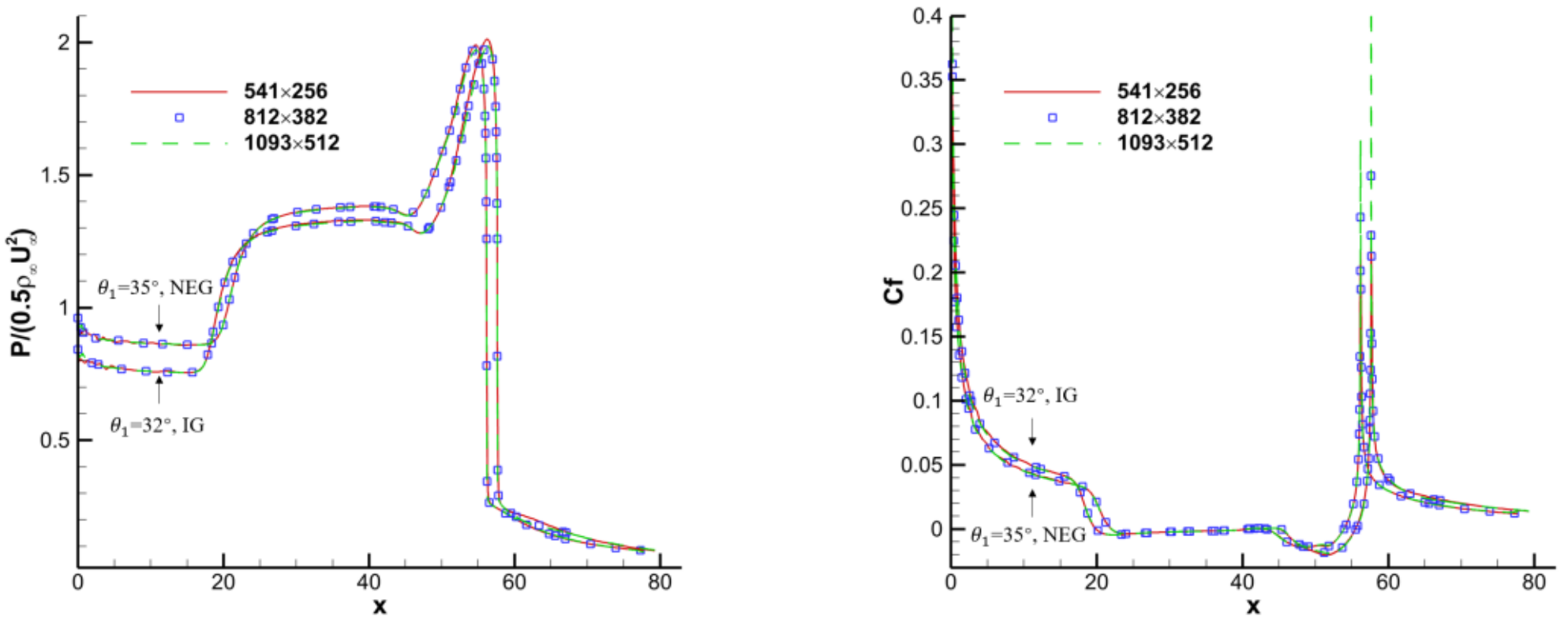

(a) Wall pressure coefficients (b) Skin friction

Fig. 7 Distributions of wall aerodynamic forces using different gas models on three different grids

Fig. 8 depicts the convergence curves of the drag coefficient of the two gas models on medium grids to illustrate the convergence of steady-state calculations. It indicates that as the number of iterations increases, the distribution of $C_D$ initially decreases, then oscillates slightly before leveling off. This behavior demonstrates the convergence characteristics of the calculations, indicating that a steady-state solution can be obtained after sufficient iterations.

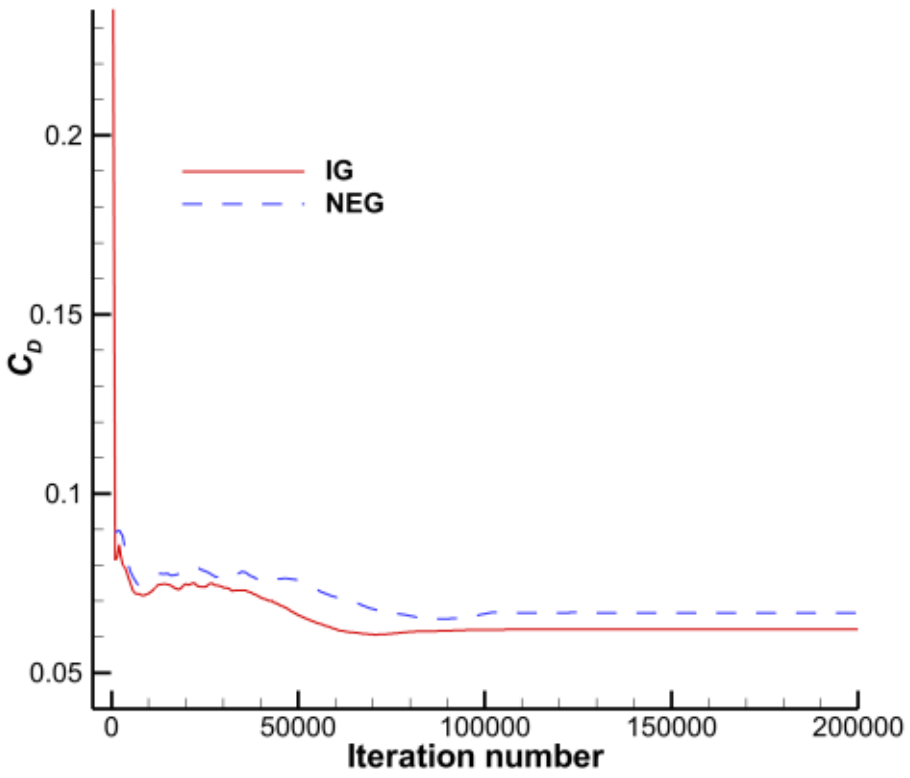

Fig. 8 Converging histories of $C_D$ of two gas models at $\theta_1 = 32°$ (IG) and $\theta_1 = 35°$ (NEG) in grid convergence studies on the chosen $812 \times 382$ grids

In addition to using wall parameters, the grid convergence is also qualitatively assessed from the perspective of flow structures. Fig. 9 shows the numerical schlieren of the IG model on three different grids, where the red solid lines denote sonic lines. The shock outlines from the medium grids ($812 \times 382$) are extracted as a reference and superimposed with solid circles in the other results to compare structural differences. The coarse and fine grids, in terms of shock structures, exhibit similar structures to the medium grids, with the intersection points of the separation shock and leading-edge shock, as well as the triple-point locations, being nearly identical. Regarding the shape of the sonic lines, the medium and fine grids show qualitatively consistent shapes behind the bow shock, while the coarse grids' results differ somewhat. Therefore, grid convergence is achieved for the IG model at the medium grid resolution of $812 \times 382$.

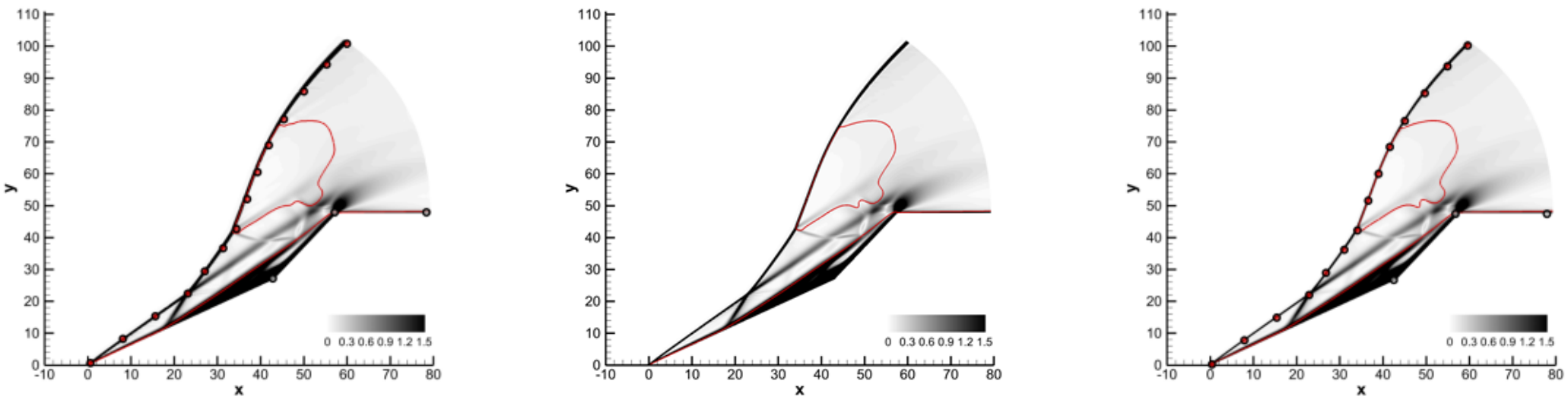

(a) 541 × 256 (b) 812 × 382 (c) 1093 × 512

Fig. 9 Numerical schlieren of IG model on three different grids (Red solid lines denote sonic lines, and solid circles represent the shock outlines extracted from the medium grids)

Fig. 10 illustrates the numerical schlieren of the NEG model on three different grids, with solid circles marking the shock outlines extracted from the medium grids. The differences in interaction structures and sonic line distributions among the three grids are minimal. The medium grids were selected for the following calculations, considering the need to resolve complex flow structures in the main study.

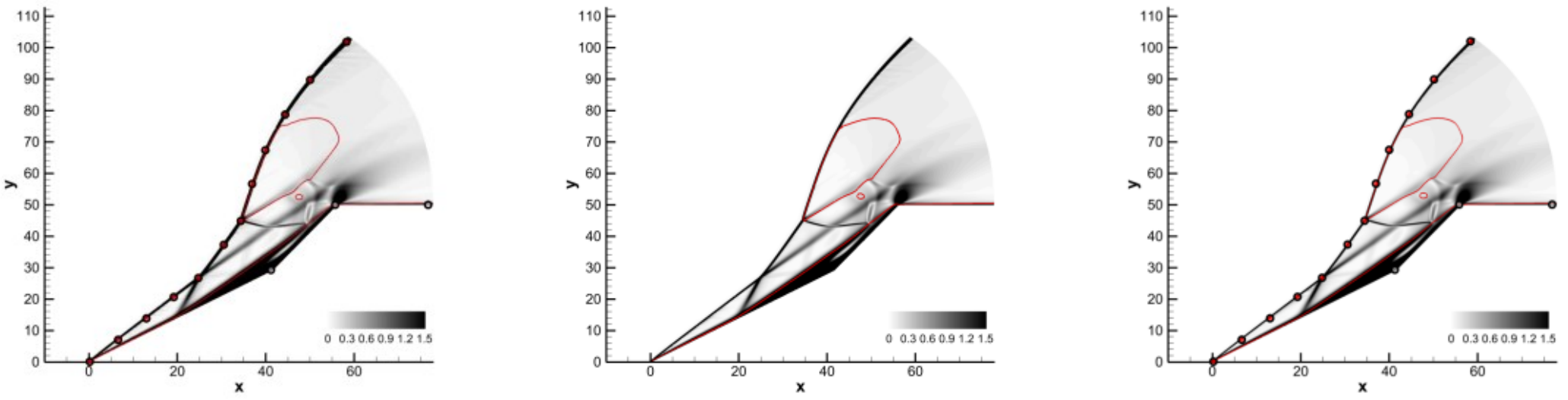

(a) 541 × 256 (b) 812 × 382 (c) 1093 × 512

Fig. 10 Numerical schlieren of the NEG model on three different grids (Red solid lines denote sonic lines, and solid circles denote the shock outlines extracted from the medium grids)

Based on the above grid convergence study, the grid number for the following variable $\theta_1$ configurations set to $n_{\text{streamwise}} \times n_{\text{normal}} = 812 \times 382$, with the first normal grid interval set to 0.001 $mm$ and an interval growth rate of 1.12. Numerical simulations were conducted for the specified geometry, grids, and conditions, and the results are discussed in the following sections.

## 3.2 Evolution of steady interaction characteristics and sudden structural change

### 3.2.1 Steadiness vs. unsteadiness in the evolution of interactions with varying $\theta_1$

Calculations for double wedge configurations with $\theta_1$ ranging from 0° to 40° at 5° intervals were first performed based on the computational conditions described ahead. The study revealed that the flow becomes unsteady at $\theta_1 = 30°$. Additional cases were generated at 1° intervals within the ranges $\theta_1 = 25°$ to 30° and $\theta_1 = 30°$ to 35° to further identify the range of $\theta_1$ where unsteadiness occurs. A diagram indicating whether the flow will be steady for the two gas models can be obtained based on the computational results, as shown in Fig. 11. It demonstrates that the minimum boundary angle for unsteadiness in the IG model is $\theta_{1,uns}^{min} = 28°$, and the maximum boundary angle is $\theta_{1,uns}^{max} = 30°$. For the NEG model, $\theta_{1,uns}^{min} = 30°$ and $\theta_{1,uns}^{max} = 33°$. When $\theta_1 < \theta_{1,uns}^{min}$ or $\theta_1 > \theta_{1,uns}^{max}$, the flow remains steady.

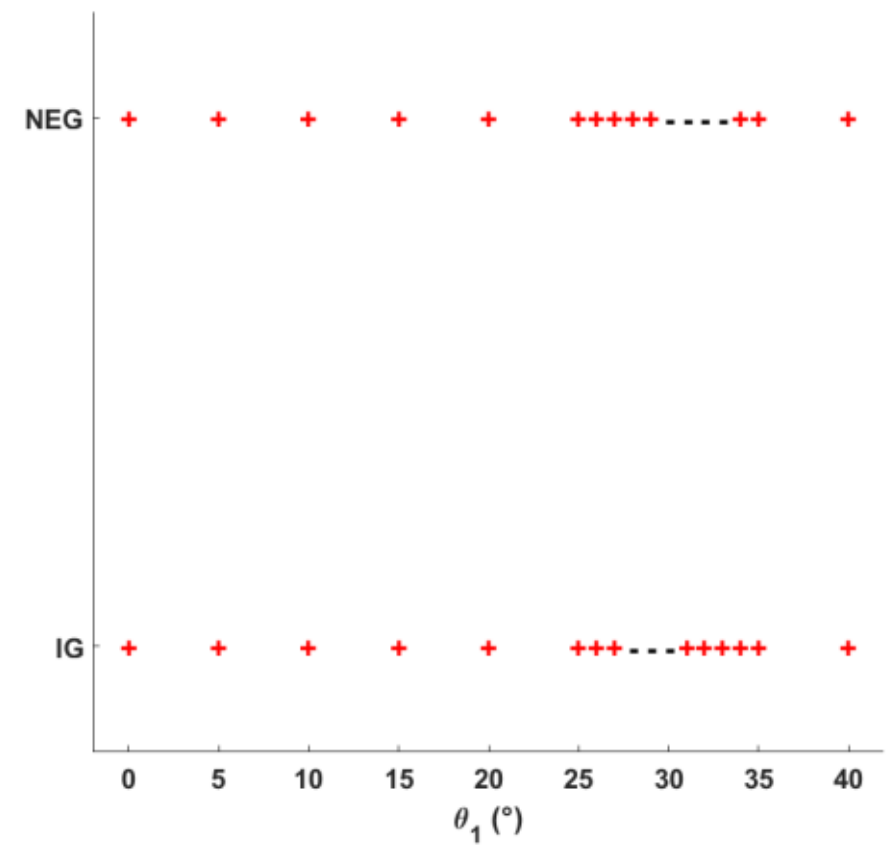

Fig. 11 Flow steadiness vs unsteadiness of two gas models at different $\theta_1$ ("+" depicts steady flow; "-" depicts unsteady flow)

The IG results were initially utilized as an example to show the changes in interaction structures for $\theta_1 = 26° \rightarrow 27°$ and $\theta_1 = 32° \rightarrow 31°$ and qualitatively illustrate the changes in the flow field as $\theta_1$ increases toward $\theta_{1,uns}^{min}$ and decreases toward $\theta_{1,uns}^{max}$, as depicted in Fig. 12. When $\theta_1 = 26°$ for the case of $\theta_1 \rightarrow \theta_{1,uns}^{min}$, the transmitted shock originating from the triple-point passes through the slip line and expansion waves around the expansion corner, eventually impinging the third wedge. The impingement position of the transmitted shock and the wedge surface shift leftward and stop near the expansion corner's right side during the transition from $26° \rightarrow 27°$. Further increasing $\theta_1$ leads to the onset of unsteadiness in the flow field. When $\theta_1 = 32°$ for the case of $\theta_1 \rightarrow \theta_{1,uns}^{max}$, the transmitted shock impinges the shear layer of separation on the second wedge surface. The triple-point moves backward, and the impingement position shifts toward the expansion corner, reaching its closest proximity to the corner when $\theta_1 = 31°$ during the transition from $32° \rightarrow 31°$. The results indicate that the onset of unsteadiness correlates to the impingement location of the transmitted shock. When the transmitted shock approaches the expansion corner from both sides, the flow structure loses stability, causing the transition from steady to unsteady flow.

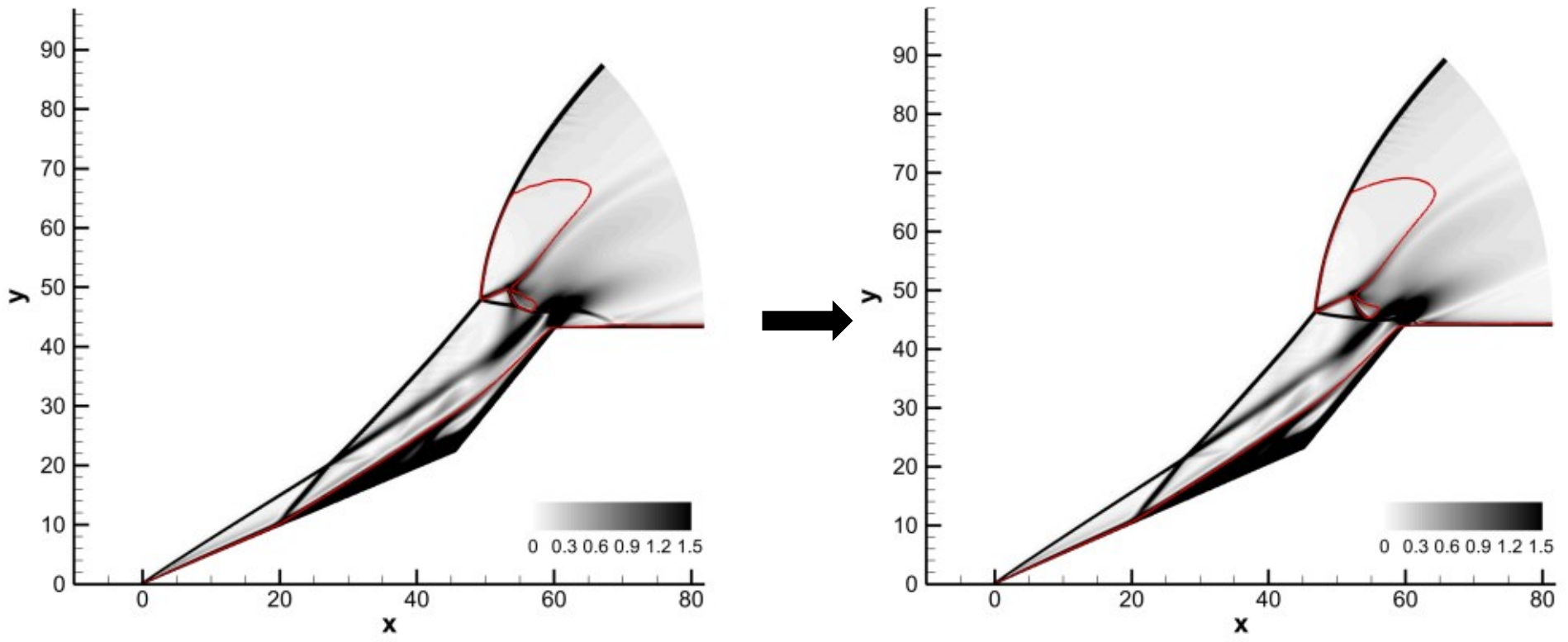

(a) $\theta_1 = 26° \rightarrow 27°$

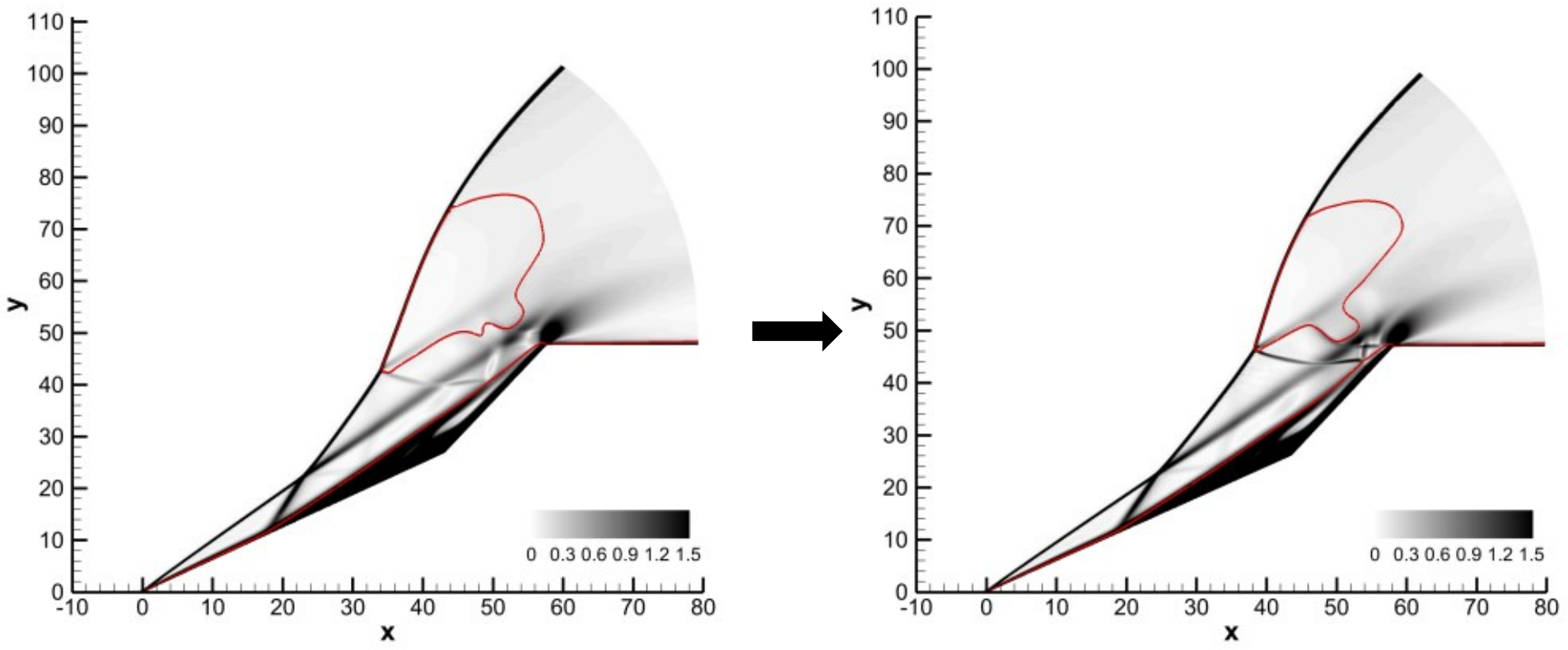

(b) $\theta_1 = 32° \rightarrow 31°$

Fig. 12 Evolution of the interaction structures as $\theta_1$ approaches the boundary angles for unsteadiness from both sides in the IG case

Fig. 13 shows the changes in interaction structures for the NEG model at $\theta_1 = 28° \rightarrow 29°$ and $\theta_1 = 35° \rightarrow 34°$. The impingement point of the transmitted shock approaches the expansion corner from both sides during the angular variation. When the distance between the point and the expansion corner shortens to a certain extent, the flow field becomes unsteady, which is consistent with the behavior observed in the IG flow.

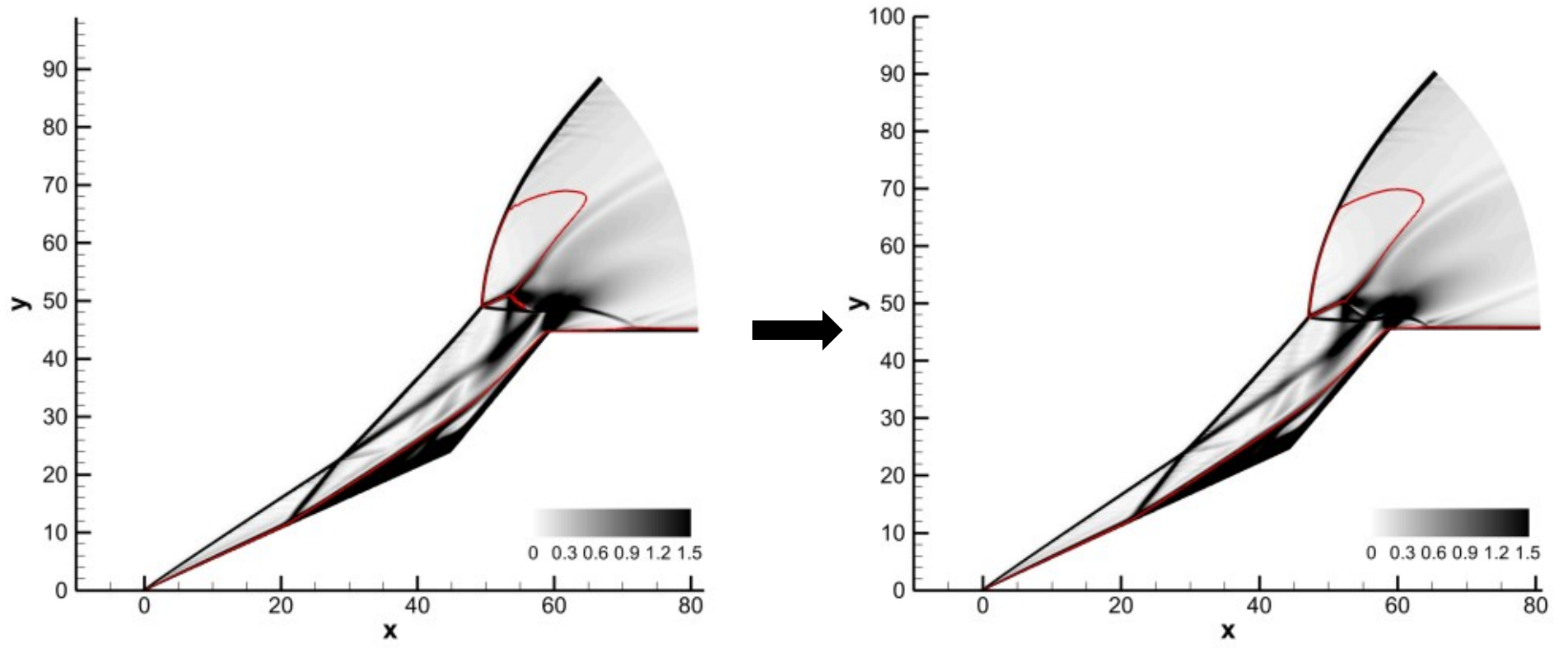

(a) $\theta_1 = 28° \rightarrow 29°$

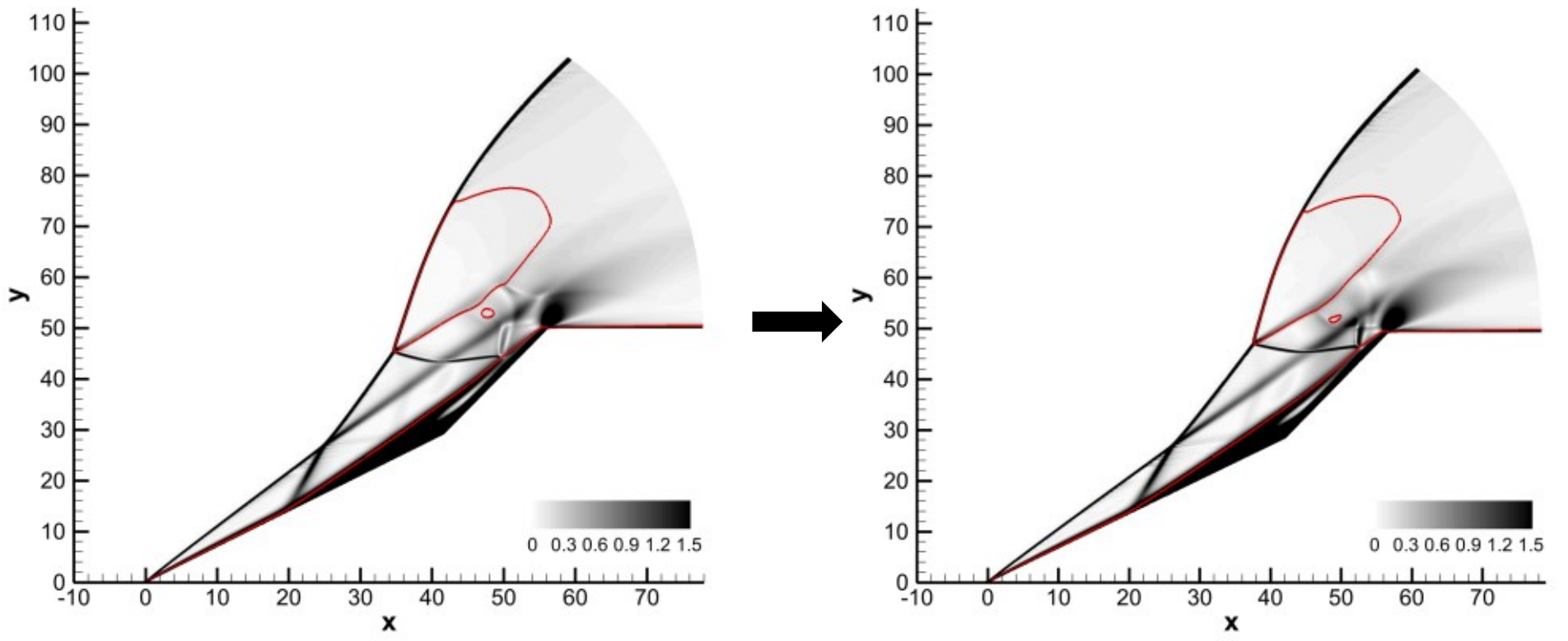

(b) $\theta_1 = 35° \rightarrow 34°$

Fig. 13 Evolution of the interaction structures as $\theta_1$ approaches the boundary angles for unsteadiness from both sides in the NEG case

### 3.2.2 Evolution of global interaction characteristics and sudden structural change

Previous sections investigated the characteristics of the flow field evolving toward unsteadiness. As discussed, unsteadiness occurs when $\theta_{1,uns}^{min} \leq \theta_1 \leq \theta_{1,uns}^{max}$, and its onset is related to the position of the transmitted shock impingement. The following studies the interaction characteristics of the flow evolution in steady state, focusing on the following two aspects: First, representative structures from different gas models with the increase of $\theta_1$ are briefly described to gain a more intuitive understanding of the evolution of interaction structures with varying $\theta_1$. Second, the shock waves and separation structures are analyzed as the primary subjects, and the corresponding flow fields are examined to investigate the changes in interaction characteristics with $\theta_1$.

(1) Overview of the evolution of interaction structures with $\theta_1$

This study selects the cases of NEG with $\theta_1 = 0°$, 20°, 27°, 34°, 40° and IG with $\theta_1 = 40°$ using the current computations to illustrate typical structures present in the flows. Fig. 14 depicts the density contours of the corresponding angles, where the black solid lines demote sonic lines, and IP1, IP2, and IP denote shock intersection points, which will be explained later. The results reveal the following: (a) At $\theta_1 = 0°$, the separation zone is at its maximum extent; the separation shock (SS) intersects the leading-edge shock (LS) at IP1. The combination of two shock waves (CS) extends to the right. Eventually, CS intersects the reattachment shock (RS) at IP2, generating a new shock; (b) At $\theta_1 = 20°$, the separation zone shrinks. The bow shock (BS), RS, oblique shock (OS) and CS intersect at IP2, with a subsonic region appearing downstream of IP2; (c) At $\theta_1 = 27°$, IP2 transforms into a triple-point, and RS degenerates into a compression wave (CW). The transmitted shock (TS) emitted from IP2 passes through CW and expansion waves, landing on the third wedge surface. Another small subsonic region forms after the intersection of TS and CW; (d) At $\theta_1 = 34°$, the subsonic region behind BS expands further, and the distance between IP2 and IP1 shortens. The flow field develops two slip lines, SL1 and SL2, originating from IP1 and IP2. TS crosses SL1 and intersects the end of separation on the aft wedge surface; (e) In the NEG model, at $\theta_1 = 40°$, CS disappears, and IP1 and IP2 merge into IP. A supersonic channel forms between the subsonic region downstream of IP and the boundary layer; (f) Fig. 14(f) indicates that another pattern is observed in the IG flow at $\theta_1 = 40°$, where BS moves to the tip of the double wedge, and the subsonic region covers most of the flow field.

The NEG results exhibit that when $\theta_1$ varies from 0° to 34°, the main interaction structures remain similar, including CS, RS (or CW), and BS. During this process, the structures evolve relatively continuously, and changes such as the upstream movement of the triple-point and the reduction in the separation zone scale can be expected as $\theta_1$ varies. However, when $\theta_1$ changes from 34° to 40°, the profile of the boundary layer and the wave system of the interactions differ significantly from those of the previous cases. The main interaction structures, composed of LS, BS, TS, and the supersonic channel, arise during the development of the flows. This phenomenon refers to a sudden structural change where the main interaction changes and new structures emerge. Additional computations were performed at 1° intervals within the range $\theta_1 = 35°$ to 40° to clarify the evolution of interaction patterns during this structural transition. These results will be discussed in detail in the next section.

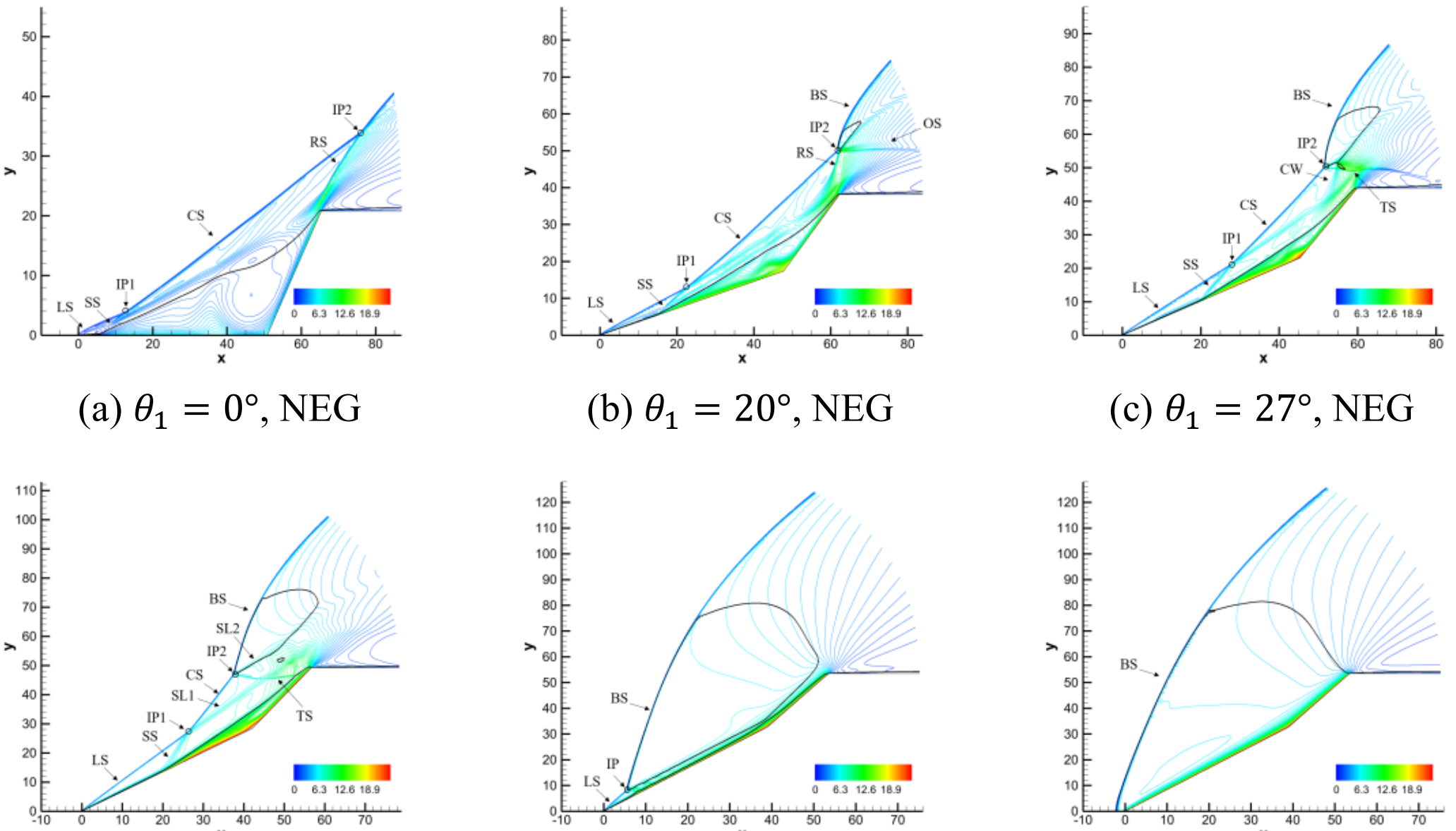

(a) $\theta_1 = 0°$, NEG (b) $\theta_1 = 20°$, NEG (c) $\theta_1 = 27°$, NEG

(d) $\theta_1 = 34°$, NEG (e) $\theta_1 = 40°$, NEG (f) $\theta_1 = 40°$, IG

Fig. 14 Density contours for selected $\theta_1$ in NEG and IG results (Black solid lines denote sonic lines, IP1, IP2, and IP denote shock intersection points, LS denotes leading-edge shock, SS is the separation shock, CS is the combination of two shock waves, RS is the reattachment shock, BS denotes bow shock, OS is the oblique shock, CW is the compression wave, TS is the transmitted shock, and SL1 and SL2 are slip lines)

This study separately analyzes shock structures and flow separation as the primary subjects, examining their behavior under different gas models to further investigate the quantitative characteristics of interactions with varying $\theta_1$. In addition, the phenomenon of sudden structural change in structure is also studied.

(2) Quantitative variations in shock characteristics with $\theta_1$ and sudden structural change

First, the trajectories of IP1, IP2, and IP were investigated. A new coordinate system $X'O'Y'$ was established with the first wedge surface as the horizontal axis, while the coordinate origin was fixed as the original $O'$ to represent the relative position relationship between the shock intersection points and the wedge surfaces:

$$x' = x\cos(\theta_1) - y\sin(\theta_1), y' = x\sin(\theta_1) + y\cos(\theta_1) \tag{5}$$

The trajectories of IP1, IP2, and IP in these transformed coordinates are shown in Fig. 15, where dashed lines denote unsteady flow. Some trajectory points are labeled with their corresponding $\theta_1$ for illustration. When $\theta_1 \leq 25°$, the angle difference between adjacent trajectory points is 5°, while for $\theta_1 > 25°$, the angle difference is 1°. Fig. 15(a) indicates that as $\theta_1$ increases, IP2 primarily moves from the upper right to the lower left along the negative directions of both coordinate axes, with the trajectory positions of IG having higher $y'$ compared to those of NEG. Fig. 15(b) demonstrates that a magnified view of the region $x' \in [6, 36]$ is provided. IP1 moves to the right as $\theta_1$ increases when $\theta_1 < \theta_{1,uns}^{min}$, and moves to the left as $\theta_1$ increases when $\theta_1 > \theta_{1,uns}^{max}$, resulting in a loop-shaped trajectory. The trajectory of NEG with $\theta_1$ is consistent with those of IG but with a larger trajectory range indicated. The trajectory of IP, formed by the merging of IP1 and IP2, further moves toward the origin $O'$ as $\theta_1$ increases in the results of two gas models.

After the flow transitions from unsteadiness to steadiness in the case of IG, the increase of 1°of $\theta_1$ will yield a larger distance between trajectory points of IP1 and IP2, respectively, than those when $\theta_1 < \theta_{1,uns}^{min}$. This indicates that the shapes and positions of LS, SS, BS, and TS in the flow field are changing rapidly. This phenomenon is called sudden change and marks the rapidly changing regions of trajectory points in Fig. 15(a) and (b) with "Sudden change" and arrows, covering the range of $\theta_1$ from 31° to 34°. During $\theta_1 = 33° \to 34°$, IP1 and IP2 merge, and the flow undergoes the previously mentioned sudden structural change, resulting in the structure illustrated in Fig. 14(e). The details of this transition will be discussed in Section 3.3. Based on the trajectory results, for IG cases, the angle range of the sudden change is $\theta_1 = 31° - 34°$, while the sudden structural change occurs during $\theta_1 = 33° \to 34°$; for NEG cases, the range is $\theta_1 = 34° - 38°$, and the sudden structural change occurs when $\theta_1 = 37° \to 38°$.

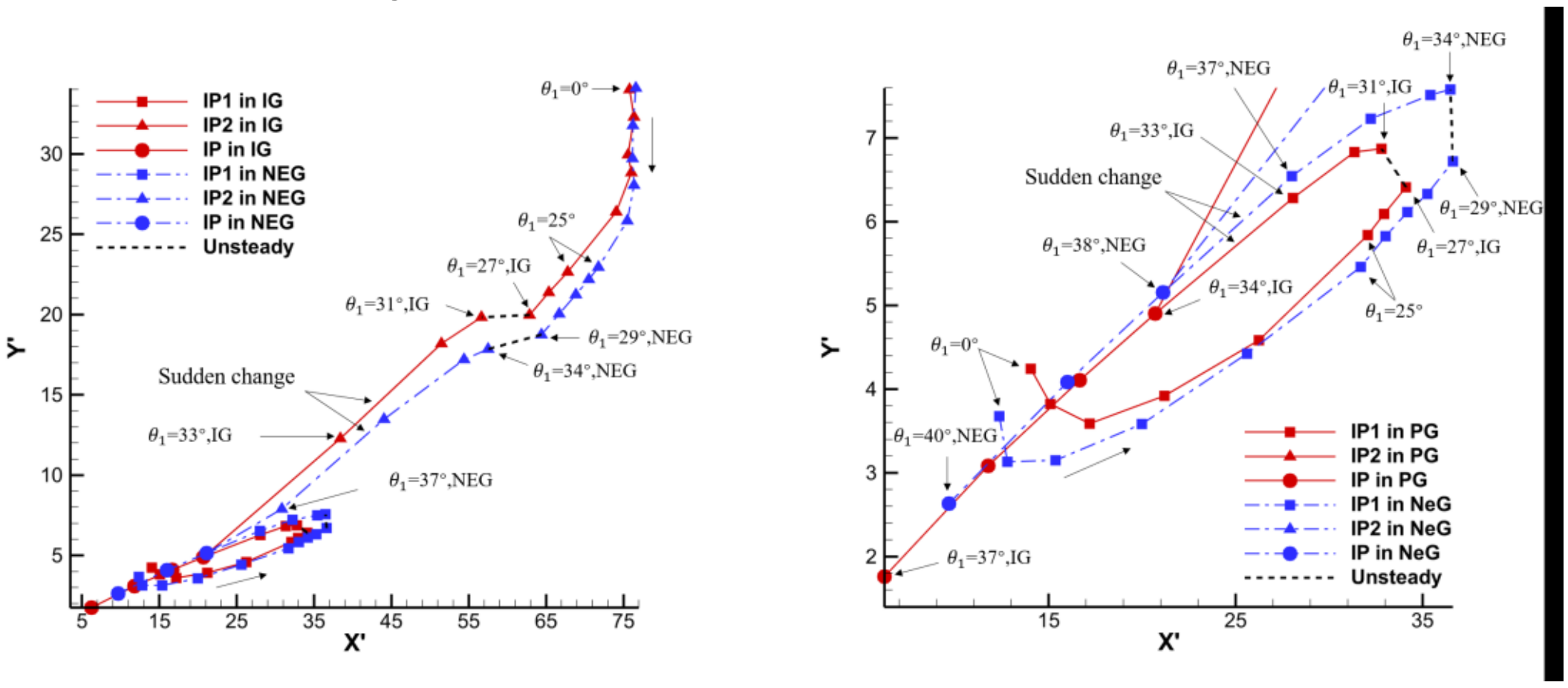

(a) Overall view (b) Enlarged view

Fig. 15 Trajectories of IP1, IP2, and IP with varying $\theta_1$ in different gas models (some trajectory points are labeled with their corresponding $\theta_1$)

Fig. 16 shows the variation of the distance between trajectory points of IP1 and IP2 with $\theta_1$, with the regions of sudden change marked in the figure to quantitatively illustrate the characteristics of the sudden change. When $\theta_1 < \theta_{1,uns}^{min}$, the entire curve decreases as $\theta_1$ increases, with the absolute value of the slope gradually increasing. When $\theta_1 > \theta_{1,uns}^{max}$, within the regions of sudden change, the absolute value of the slope increases, and the distances rapidly decrease to zero, indicating the coalescence of IP1 and IP2 into IP. The degree of change is significantly different from that observed when $\theta_1 < \theta_{1,uns}^{min}$.

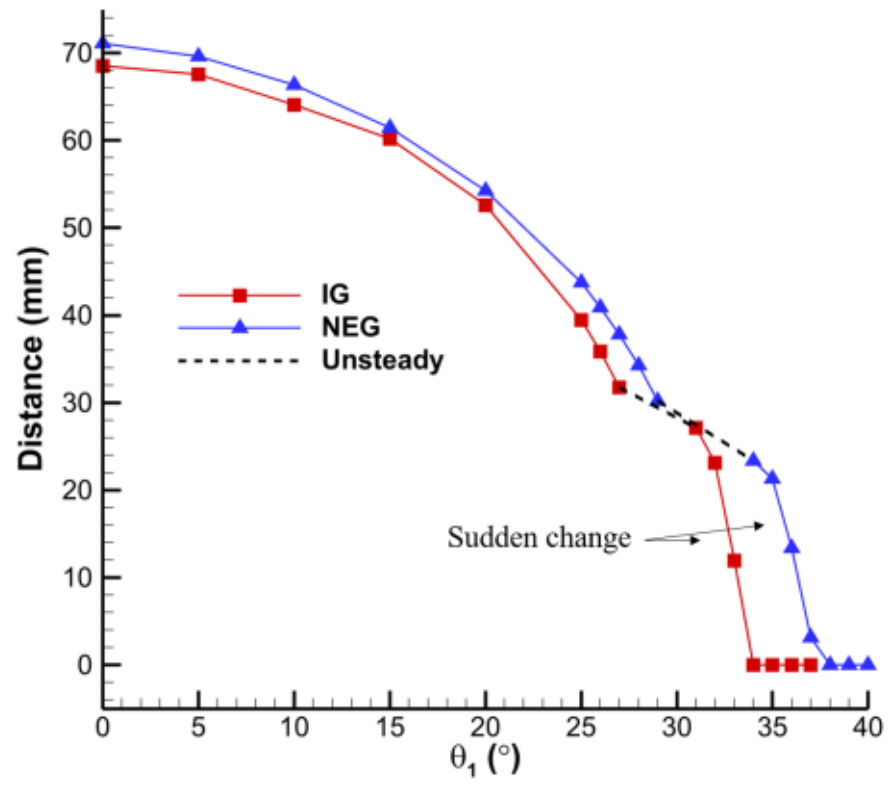

Fig. 16 Variation of distances between trajectory points of IP1 and IP2 with $\theta_1$

The following discusses the variation of shock angles with $\theta_1$. Fig. 17 illustrates the distributions of angles of LS and SS with $\theta_1$ in two gas models. First, those of LS are concerned. The angle of LS in the IG case decreases with $\theta_1 = 0 \to 20°$, reaching a minimum at $\theta_1 = 20°$, and then increases from $\theta_1 = 20° \to 27°$ until the flow becomes unsteady. When $\theta_1 \geq 31°$, the curve rises and reaches a maximum at $\theta_1 = 37°$. Afterward, with a further increase of 1°, LS disappears, and the flow field is dominated solely by a bow shock. For simplicity, the subsequent angles are referred to as 0°. The trend of the NEG distribution is consistent with that of IG but with smaller angles. In addition, the theoretical prediction of the LS angle under inviscid conditions is derived for comparison. The results show that when $\theta_1 \geq 20°$, the trend of the inviscid curve resembles that of the two gas models despite quantitative differences attributed to the influence of the boundary layer. Only the angles before the sudden structural change are measured for SS when large-scale vortices still exist in the flow field. The curve trends for $\theta_1 < \theta_{1,uns}^{min}$ and $\theta_1 > \theta_{1,uns}^{max}$ are similar to those of LS, but SS reaches its minimum angle at $\theta_1 = 15°$ for IG and $\theta_1 = 10°$ for NEG.

Fig. 18 shows the variation of the CS angle, which includes the absolute angle between CS and the incoming flow and the relative angle between CS and the wedge surface. It indicates that the absolute angle increases with $\theta_1$ until CS disappears, and for simplicity, its angle is represented as 0°. The angles for NEG are generally smaller than those for IG. The relative angle shows slight variation when $\theta_1 < \theta_{1,uns}^{min}$, with its distribution nearly horizontal. However, when $\theta_1 > \theta_{1,uns}^{max}$, due to the influence of the sudden change, the relative angle distribution begins to increase, and the magnitude of the angle change becomes significantly larger.

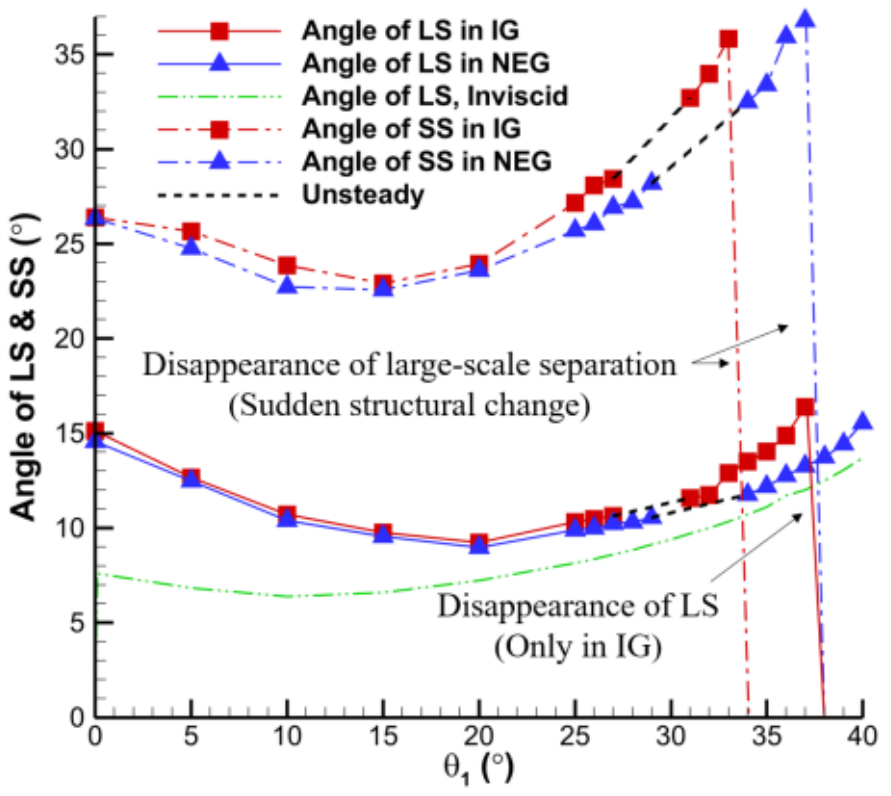

Fig. 17 Distributions of angles of LS and SS with $\theta_1$ in different gas models

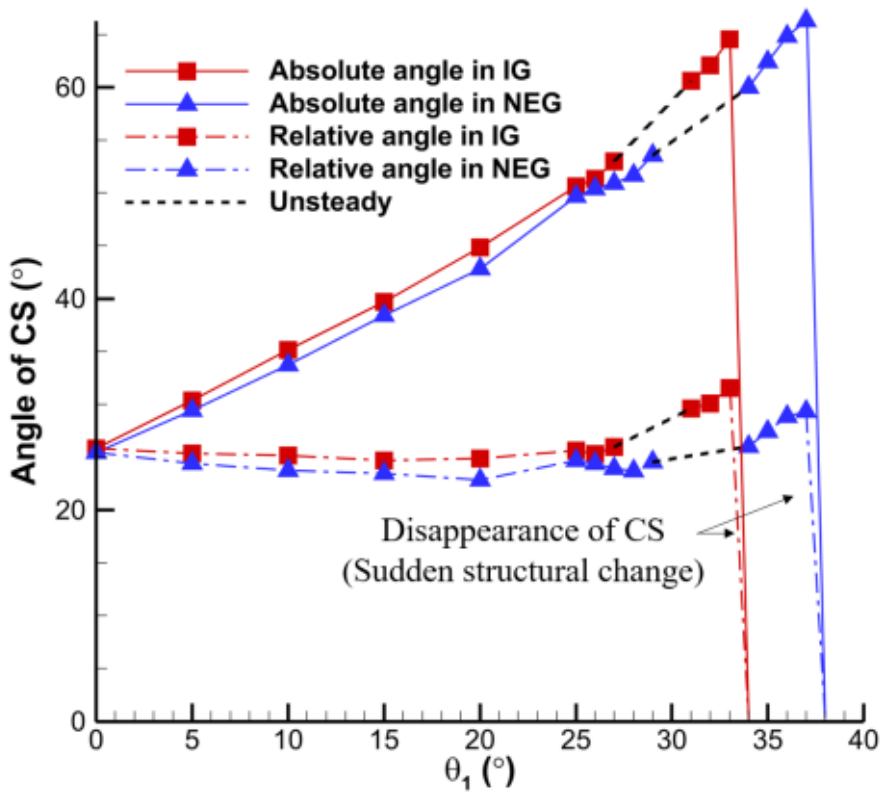

Fig. 18 Distributions of absolute angles and relative angles of CS with $\theta_1$ in different gas models

(3) Quantitative variations in separation characteristics with $\theta_1$ and sudden structural change

Previously, this study identified the range of angles where sudden change occurs using shock intersection trajectories and shock angles as the analysis subjects. Parameters related to the interaction structures within this range undergo rapid changes, culminating in the sudden structural change. This section focuses on large-scale separation characteristics, with the selected parameters including the distance from the separation point to the compression corner ($D_{sc}$), the distance from the reattachment point to the compression corner ($D_{rc}$), the distance between the separation and reattachment points ($D_{sr}$), and the separation angle (SA) and reattachment angle (RA). Figs. 19(a) and (b) exhibit the variations of $D_{sc}$ and $D_{rc}$ with $\theta_1$. When $\theta_1 < \theta_{1,uns}^{min}$, $D_{sc}$ and $D_{rc}$ decrease as $\theta_1$ increases, with the separation and reattachment points moving closer to the compression corner. At the same angles, IG's separation and reattachment points are closer to the compression corner than in NEG. When $\theta_1 > \theta_{1,uns}^{max}$, the values of $D_{sc}$ and $D_{rc}$ are greater than their minimum values at $\theta_1 = \theta_{1,uns}^{min} - 1°$, indicating that the separation and reattachment points move away from the compression corner during the transition back to steady flow. After the sudden structural change, the large-scale vortices disappear, and for simplicity, $D_{sc}$ and $D_{rc}$ are set to 0.

Fig. 19(c) shows the variation curve of $D_{sr}$ to characterize changes in the separation zone scale. When $\theta_1 < \theta_{1,uns}^{min}$, $D_{sr}$ decreases as $\theta_1$ increases; when $\theta_1 > \theta_{1,uns}^{max}$, $D_{sr}$ begins to increase with $\theta_1$, showing a trend opposite to that observed when $\theta_1 < \theta_{1,uns}^{min}$.

Fig. 19(d) depicts the variations of SA and RA obtained by measuring the angles between the tangents and streamlines within the separation zone and the wedge surface. Since the angles in IG are generally larger, the shape of the separation zone in IG is more pronounced. When $\theta_1 < \theta_{1,uns}^{min}$, SA and RA decrease as $\theta_1$ increases, corresponding to a gradual flattening of the separation zone shape. When $\theta_1 > \theta_{1,uns}^{max}$, SA remains relatively stable with increasing $\theta_1$, while RA exhibits significant fluctuations, first sharply increasing to a peak and then gradually decreasing.

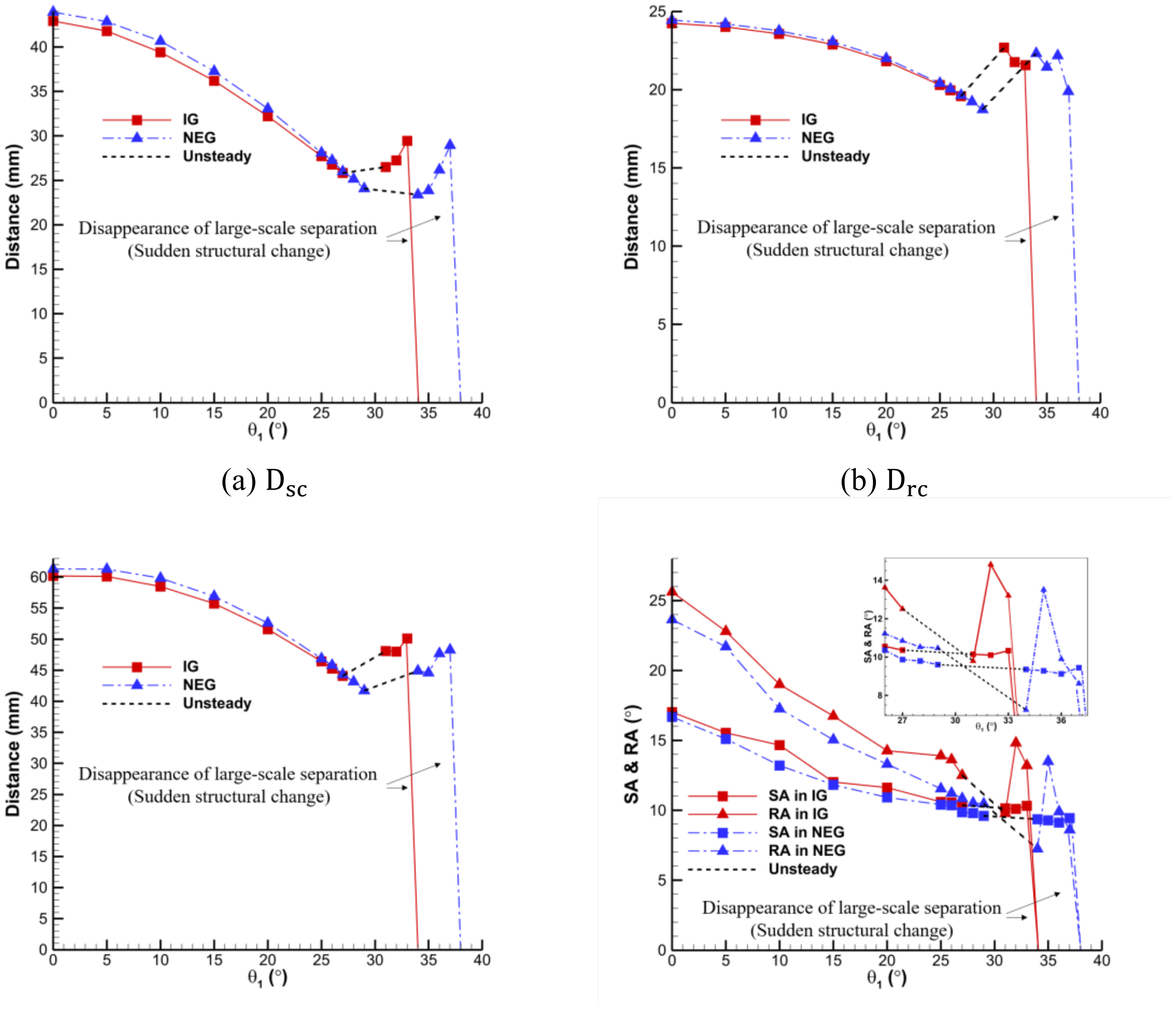

(a) $D_{sc}$ (b) $D_{rc}$

(c) $D_{sr}$ (d) SA/RA

Fig. 19 Variations in the separation characteristics with $\theta_1$ in different gas models

Accordingly, a brief overview of the evolution of the overall interaction structures was first provided, followed by an analysis of the evolution of interaction characteristics from the perspectives of shock structures and flow separation. The study revealed the presence of sudden changes in the flow field, during which the wave system structures rapidly shift. As the sudden change approaches its conclusion, the flow field undergoes a further sudden structural change. The following section discusses the evolution of the flow field interaction structures in greater detail, focusing on the changes during the sudden change process from the perspective of interaction patterns.

### 3.3 Analysis of interaction structures at different $\theta_1$

The previous section conducted a series of comparisons and analyses focusing on the evolution of the geometric characteristics of interactions with different $\theta_1$. This section investigates the evolution of flow field interaction patterns and the specific characteristics of sudden structural change. This study explores the structures, flow field parameters, and their variations within these patterns by dividing the range of angles corresponding to different interaction patterns.

A brief description of the characteristics of several typical interaction patterns involved in this study is provided to facilitate subsequent discussions. The first is Type VI interaction, as shown in Fig. 20(a). Shocks on the same side intersect to produce a combined shock (an oblique shock) and a slip line (SL). In addition, a transitional shock (red line) or expansion waves (dashed lines) can emerge from the intersection, or no significant structure can form. Thus, Type VI interaction has two sub-modes. Next is Type V interaction, illustrated in Fig. 20(b). Two incident shocks, BS and TS, form a multi-wave point in the flow field, from which a supersonic jet is emitted. The flow above and below the jet is subsonic, but below the lower subsonic region, the flow becomes supersonic. Type IV interaction consists of one incident shock, one oblique shock (or normal shock), BS, and TS, with its structure and velocity distribution shown in Fig. 20(c). TS reflects SL to form a supersonic jet, with subsonic flow above and below the jet. Finally, Type III interaction, as shown in Fig. 20(d), mainly comprises one incident shock, BS, and TS. Differing from the above typical interaction patterns, Olejniczak et al. [15] proposed two variants of these patterns. One is a transitional state during the transformation from Type VI to Type V interaction, characterized by a shock structure largely consistent with Type V but without forming a subsonic region below the supersonic jet (which can remain supersonic). This is called the Type VI–Type V transition process (which is abbreviated as Type VI → V interaction in this study). The other is a structure resulting from the further evolution of Type IV interaction, where TS directly intersects the wall and generates a subsequent jet, lacking the subsonic flow below the jet typical of standard interactions. The author termed this Type $IV_r$ interaction (Fig. 1). Both patterns will be referenced in the results.

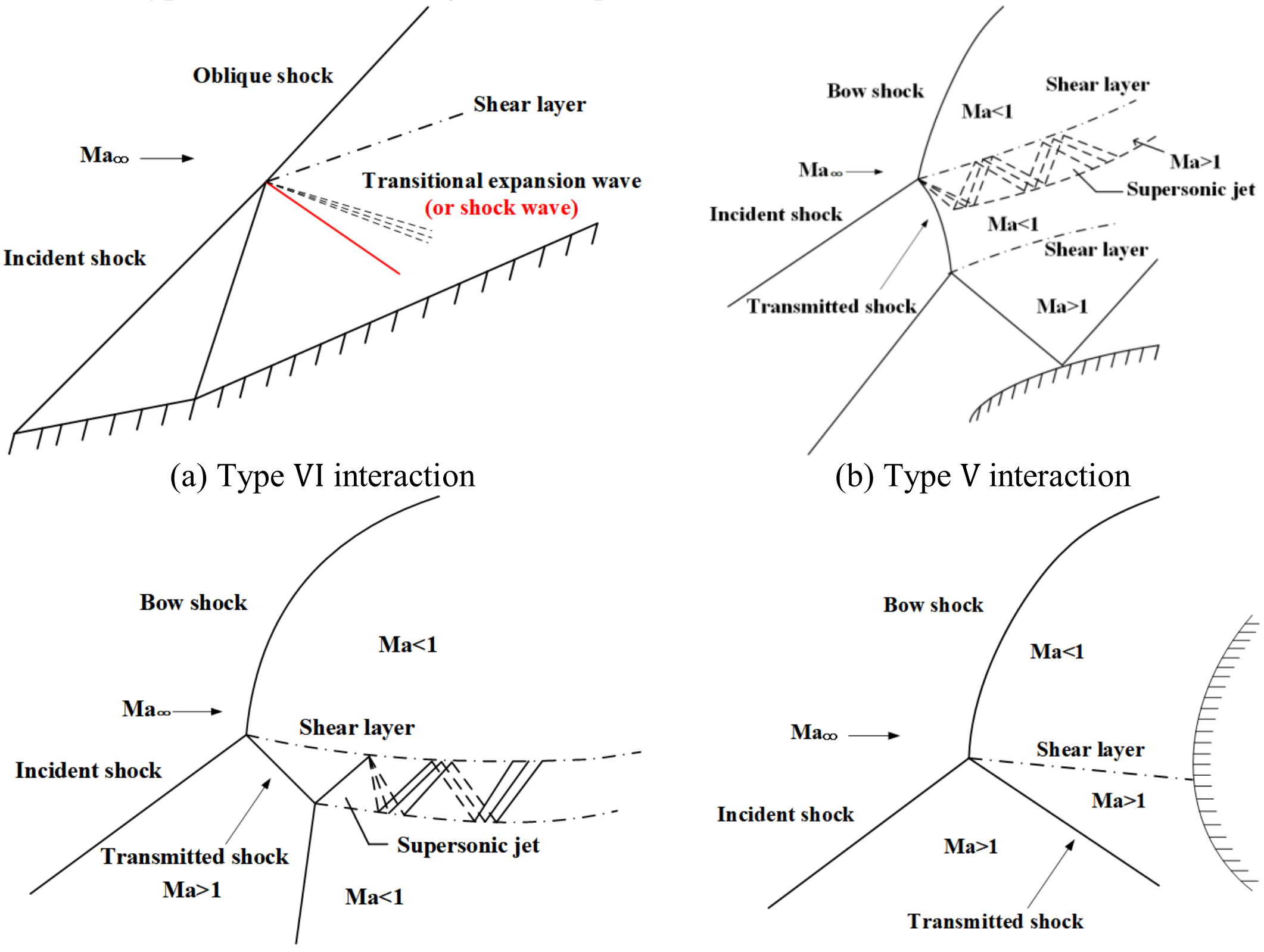

Fig. 20 Schematic diagrams of typical shock interactions, adapted from Edney [28]

Based on the four fundamental interaction patterns introduced above and the calculations conducted, the flow field interaction patterns can be classified. Additional calculations were performed for specific $\theta_1$ to determine the boundaries ($\theta_1$) for each interaction type and the resulting distribution of interaction patterns with respect to $\theta_1$ is presented in Fig. 21. This study describes the results in order of increasing $\theta_1$ as follows: (a) For small $\theta_1$, the interaction pattern is Type VI. "•" indicates the presence of transitional expansion waves or no significant structure between the shock wave, SL, and the wall (Fig. 20(a)), while "o" indicates the presence of transitional shocks. Transitions between these two cases occur within Type VI interaction. (b) As $\theta_1$ increases further, Type V transitions to Type VI → V interaction, denoted by "▲". (c) When $\theta_1 \in [\theta_{1,uns}^{min},\ \theta_{1,uns}^{max}]$, the flow is unsteady and represented by black dashed lines. (d) When $\theta_1 > \theta_{1,uns}^{max}$, the flow field returns to a steady state, and the interaction structure transitions to Type III and the following mentioned Type $III_r$ interaction, denoted by "▼". (e) As $\theta_1$ continues to increase, Type III transitions to a new interaction pattern similar to Type $IV_r$. For convenience, the study still refers to it as Type $IV_r$ interaction and denote it with "♦". The double wedge configurations and inflow conditions in this study differ from those in [15], primarily due to the consideration of flow viscosity. Although the new structure resembles Type $IV_r$, there are differences in detail. (f) In IG, when $\theta_1 \geq 38°$, only BS remains in the flow field, denoted by "×". The following discusses and analyzes the interaction patterns observed in the flow and their evolutionary mechanisms.

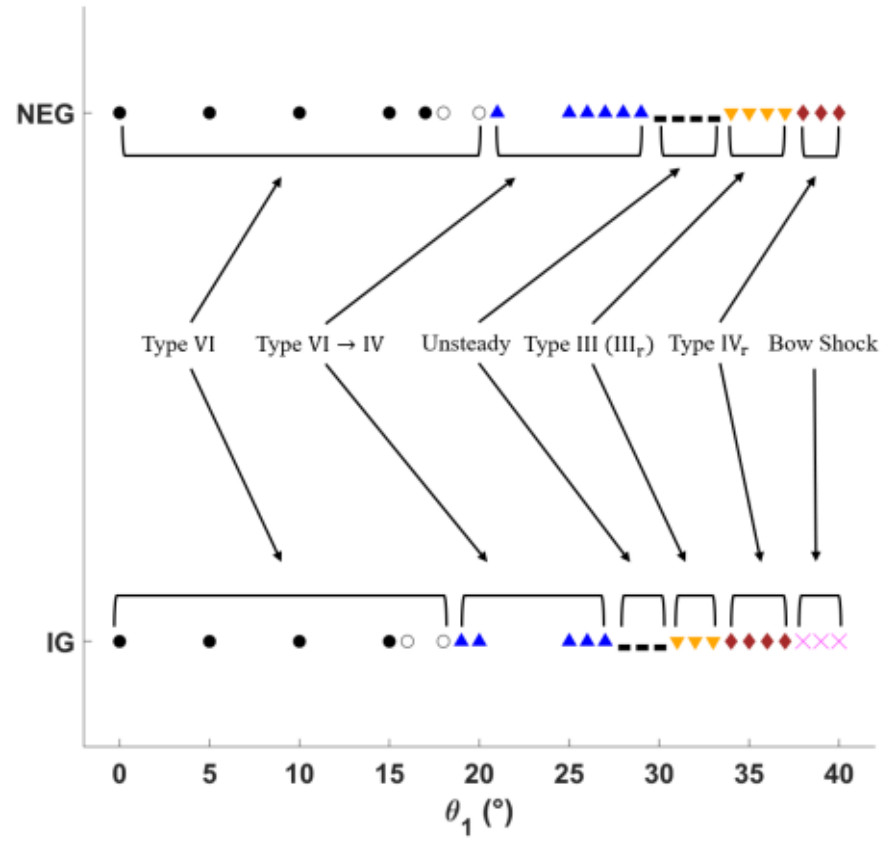

Fig. 21 Distribution of interaction patterns in the flow fields in different gas models and $\theta_1$ ("•" denotes Type VI interaction with transitional expansion waves, "o" denotes Type VI interaction with transitional shocks, "▲" denotes Type VI → V interaction, dashed lines denote unsteady flow, "▼" denotes Type III/Type $III_r$ interaction, "♦" represents Type $IV_r$ interaction, and "×" illustrates the flow field with only BS remaining)

3.3.1 Type VI interaction — corresponds to $\theta_1 = 0° - 18°$ (IG)/$\theta_1 = 0° - 20°$ (NEG)

Fig. 21 indicates that the IG flow field exhibits Type VI interaction when $\theta_1 = 0° - 18°$, with the transitional expansion waves transforming into transitional shock waves at $\theta_1 = 16°$. In NEG, Type VI interaction occurs at $\theta_1 = 0° - 20°$, with the transitional expansion waves transforming at $\theta_1 = 18°$. Fig. 22 presents pressure contours at $\theta_1 = 0°$, 10°, and 15°, where the black solid lines denote sonic lines to illustrate the evolution of the interaction structure before the transformation of the transitional expansion waves. LS and SS in the pressure contours intersect to produce CS, which then interacts with RS formed by CW, combining to form a new shock $CS_1$. This interaction pattern is consistent with the Type VI interaction shown in Fig. 20(a). As $\theta_1$ increases from 0° to 15°, the scale of the separation zone associated with the subsonic region decreases and flattens, and $CS_1$ transitions toward BS. However, the two gas models have no significant differences in shock shapes, intersection positions, or separation zone scale.

| $\theta_1 = 0°$ | $\theta_1 = 10°$ | $\theta_1 = 15°$ |
| --- | --- | --- |

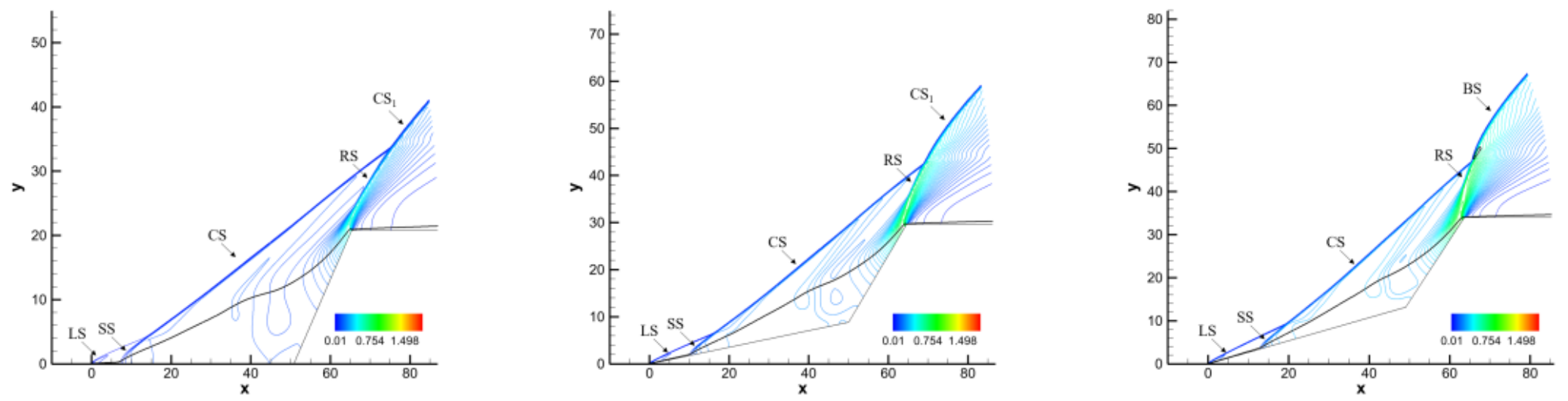

(a) IG

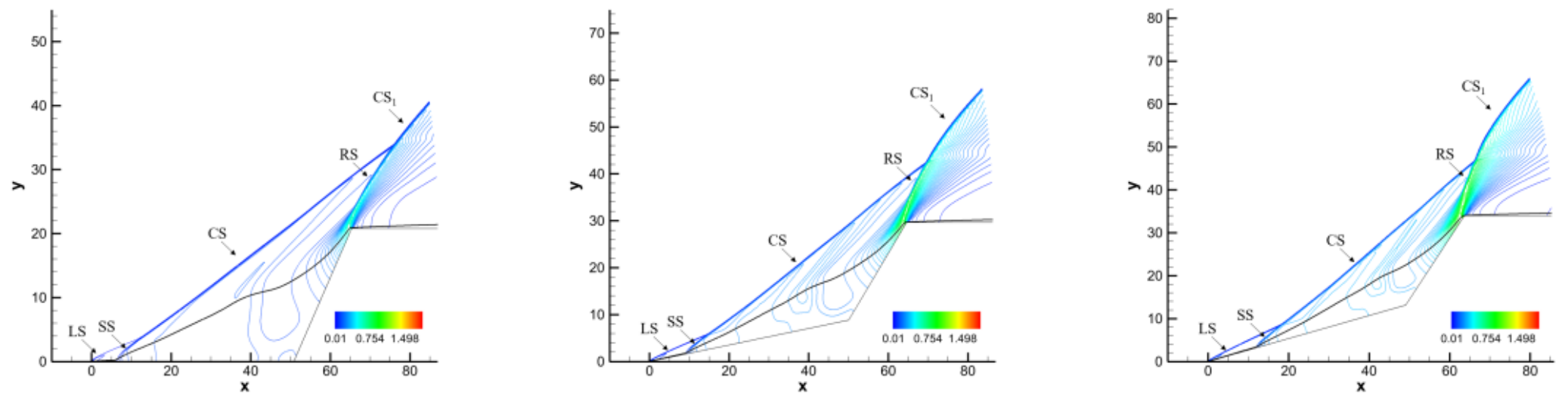

(b) NEG

Fig. 22 Pressure contours of Type VI interaction flow fields in different gas models and $\theta_1$ (Black solid lines denote sonic lines, LS represents leading-edge shock, SS is the separation shock, CS and $CS_1$ is the combination of two shock waves, RS is the reattachment shock, and BS is the bow shock)

Fig. 23 depicts numerical schlieren for $\theta_1 = 0° - 15°$ to visualize the flow field structures, where green solid lines with arrows denote streamlines, and red solid lines denote sonic lines. The streamlines in the numerical schlieren reveal the primary structures within the separation zone, including the main separation vortex on the right side of the corner, the secondary separation vortex near the wall, and the tertiary separation vortex on the left side of both. The scales of all three vortices gradually decrease as $\theta_1$ increases. In addition to the vortices, the numerical schlieren also displays CW over the separation zone, SL in the Type VI interaction, and white striped expansion waves (EW). CW and CS intersect without producing significant disturbances.

| $\theta_1 = 0°$ | $\theta_1 = 10°$ | $\theta_1 = 15°$ |
| --- | --- | --- |

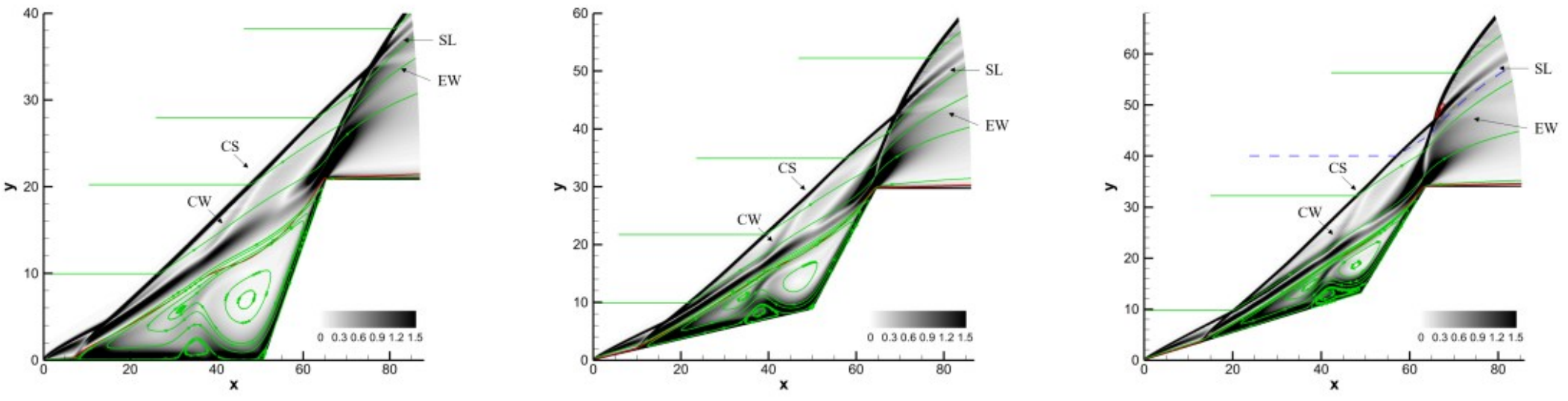

(a) IG

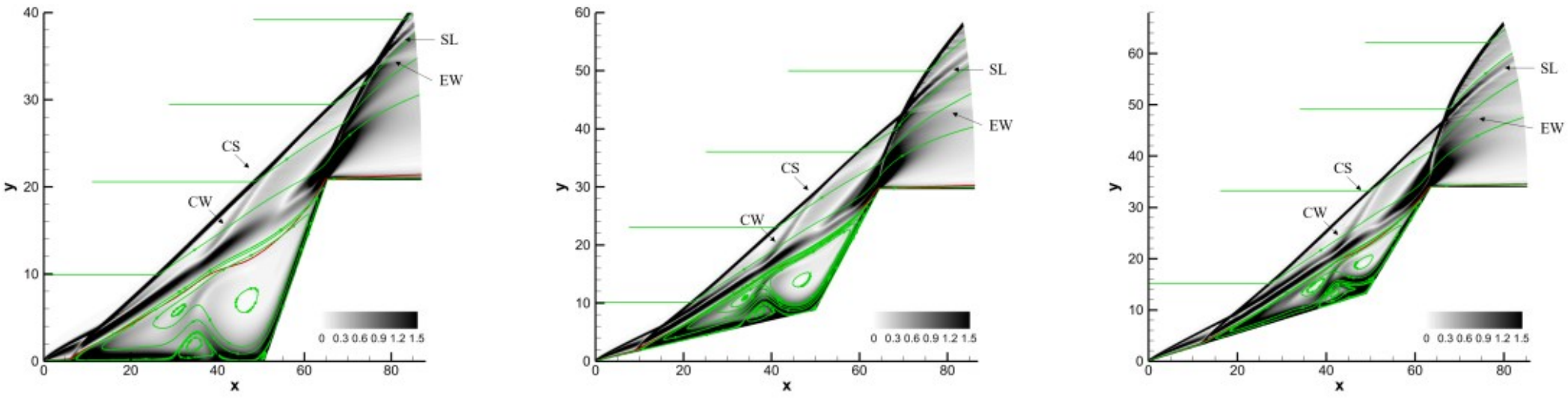

(b) NEG

Fig. 23 Numerical schlieren of Type VI interaction flow fields in different gas models and $\theta_1$ (Green solid lines denote streamlines, red solid lines denote sonic lines, CW is the compression wave, CS is the combination of two shock waves, SL is the slip line, EW is the expansion wave, and the blue dashed line indicates the streamline position used for data extraction)

As mentioned earlier, in Type VI interaction, further increasing $\theta_1$ causes transitional expansion waves to transform into transitional shocks. Figs. 24 and 25 illustrate numerical schlieren for IG and NEG at the moment of transition to illustrate this change. The results show that the white-striped structures in Fig. 23 become black streaks after the transition. The states before and after the transition for IG at $\theta_1 = 15°$ and 16° were selected to clarify the relationship between the streak structures and shocks. Streamlines passing through the expansion waves and shock waves are chosen (their positions are marked with blue dashed lines in the numerical schlieren), and the pressure distributions along these streamlines are extracted, as shown in Fig. 26. Focusing on the region after the pressure peak indicates that at $\theta_1 = 16°$, the curve exhibits a small rise during its descent. Measurements reveal that the rise corresponds to the intersection of the streamline and the black streak, a phenomenon absent at $\theta_1 = 15°$. Therefore, the rise is attributed to the influence of a weak shock, indicating the emergence of a weak transitional shock. The mechanism of the transition can be explained as follows: as $\theta_1$ increases from 15° to 16°, the subsonic region behind BS expands, and the pressure increases. The flow requires a weak shock for transition to achieve pressure balance across the SL.

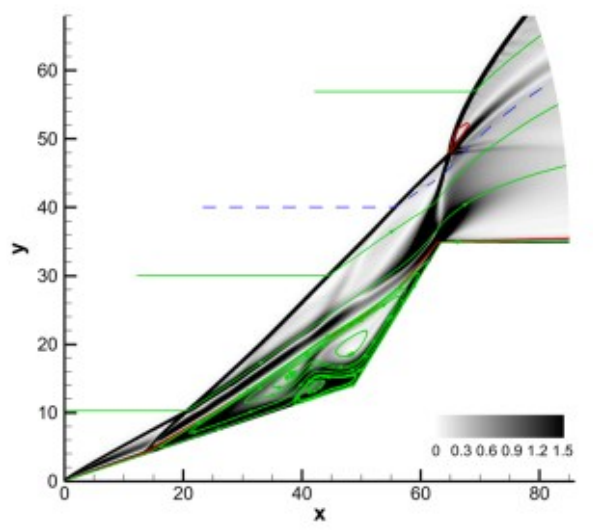

Fig. 24 Numerical schlieren of the interaction structure in IG at $\theta_1 = 16°$ after the transition (green solid lines denote streamlines, red solid lines denote sonic lines, and blue dashed line indicates the streamline position used for data extraction)

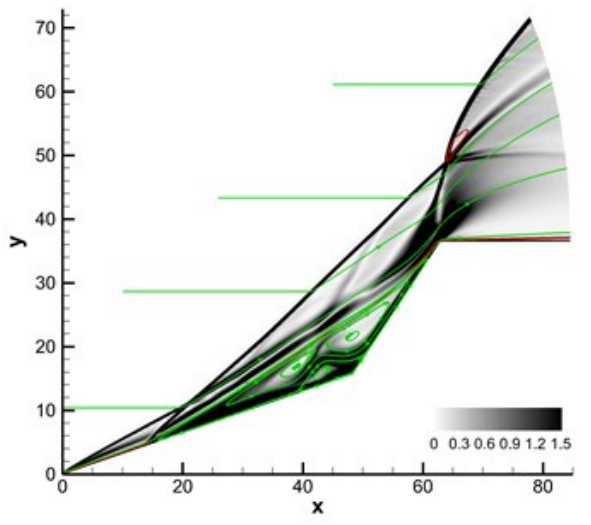

Fig. 25 Numerical schlieren of the interaction structure in NEG at $\theta_1 = 18°$ after the transition (green solid lines denote streamlines, and red solid lines denote sonic lines)

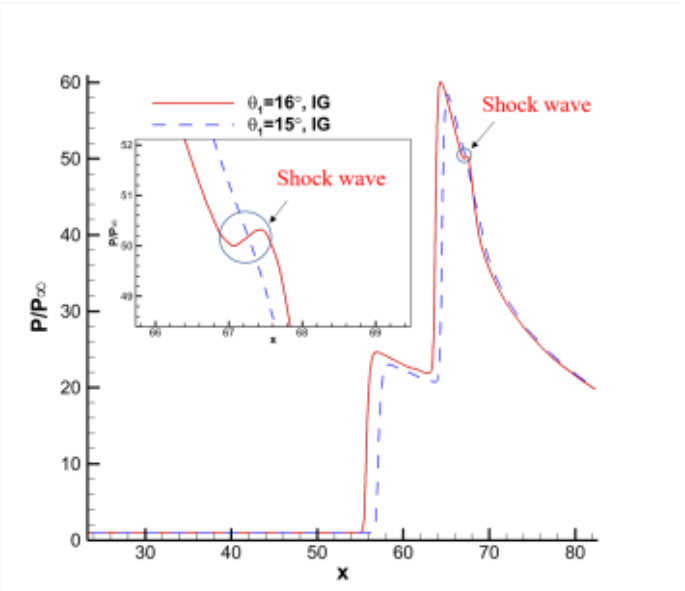

Fig. 26 Comparison of pressure distributions along the streamlines passing through expansion waves or shock waves in IG at $\theta_1 = 15°$ and 16°

3.3.2 Type VI → V interaction — corresponds to $\theta_1 = 19° - 27°$ (IG)/$\theta_1 = 21° - 29°$ (NEG)

As $\theta_1$ increases, the interaction pattern of the flow field transitions and generates new structures, as shown by the pressure contours in Fig. 27. Compared to Type VI interaction, the new interaction structure includes an additional TS, but unlike the standard Type V interaction, it lacks a supersonic jet/shear layer with a subsonic region downstream of the multi-wave point, or this region is underdeveloped. As previously noted, Olejniczak et al. [15] considered this a transitional state as Type VI–Type V transition process, and this study will also adopt this terminology. For IG, the VI → V interaction appears at 19° and is about to disappear at $\theta_1 = 27°$. In NEG, this interaction exists within the angle range $\theta_1 = 21°$ to 29°. The following conducts a detailed analysis of the flow field in the VI → V interaction.

This study selects and presents pressure contours for IG in Fig. 27 at $\theta_1 = 19°$, 27°, and for NEG at $\theta_1 = 21°$, 29°, along with results at $\theta_1 = 25°$ to compare differences at the same $\theta_1$. The results show that CS, TS, and BS intersect at the triple-point IP2, and the subsonic region downstream of IP2 expands significantly compared to Fig. 23(a). In addition, TS intersects with RS to generate a new transmitted shock $TS_1$ (Figs. 27(a.1) and (b.1)). When $\theta_1$ increases to 25° in IG, the changes in the flow field include: (a) RS degenerates into CW, and a small subsonic region exists downstream of its intersection with TS, showing a tendency to merge with the larger subsonic region; (b) IP2 begins to move forward, causing TS and $TS_1$ to shift leftward, with $TS_1$ intersecting the third wedge surface. The trends in the NEG flow field from $\theta_1 = 21°$ to 25° are similar to IG, differing in the much smaller extent of the subsonic region, a more rearward triple -point position, and $TS_1$ being farther from the third wedge surface. At $\theta_1 = 27°$ in IG, the two subsonic regions connect, and TS/ $TS_1$ passes through EW before landing on the third wedge surface near the expansion corner. Similarly, at the boundary angle $\theta_1 = 29°$ in NEG, the subsonic region downstream of the intersection between CW and TS disappears. This is likely due to the weakening of CW formed by reattachment as $\theta_1$ increases, making it insufficient to generate a subsonic region after interacting with TS.

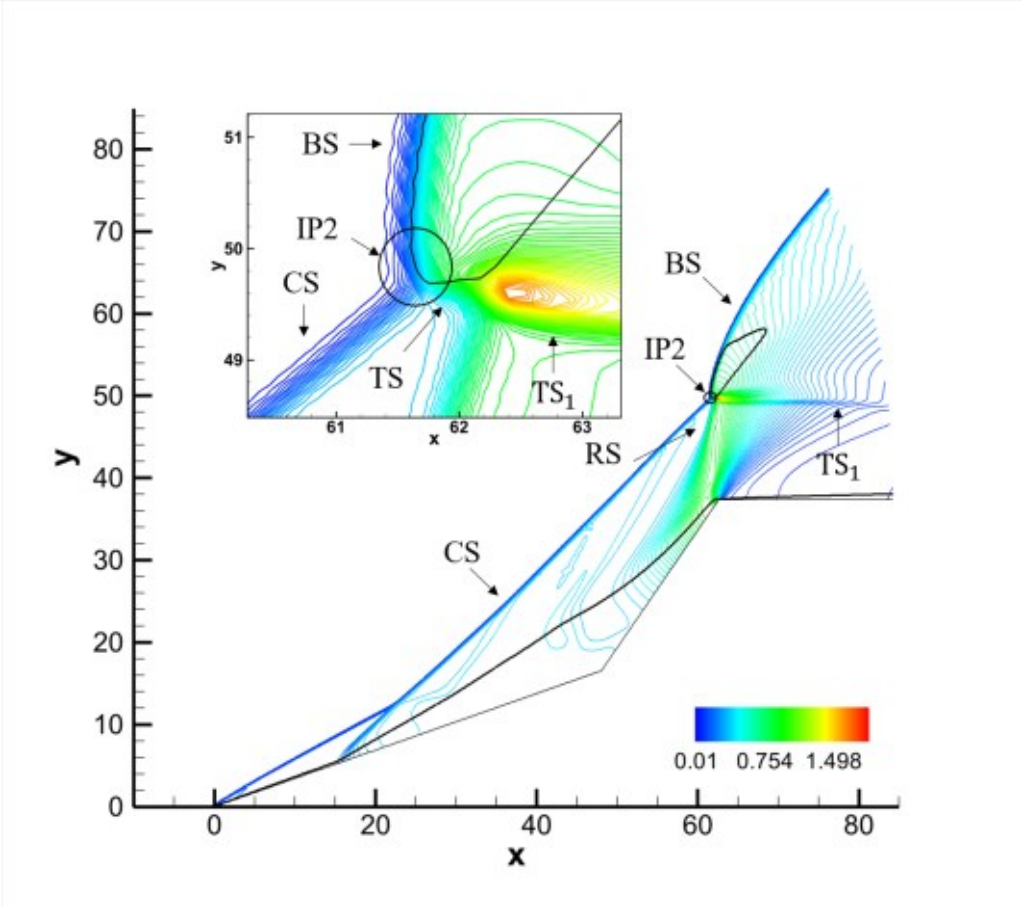

(a.1) $\theta_1 = 19°$

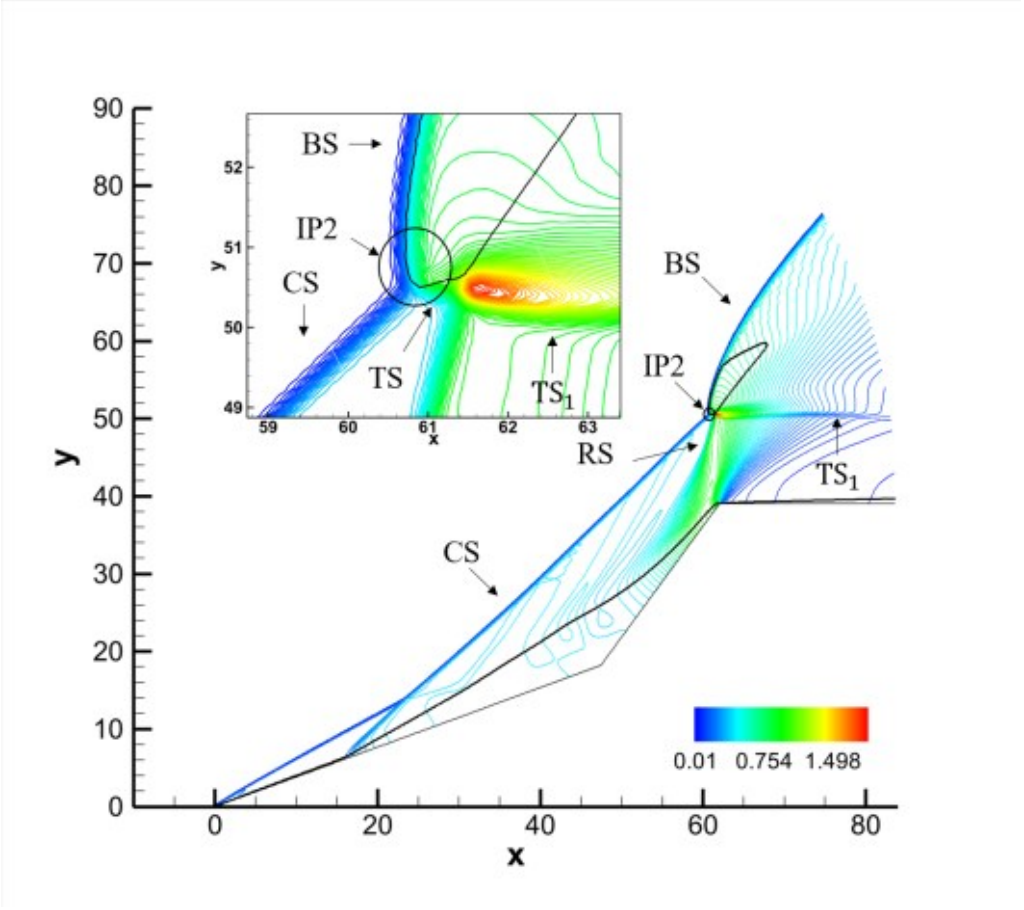

(b.1) $\theta_1 = 21°$

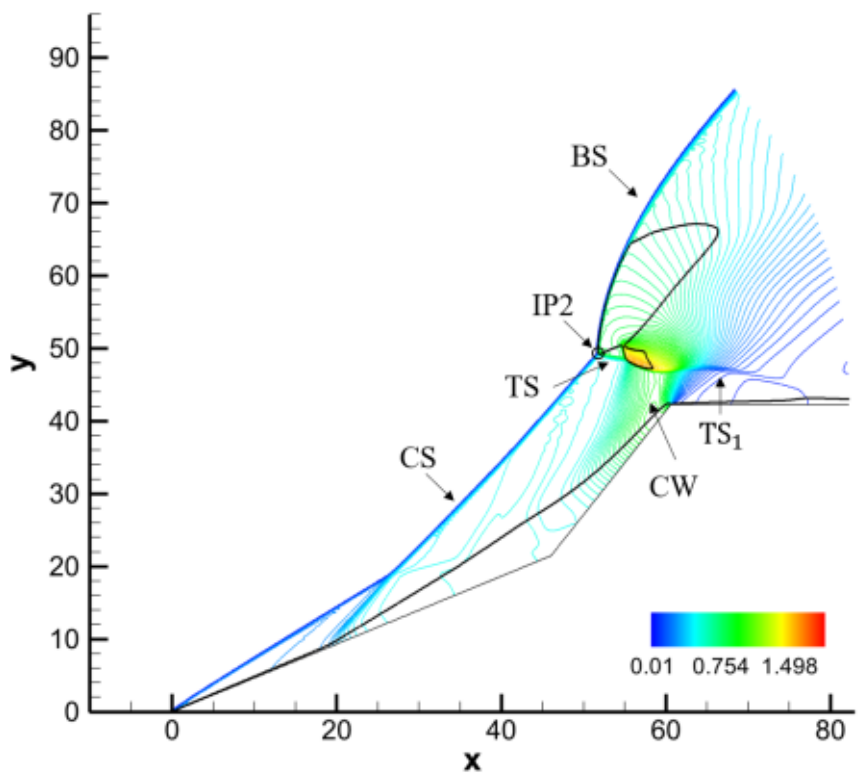

(a.2) $\theta_1 = 25°$

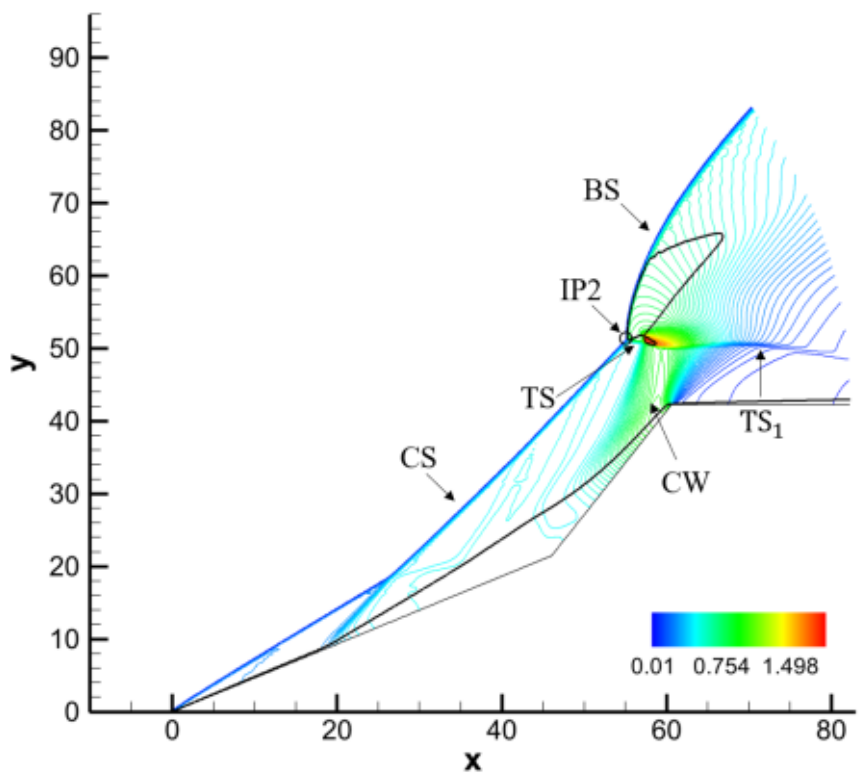

(b.2) $\theta_1 = 25°$

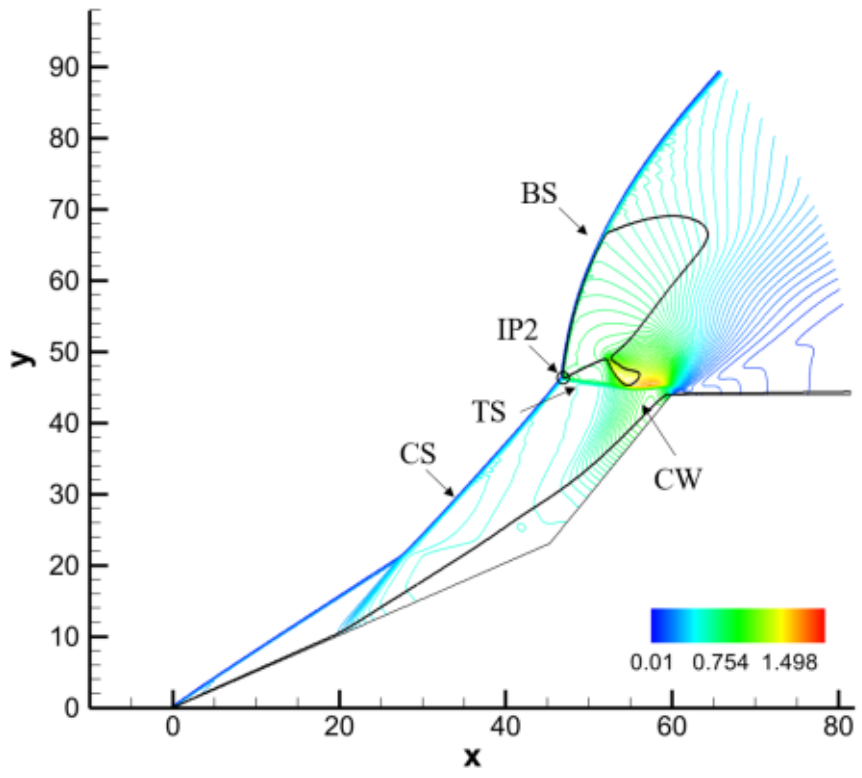

(a.3) $\theta_1 = 27°$

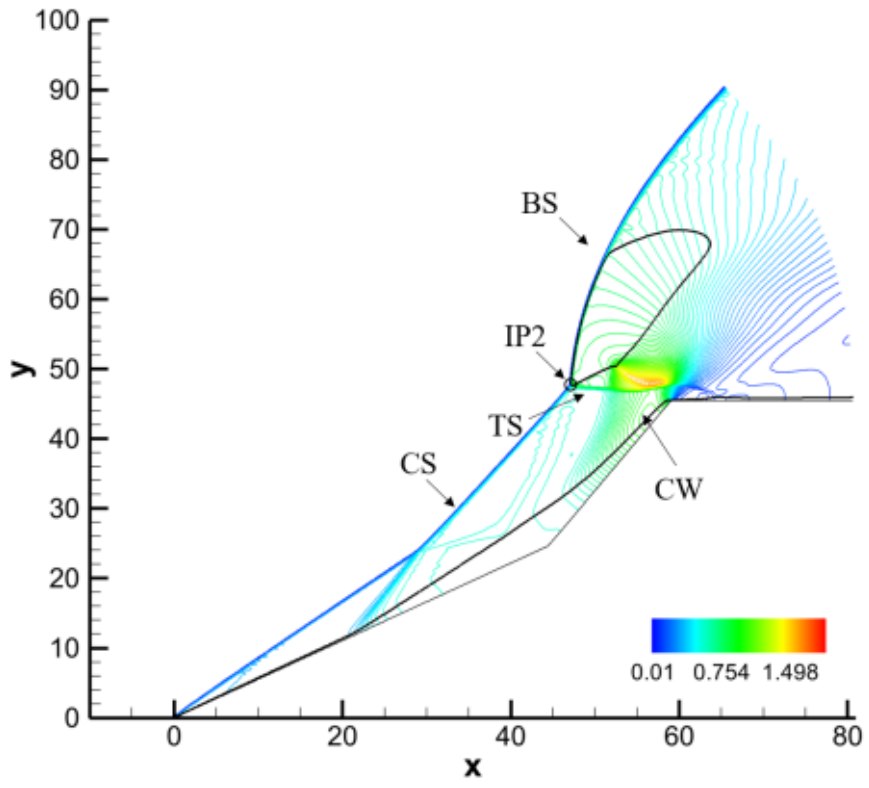

(b.3) $\theta_1 = 29°$

(a) IG (b) NEG

Fig. 27 Pressure contours of Type VI → V interaction flow fields in different gas models and $\theta_1$ (The left side shows the results for IG, the right side shows the results for NEG, black solid lines denote sonic lines, IP2 is the triple-point, CS is the combination of two shock waves, TS and $TS_1$ are transmitted shocks, BS is the bow shock, RS is the reattachment shock, and CW is the compression wave)

The corresponding numerical schlieren in Fig. 28 displays the details of separation and vortices, where streamlines and sonic lines are represented by green and red solid lines, respectively. The structures indicate that as $\theta_1$ increases, the trends in the separation zones of both gas models are similar, with the separation zones becoming more flattened and the tertiary separation vortex gradually merging with the primary separation vortex until it disappears. Current computational results show that, compared to NEG, the interaction pattern in IG at the same $\theta_1$ features a more forward triple-point and a larger subsonic region, demonstrating a faster evolution rate of the interaction pattern.

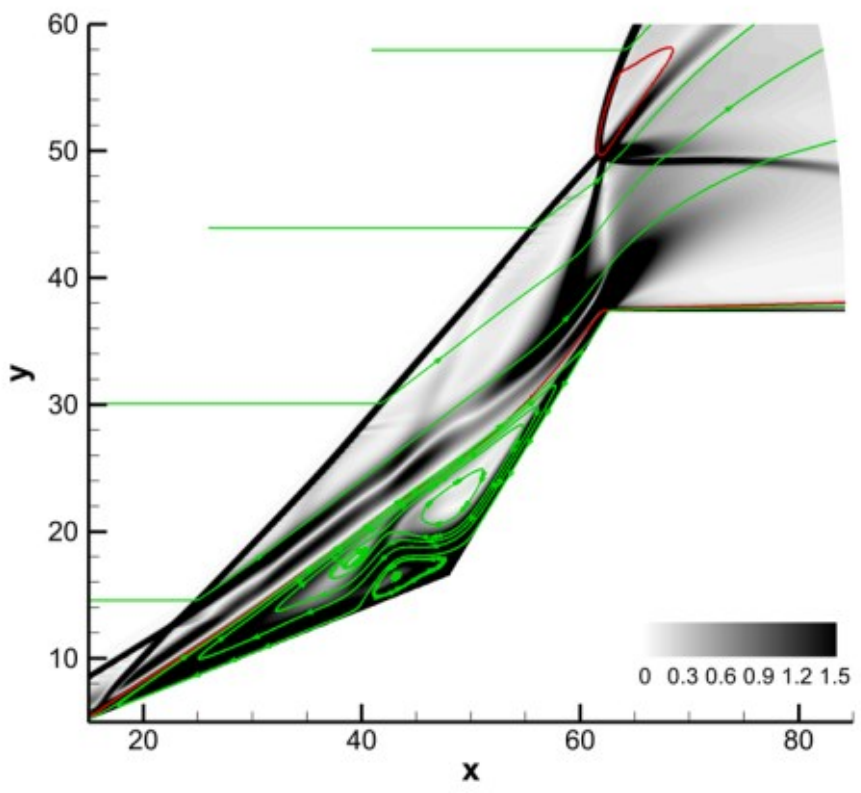

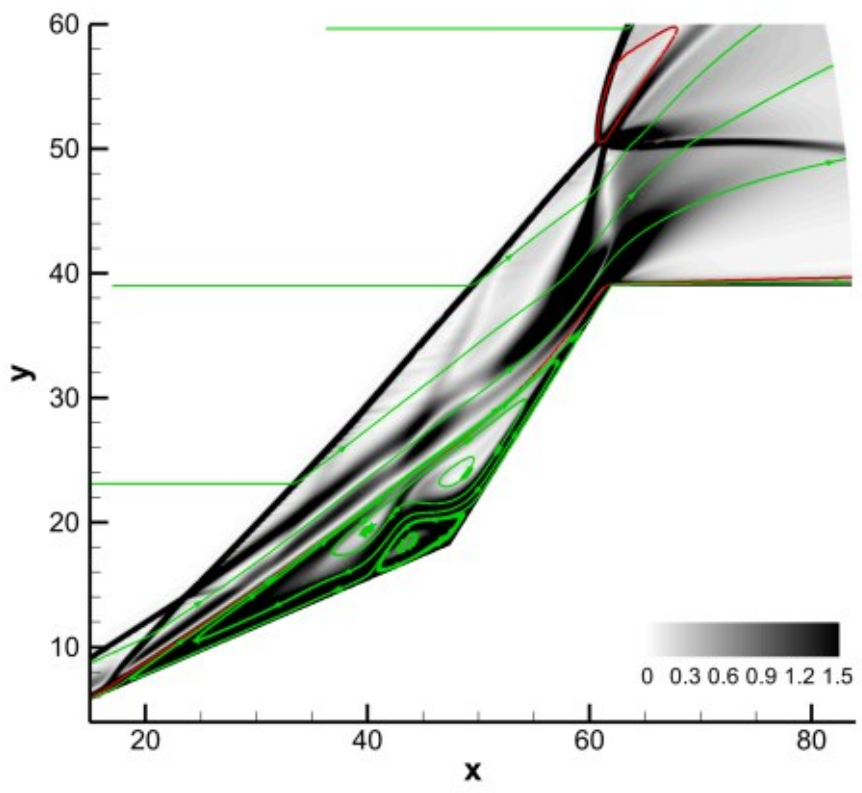

(a.1) $\theta_1 = 19°$ (b.1) $\theta_1 = 21°$

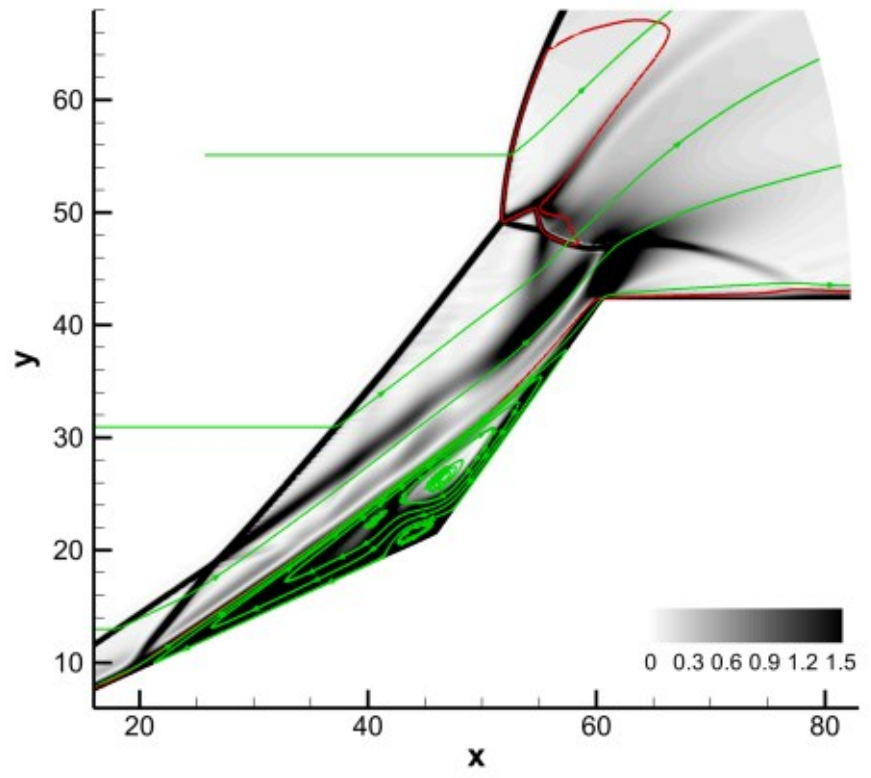

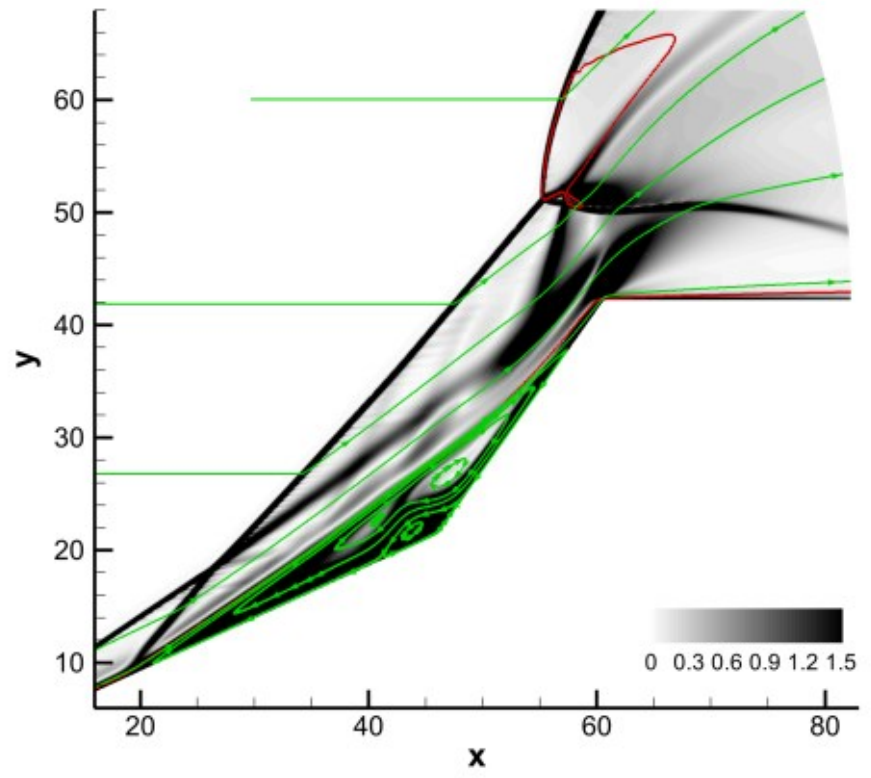

(a.2) $\theta_1 = 25°$ (b.2) $\theta_1 = 25°$

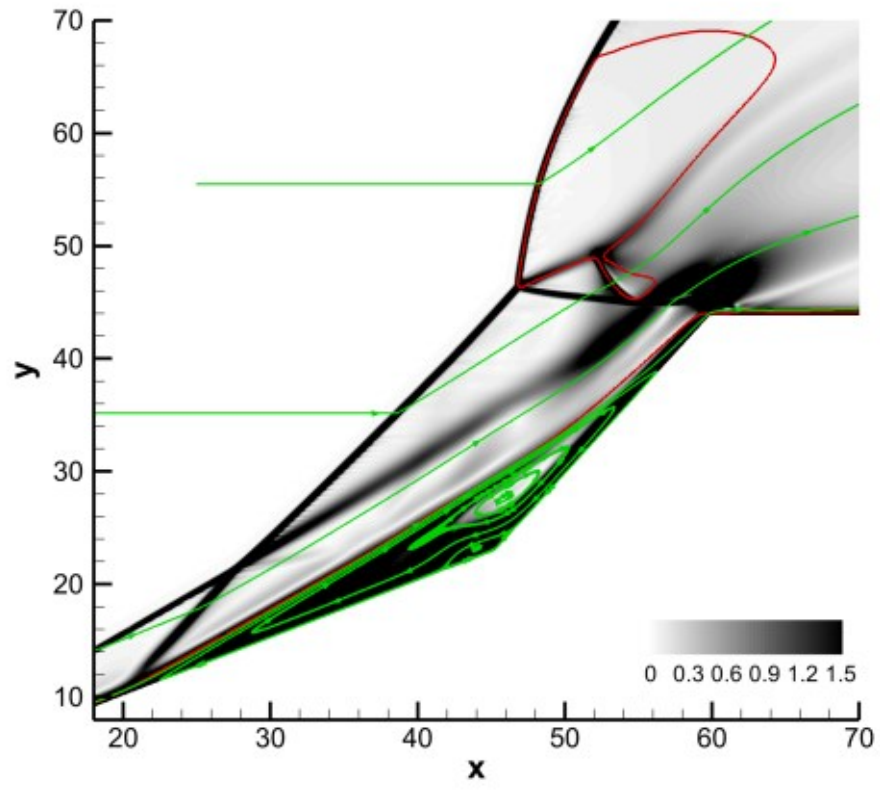

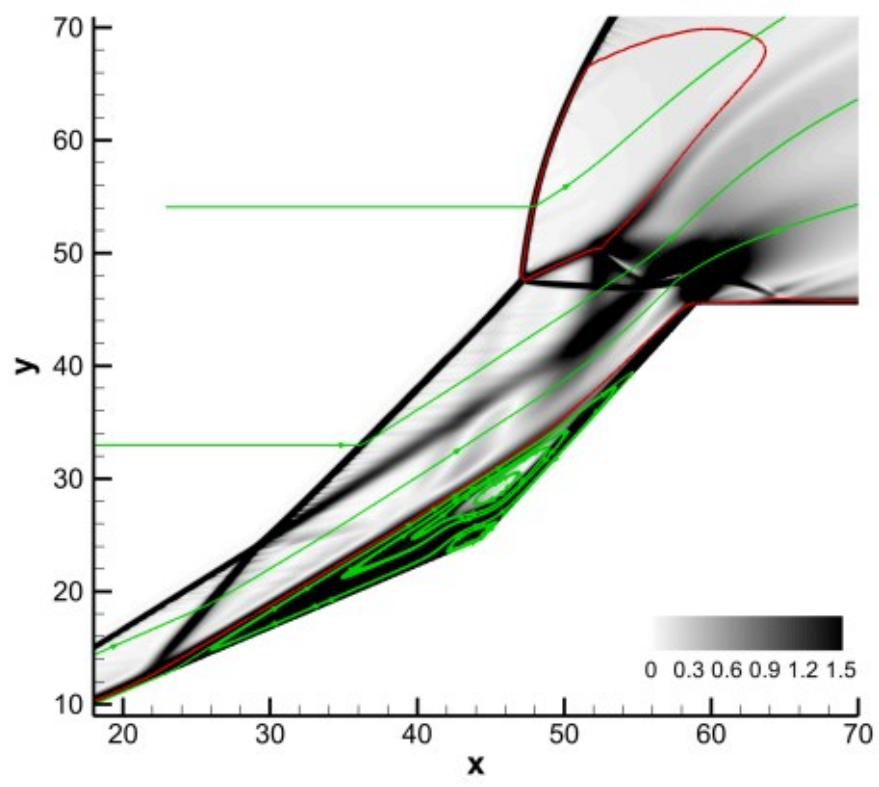

(a.3) $\theta_1 = 27°$ (b.3) $\theta_1 = 29°$

(a) IG (b) NEG

Fig. 28 Numerical schlieren of Type VI → V interaction flow fields in different gas models and $\theta_1$ (The left side shows the results for IG, the right side shows the results for NEG, green solid lines denote streamlines, and red solid lines denote sonic lines)

### 3.3.3 Type III interaction — corresponds to $\theta_1 = 31° - 33°$ (IG)/$\theta_1 = 34° - 37°$ (NEG)

As discussed earlier, when $\theta_1 > \theta_{1,uns}^{max}$, the flow field transitions from unsteady to steady, and the interaction pattern undergoes further evolution, as shown in Fig. 29. Compared to Fig. 20(d), CS in the computational flow field now acts as the incident shock, interacting with BS to produce TS, which aligns with the essential characteristics of Type III interaction. Therefore, the current interaction pattern can be identified as Type III. In IG, Type III interaction occurs within the angle range $\theta_1 = 31°$ to $33°$, while in NEG, it occurs within $\theta_1 = 34°$ to $37°$. The following conducts a detailed analysis of Type III interaction.

First, pressure contours in different models and $\theta_1$ are depicted. The results for the selected $\theta_1$ and pressure contours in IG and NEG are illustrated in Fig. 29, where black solid lines denote sonic lines. The results indicate that TS reflects on the boundary layer and generates EW, as shown in Figs. 29(b.2) and (b.3), in addition to exhibiting the characteristics of Type III interaction, as shown in Fig. 20(d). EW reflects on the subsonic region downstream of IP2 to generate a shock wave (SW), reflecting on the boundary layer. At $\theta_1 = 31°$ (IG) and $34°$ (NEG), the former features a quasi-normal shock (QNS) structure ahead of the subsonic region behind the wave. In contrast, the latter contains two independent subsonic regions, with a QNS structure also present ahead of the smaller subsonic region. This structure was analyzed in the literature [19] and will not be discussed in detail. As $\theta_1$ increases, comparing Figs. 29(a.2) with (a.3) and (b.2) with (b.3) indicate that the subsonic region and boundary layer form a supersonic channel that begins to elongate, with a significant increase in the number of reflections within the channel.

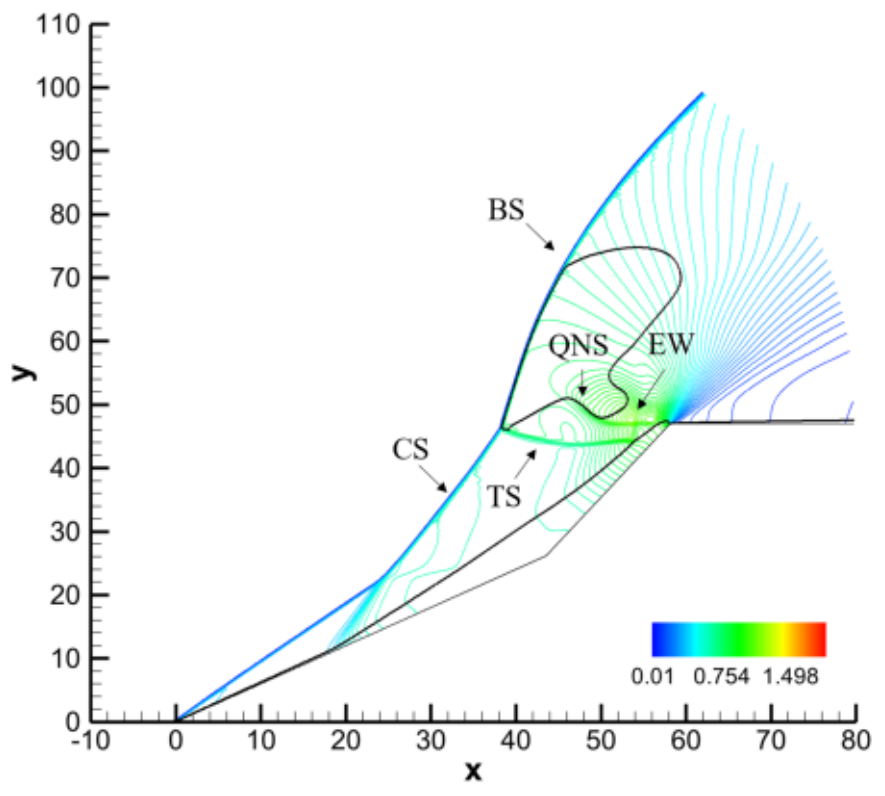

(a.1) $\theta_1 = 31°$

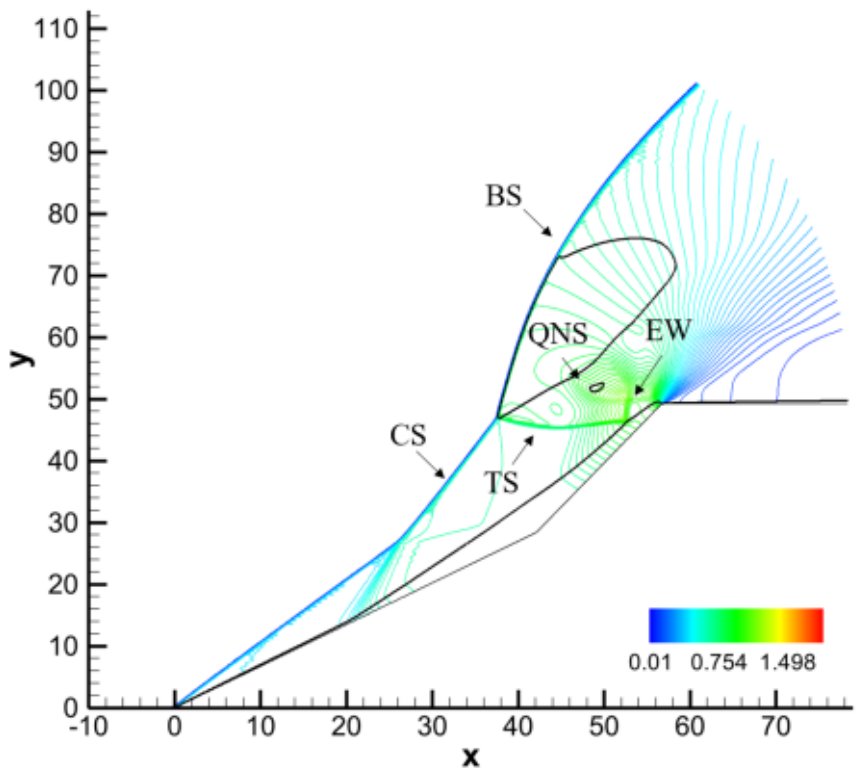

(b.1) $\theta_1 = 34°$

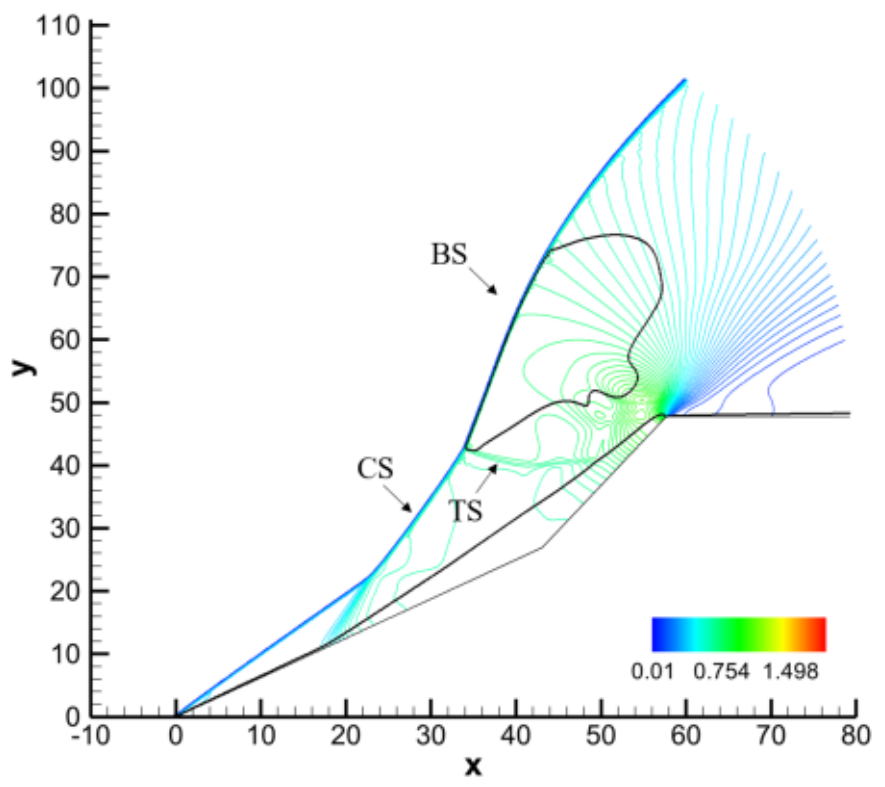

(a.2) $\theta_1 = 32°$

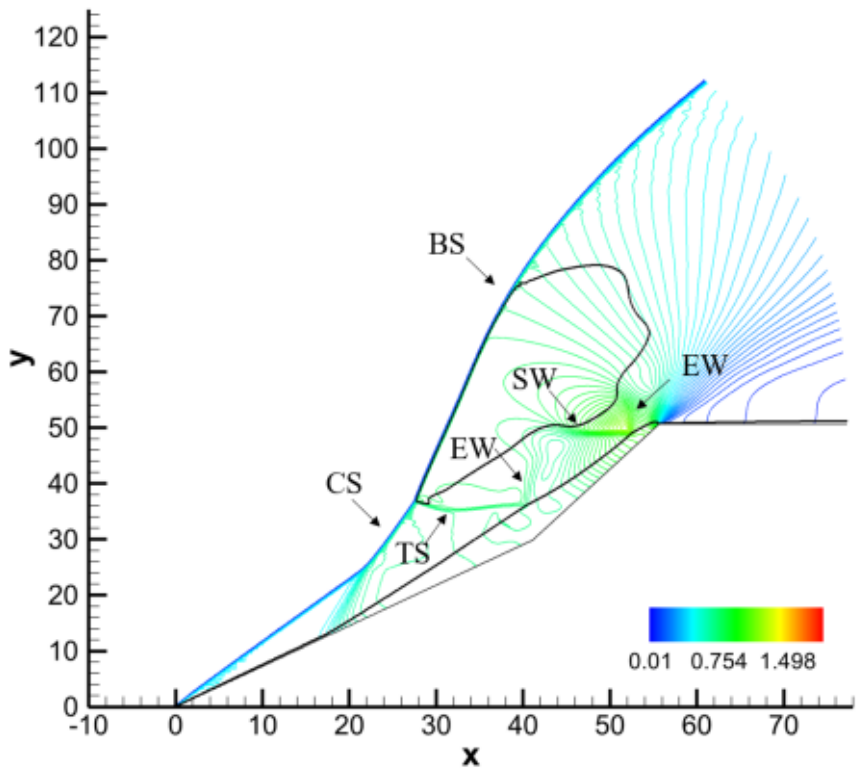

(b.2) $\theta_1 = 36°$

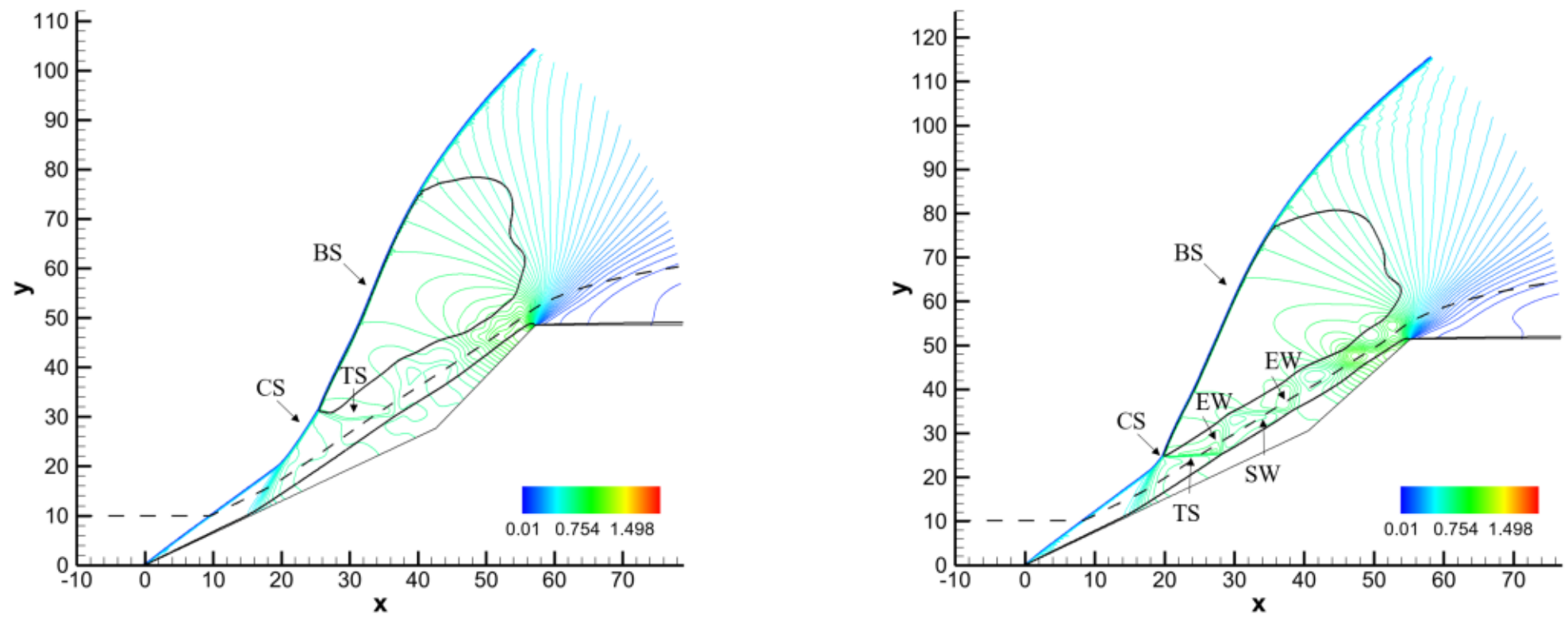

(a.3) $\theta_1 = 33°$
(a) IG

(b.3) $\theta_1 = 37°$
(b) NEG

Fig. 29 Pressure contours of Type III interaction flow fields in different gas models and $\theta_1$ (The left side shows the results for IG, the right side shows the results for NEG, black solid lines denote sonic lines, black dashed line indicates the streamlined position used for data extraction, CS is the combination of two shock waves, TS is the transmitted shocks, BS is the bow shock, QNS is the quasi-normal shock, EW is the expansion wave, and SW is the shock wave)

Next, numerical schlieren of the flow field are presented in Fig. 30, where green solid lines denote streamlines, and red solid lines denote sonic lines. The results show that as $\theta_1$ increases, tertiary separation reappears in the separation zones of Fig. 30(a.3) and (b.3). In addition, the distance between the subsonic region downstream of the triple-point and SL1 decreases as $\theta_1$ increases. Combining Figs. 29(b.2) and 30(b.2) indicate that when the EW generated by the reflection of TS is sufficiently strong, it crosses SL1 and intersects with SL2 at the lower boundary of the subsonic region, reflecting as SW. However, as $\theta_1$ increases, the EW in Fig. 30(b.3) reflects on SL1. Figs. 30(b.2), (a.3), and (b.3) show that TS undergoes unconventional reflection on the separation vortex. In earlier studies, this interaction pattern was observed only in IG [19], but this study identified it in IG and NEG.

In addition to the typical Type III interaction, Fig. 30(b.3) also reveals a new interaction pattern in the flow field. IP1 and IP2 are very close, and the SL1 and SL2 nearly overlap. TS reflects on the separation vortex, generating a series of EW and SW, resulting in a structure not explicitly identified in previous Type III interactions. This interaction pattern is quite similar to the Type $IV_r$ interaction mentioned later, with the only difference being that the reflection occurs on the separation vortex, whereas in Type $IV_r$, it primarily occurs on the first wedge. For now, this study refers to it as Type $III_r$ interaction.

As previously discussed, sudden change has already appeared in the evolution of Type III interaction structures. Considering IG as an example, this is specifically manifested as rapid changes in the interaction structure when $\theta_1$ increases by 1°, including the shortening of the distance between IP1 and IP2, the expansion of the subsonic region, and the formation of a supersonic channel. Based on the current evolutionary trend of the flow field, as $\theta_1$ continues to increase, the supersonic channel evolves into a jet-like structure, and TS reflects within it. The interaction pattern of the flow field transitions toward the Type IV interaction shown in Fig. 20(c), leading to the previously mentioned structural transition. Comparing Fig. 30(a.3) and (b.3), as the Type III interaction approaches its transition, the NEG flow field more closely resembles Type IV interaction, with a shorter distance between IP1 and IP2 and significantly more reflections within the channel.

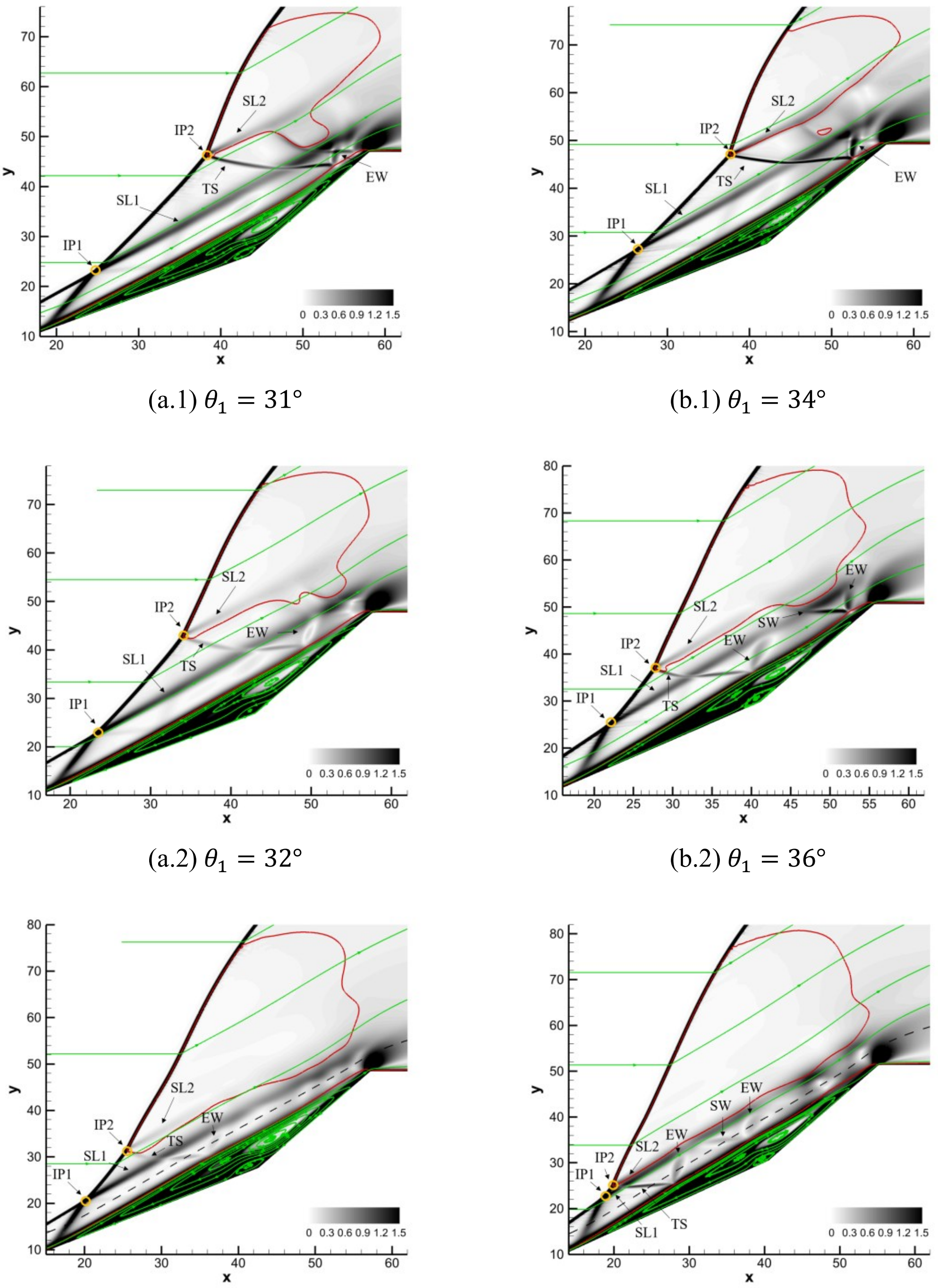

(a.1) $\theta_1 = 31°$ (b.1) $\theta_1 = 34°$

(a.2) $\theta_1 = 32°$ (b.2) $\theta_1 = 36°$

(a.3) $\theta_1 = 33°$ (b.3) $\theta_1 = 37°$

(a) IG (b) NEG

Fig. 30 Numerical schlieren of Type III interaction flow fields in different gas models and $\theta_1$ (The left side shows the results for IG, the right side shows the results for NEG, green solid lines denote streamlines, red solid lines denote sonic lines, the black dashed line indicates the streamline position used for data extraction, IP1 and IP2 are shocked intersection points, SL1 and SL2 are slip line, TS is the transmitted shocks, EW is the expansion wave, and SW is the shock wave)

This study considers the results for IG at $\theta_1 = 33°$ and NEG at $\theta_1 = 37°$ to better illustrate the quantitative characteristics of the flow and reflections in the supersonic channel before the structural transition. Streamlines passing through TS are selected, with their positions marked by black dashed lines in the numerical schlieren and pressure contours. Pressure data along these streamlines are extracted, as depicted in Fig. 31, where the positions of TS, EW, and SW are annotated in the pressure distribution. The flow for NEG is compressed after passing through LS and SS, then encounters TS at $x = 23.77\ mm$. After impacting the boundary layer, TS forms a central expansion to counteract the pressure rise behind the wave. The generated EW is reflected by SL1 at $x = 33.95\ mm$ into SW, which then reflects again on the boundary layer. At this point, 3 reflections occur within the channel. Finally, the flow is compressed and peaks during reattachment. In contrast, only 1 reflection occurs within the channel in IG.

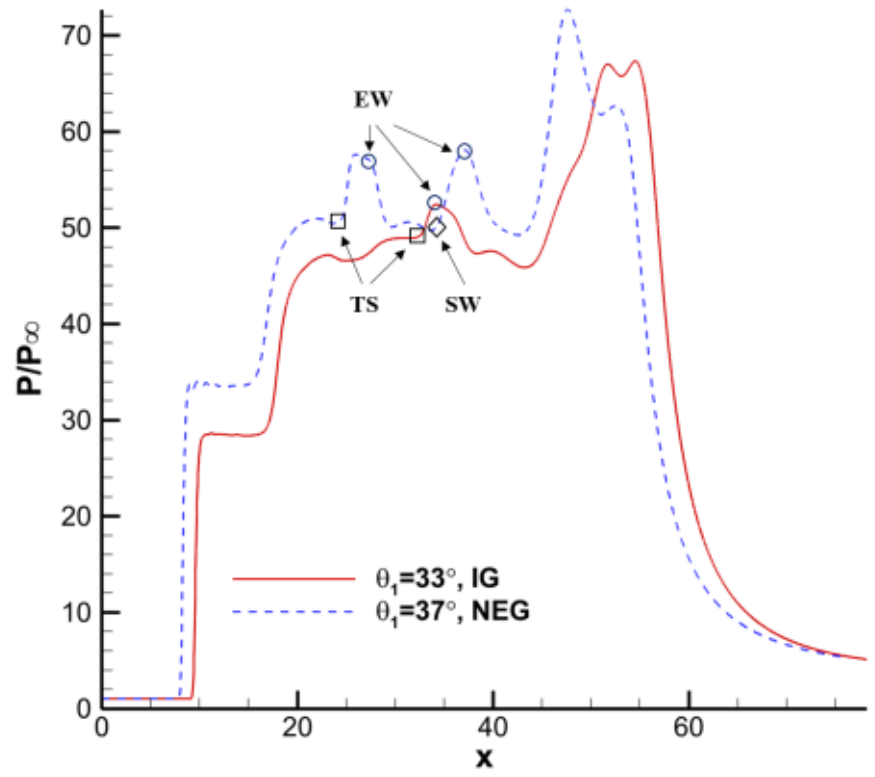

Fig. 31 Pressure distributions along the streamlines passing through TS for Type III interaction flow fields in IG at $\theta_1 = 34°$ and NEG at $\theta_1 = 37°$ (TS is the transmitted shock, EW is the expansion wave, and SW is the shock wave)

This study employes shock polar to analyze typical states to clarify the mechanisms of the interaction. It selects IG at $\theta_1 = 33°$ as an example. The sketch is shown in Fig. 32(a), with typical regions labeled by numbers. The corresponding shock polar curves are presented in Fig. 32(b), where the horizontal axis represents the flow deflection angle, and the vertical axis represents $p/p_\infty$, with clockwise flow deflection angles taken as positive. The states at each region are determined as follows: (a) The state at region 1 is defined by the shock polar in black line under freestream conditions at a fore wedge angle of 33°; (b) The state at region 2 is defined by the shock polar of region 1 (red line) at the angle of separation, namely 43°, as determined by the numerical simulation; (c) The state of region 3 is established by Prandtl-Meyer expansion, located at the intersection of the brown dashed line and the black polar curve; (d) The flow in region 2 undergoes isentropic compression and expansion to reach region 4. Since the flow deflection angle remains unchanged, the state points of regions 4 and 2 coincide; (e) Regions 5 are derived from the intersection of the brown dashed line with the freestream polar in black, taking into account the lower slip line; (f) Regions 6 and 7 are derived from the intersection of the shock polar from region 4 (pink line) with the shock polar from region 5 (green line), taking into account the lower slip line; (g) States at regions 8 and 9 are determined by the interaction of the shock polar curve of region 5 (green line) with the freestream polar in black, considering the upper slip line. When calculating the separation angle, due to slight variations in the streamlines near the separation point, the result of 43° is obtained by averaging multiple measured separation angles.

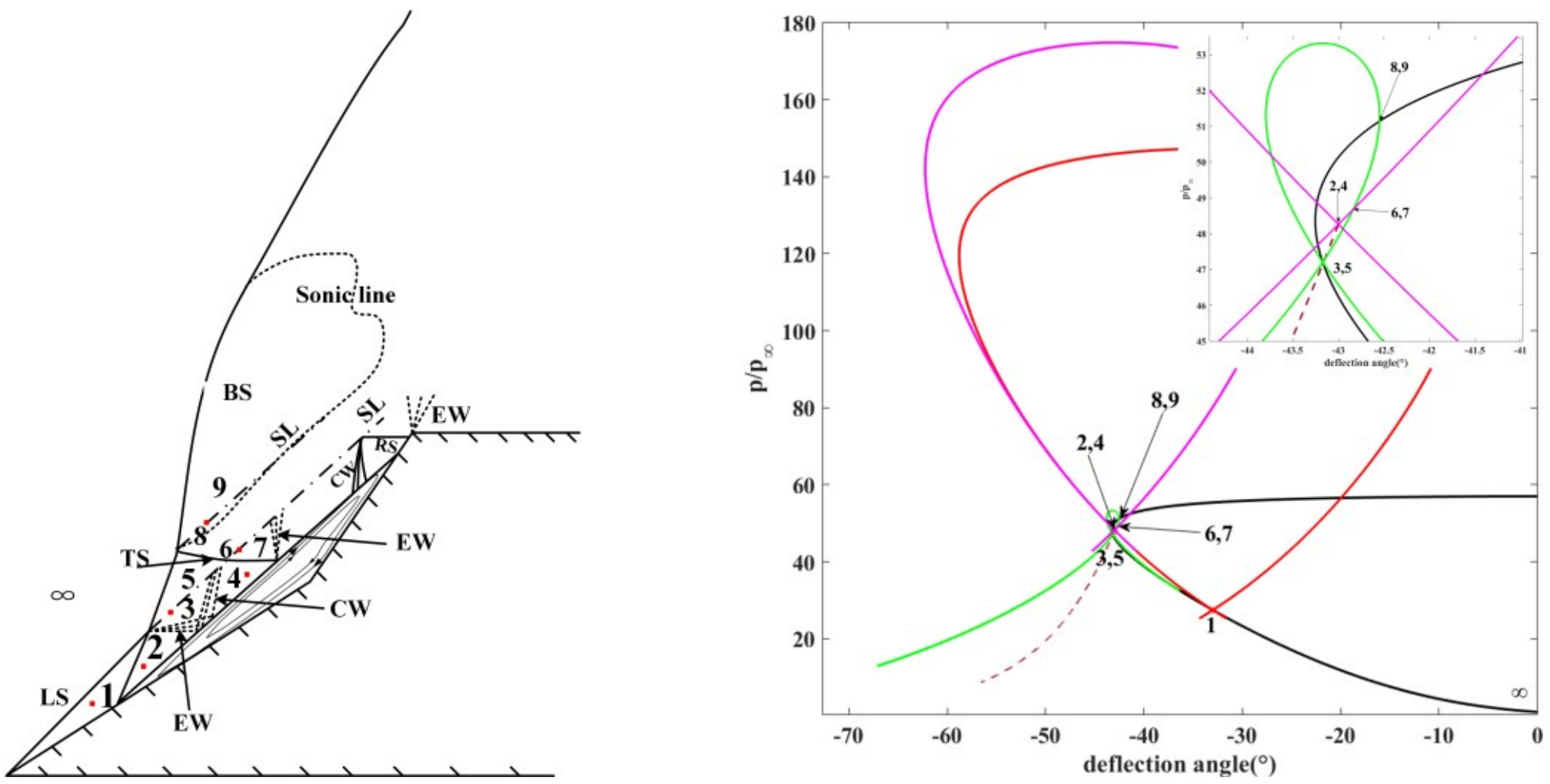

(a) Sketch of interaction (b) Shock polar

Fig. 32 Sketch of interaction and corresponding shock polar of the case of the IG model at $\theta_1 = 33°$

Based on the above analysis, the theoretical predictions (pressure) for each region in Fig. 32(a) can be obtained by comparing computational results and theoretical analysis. The points marked by red points in Fig. 32(a) are selected as representatives, and their coordinates are provided in Table 3. In addition, the measured computational pressure ratios are compared with the theoretical predictions, as listed in Table 3. The results indicate that the pressure values at the selected points in the IG calculations agree with the predictions from the polar curve method. This demonstrates that the interactions obtained in the current calculations while considering viscosity can be explained well using shock polar to clarify the flow mechanisms.

Table 3. Coordinates and affiliated regions of chosen points and pressure ratios obtained by predictions and measurements of the case of the IG model at $\theta_1 = 33°$

| Point | Coordinates | Associated region | $r_p = p/p_\infty$ | | $\frac{\lvert r_{p,m} - r_{p,sp}\rvert}{r_{p,m}}$ (%) |
|---|---|---|---|---|---|
| | | | Shock polar (sp) | Measured (m) | |
| P1 | (12.4753, 10.8577) | 1 | 27.47 | 28.44 | 3.41% |
| P2 | (21.5962, 18.3007) | 2&4 (isentropic process) | 48.26 | 46.66 | 3.43% |
| P3 | (25.1620, 22.9805) | 3&5 (slip line) | 47.21 | 46.50 | 1.59% |
| P4 | (31.5607, 31.5874) | 6&7 (slip line) | 48.75 | 52.39 | 6.94% |
| P5 | (27.3741, 33.3001) | 8&9 (slip line) | 51.04 | 51.82 | 1.51% |

### 3.3.4 Type IV$_r$ interaction — corresponds to $\theta_1 = 34° - 37°$ (IG)/$\theta_1 = 38° - 40°$ (NEG)

The discussion in the previous section revealed that as $\theta_1$ increases to the boundary of Type III interaction, the flow field develops structures similar to the supersonic jet in Type IV interaction, indicating an impending transition to this interaction type. Calculations show that when $\theta_1$ increases by just 1°, the flow field undergoes an immediate sudden structural change, a discontinuous change significantly different from the interaction structures shown in Fig. 30(a.3) and (b.3), as illustrated by the pressure contours in Fig. 33(a.1) and (b.1). When $\theta_1$ increases in IG from 33° to 34°, the structural transition includes the disappearance of CS, the immediate merging of IP1 and IP2 to form a new triple-point IP, the disappearance of the large-scale separation vortex, the forward shift of the supersonic channel to the fore wedge, and a significant increase in the number of reflections within the channel. As mentioned earlier, this new interaction pattern shares similarities with the Type IV$_r$ interaction proposed by Olejniczak et al. [15] (Fig. 1). For Type IV$_r$ interaction proposed by [15], without considering viscosity, the supersonic jet terminates at a normal shock before the compression corner to ensure the flow can pass through the discontinuous compression corner at subsonic speed and smoothly reach the aft wedge surface. A continuously varying virtual boundary is formed in the viscous calculations, allowing the supersonic jet to extend across the entire second wedge surface due to small-scale separation vortices or boundary layers near the corner. Therefore, the structure obtained in this study resembles Type $IV_r$ interaction but has distinct differences. For convenience in the following discussions, this study will continue to refer to it as Type $IV_r$ interaction.

Fig. 33(a.2) to (a.4) and (b.2) to (b.3) illustrate the changes in the interaction flow fields of the two gas models as $\theta_1$ increases. The results show that their evolutionary trends are consistent, characterized by the following features: the position of IP continues to move forward, BS approaches the tip of the double wedge, the coverage of the subsonic region behind the shock expands, and it gradually approaches and eventually merges with the shear layer; the width of the supersonic jet channel decreases, and the number of reflections within the channel first increases and then decreases. At $\theta_1 = 36°$ to $37°$ in IG, the supersonic channel on the wedge surface rapidly contracts, and the number of reflections within the channel significantly decreases; the subsonic region of the boundary layer merges with the region behind BS at $x \approx 23\,mm$, causing the supersonic channel to disappear.

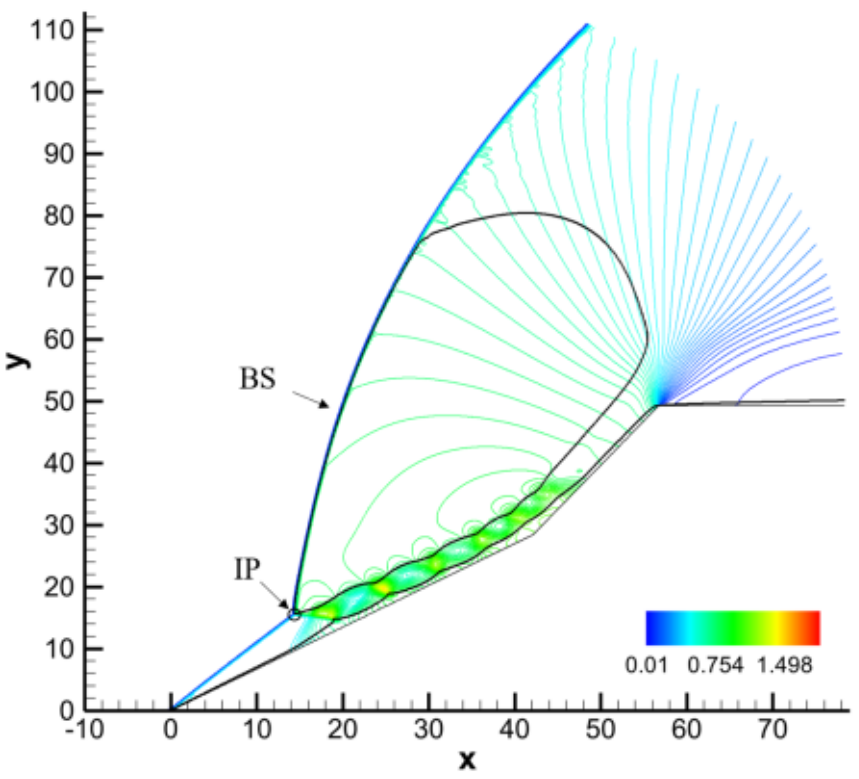

(a.1) $\theta_1 = 34°$

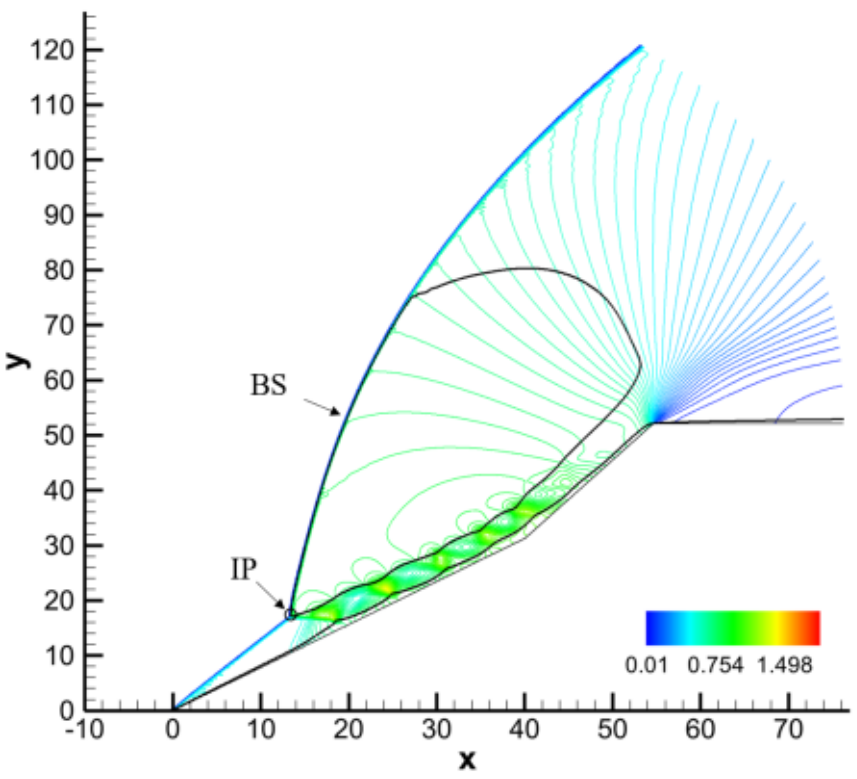

(b.1) $\theta_1 = 38°$

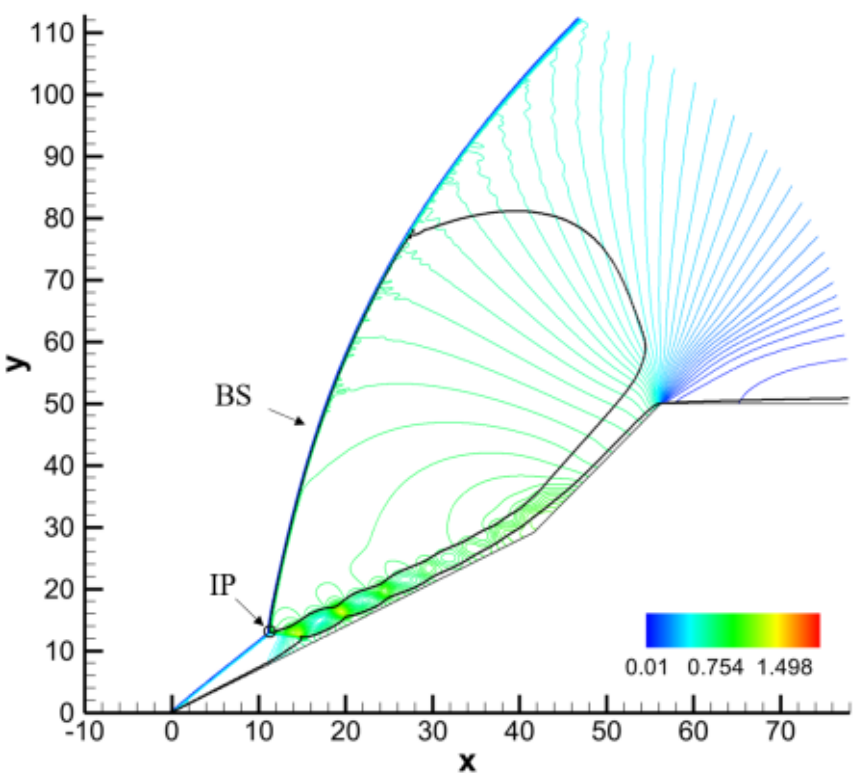

(a.2) $\theta_1 = 35°$

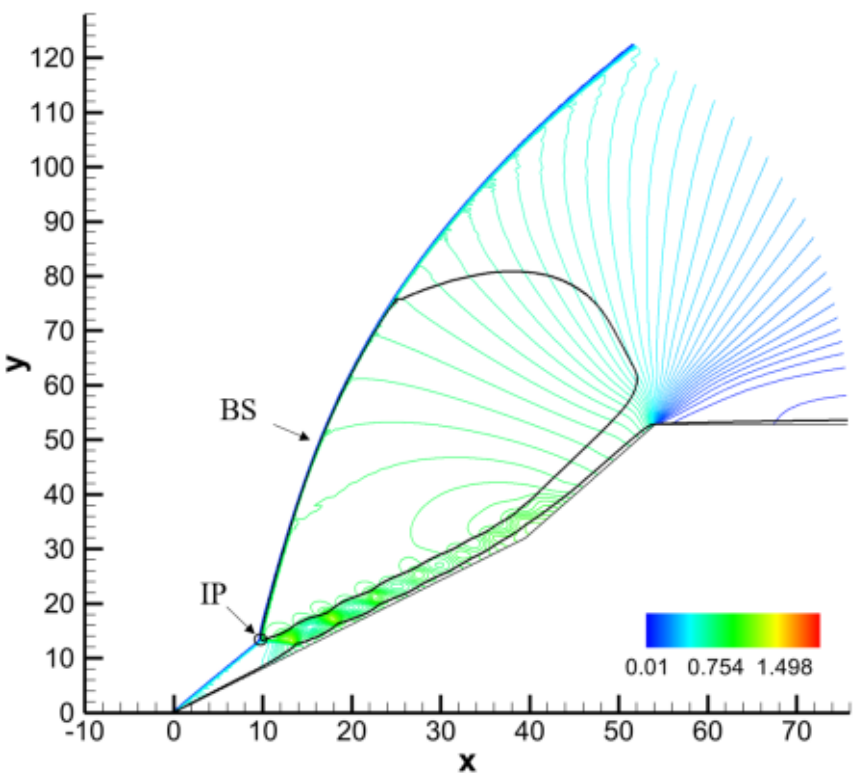

(b.2) $\theta_1 = 39°$

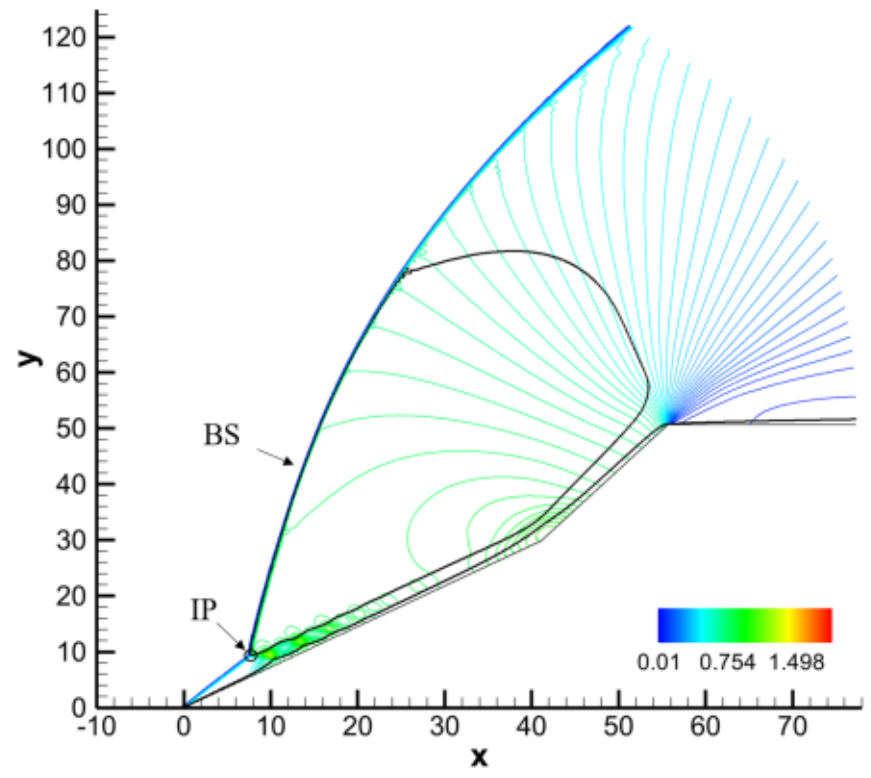

(a.3) $\theta_1 = 36°$

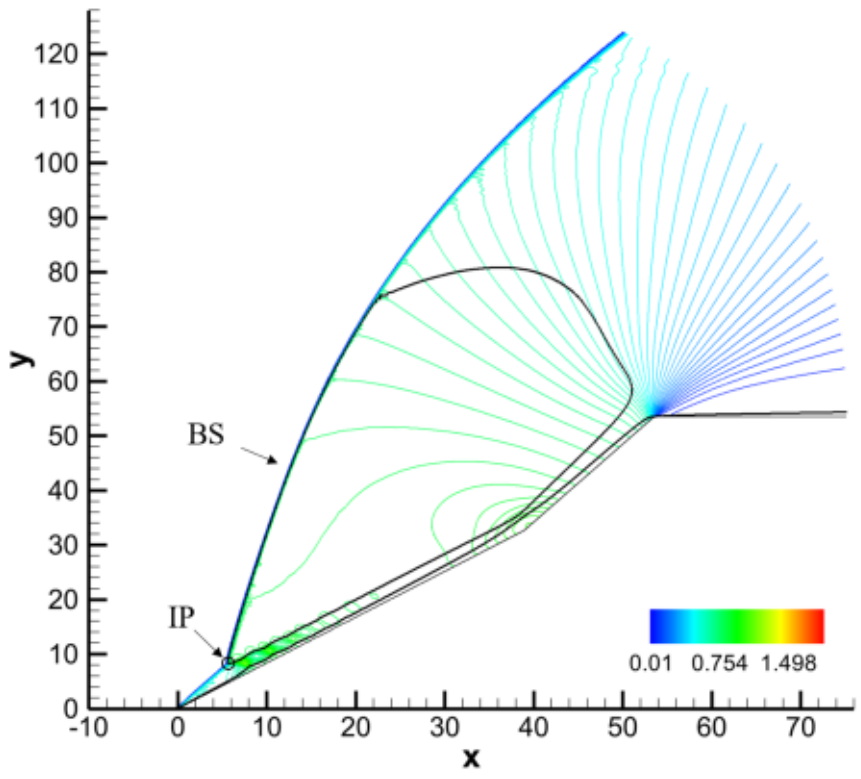

(b.3) $\theta_1 = 40°$

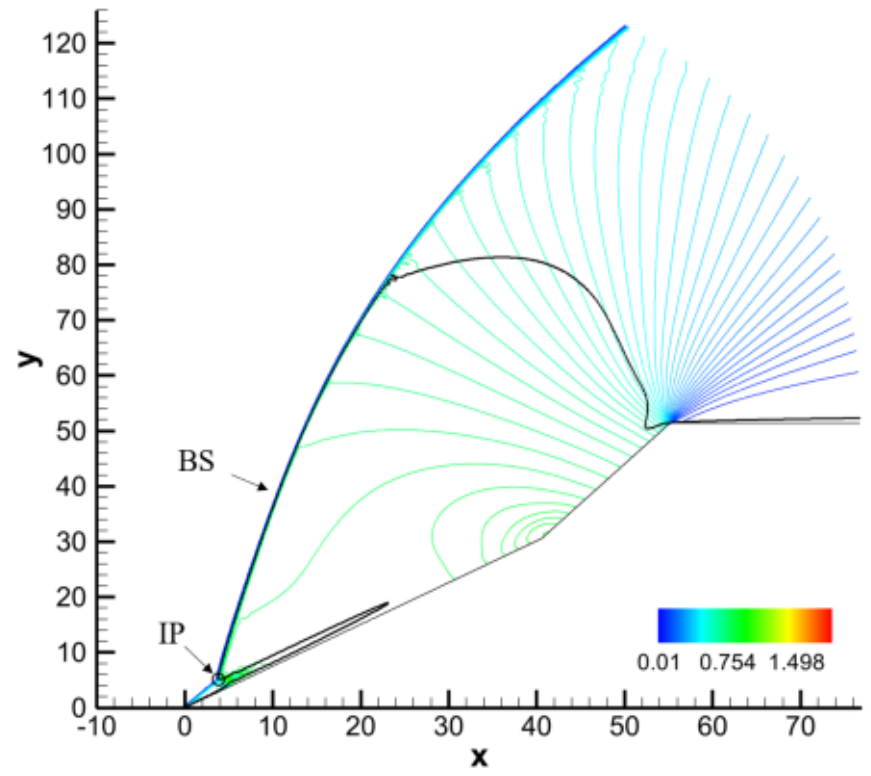

(a.4) $\theta_1 = 37°$

(a) IG (b) NEG

Fig. 33 Pressure contours of Type $IV_r$ interaction flow fields in different gas models and $\theta_1$ (The left side shows the results for IG, the right side shows the results for NEG, black solid lines denote sonic lines, IP is the shock intersection point, and BS is the bow shock)

Fig. 34 illustrates the numerical schlieren to reveal more detailed reflection structures. In the case of IG, the results first display the following fundamental structures: the TS generated by the intersection of LS and BS impinges on the fore wedge, causing boundary layer separation and producing a distinct SS. Compared to the results at $\theta_1 = 33°$ (Fig. 30(a.3)), the characteristics of the supersonic jet in the flow field at $\theta_1 = 34°$ (Fig. 34(a.1)) become significantly more pronounced. SW and EW undergo reflections within the supersonic channel, with the reflection intensity diminishing as the number of reflections increases, eventually fading away after passing the compression corner. In the case of NEG, the overall characteristics of the Type $IV_r$ interaction flow field formed after the sudden structural change is similar to those of IG. However, the latter exhibits more pronounced changes in the flow field during the formation of the new structure, as evidenced by the greater reduction in the distance between IP1 and IP2 and the increased number of reflections. In addition, small-scale separation vortices are observed near the first wedge surface and the compression corner, which gradually dissipate as $\theta_1$ increases.

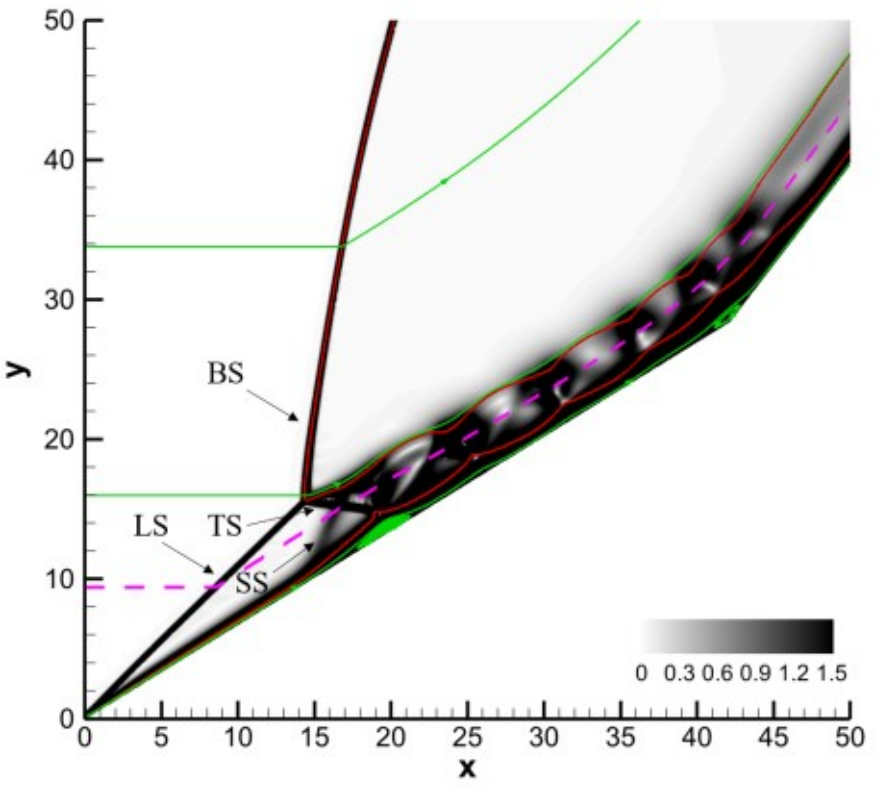

(a.1) $\theta_1 = 34°$

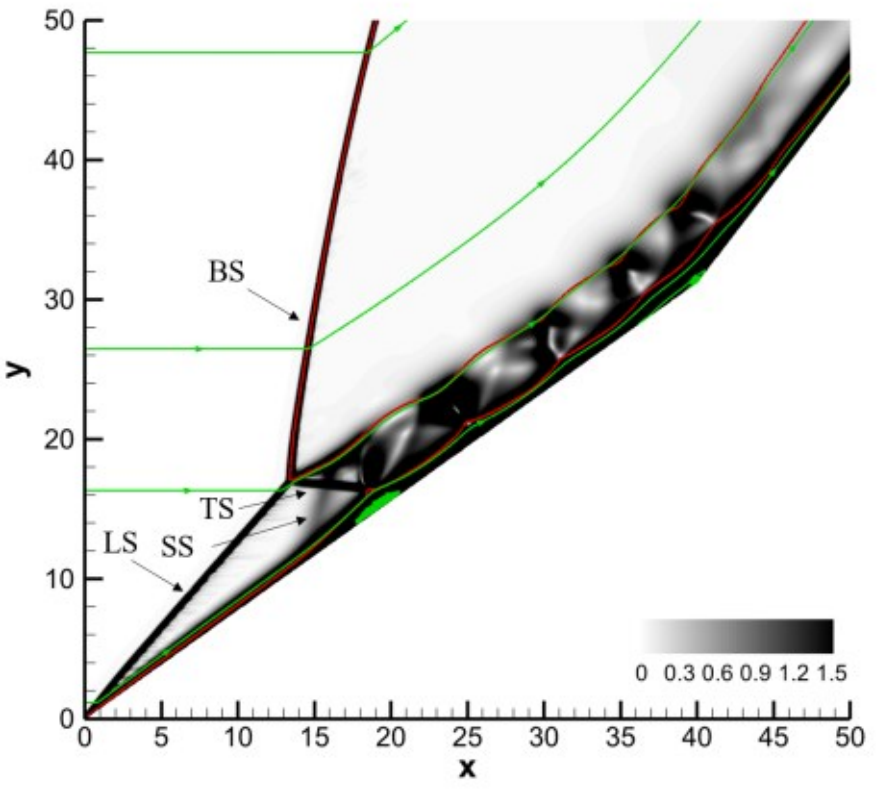

(b.1) $\theta_1 = 38°$

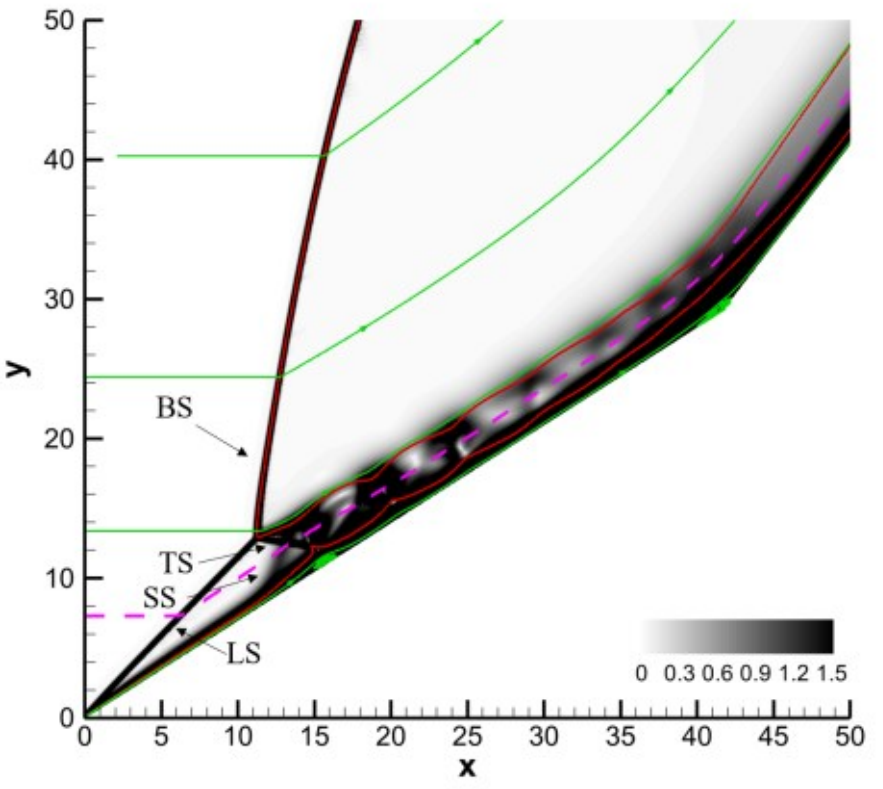

(a.2) $\theta_1 = 35°$

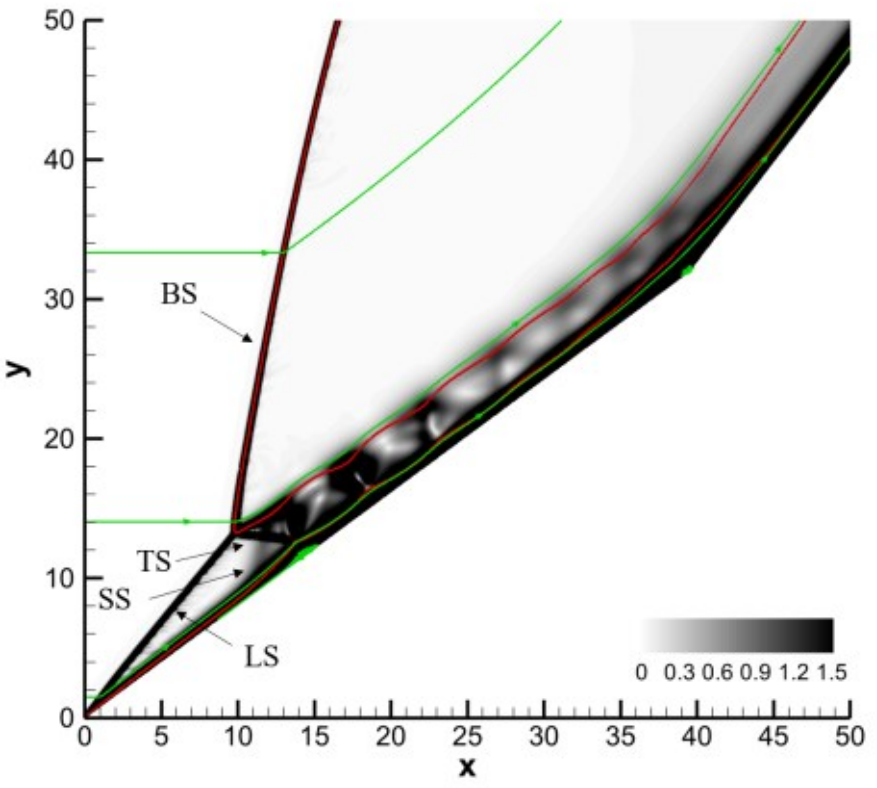

(b.2) $\theta_1 = 39°$

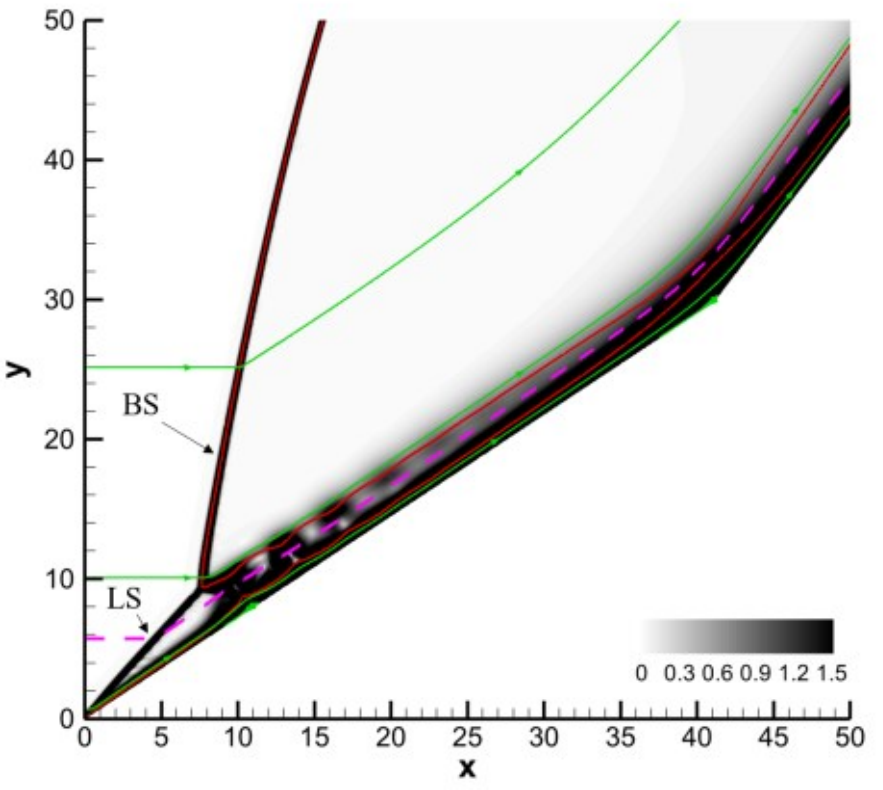

(a.3) $\theta_1 = 36°$

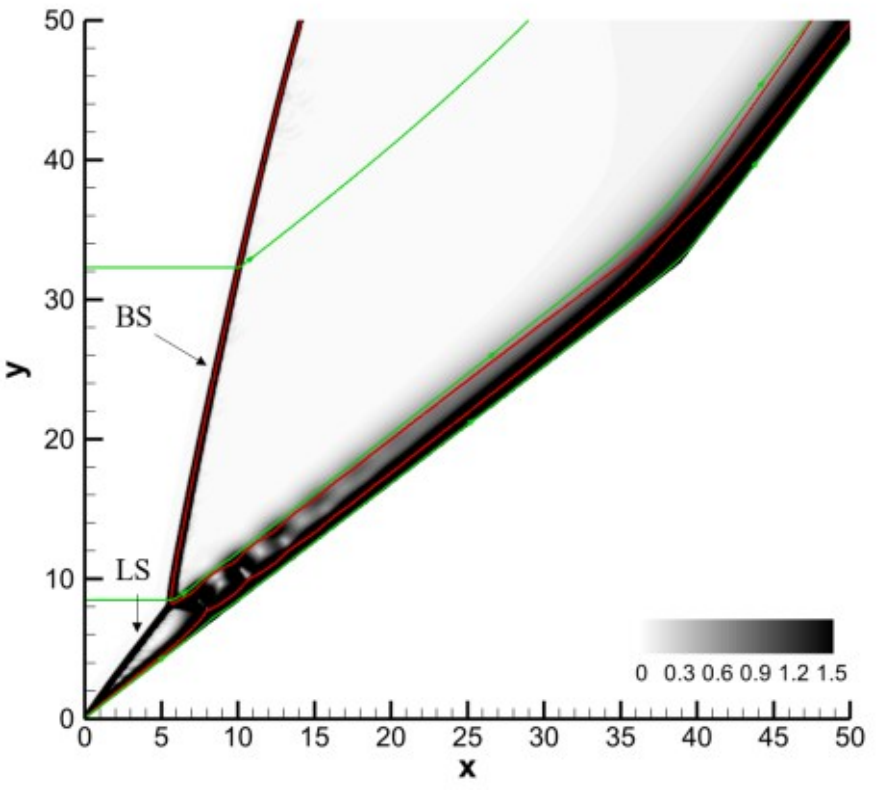

(b.3) $\theta_1 = 40°$

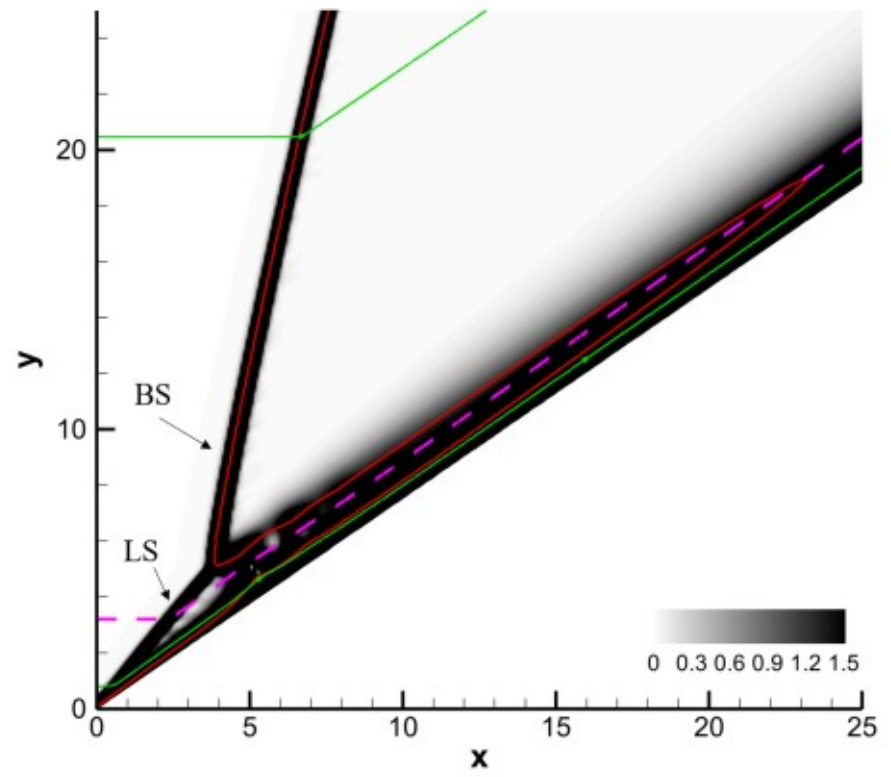

(a.4) $\theta_1 = 37°$

(a) IG (b) NEG

Fig. 34 Numerical Schlieren of Type $IV_r$ interaction flow fields in different gas models and $\theta_1$ (The left side shows the results for IG, the right side shows the results for NEG, green solid lines denote streamlines, red solid lines denote sonic lines, purple dashed line indicates the streamline position used for data extraction, LS is leading-edge shock, BS denotes bow shock, TS is transmitted shocks, and SS is separation shock)

This study takes the results of IG at four angles as an example to investigate the reflection phenomena within the supersonic channel. Streamlines passing through the TS within the supersonic channel (indicated by purple dashed lines in Fig. 34) are selected, and the corresponding pressure distributions are extracted and presented in Fig. 35. The pressure oscillations in the distribution curves reflect, to some extent, the reflections of SW and EW within the channel. The results are as follows: (a) At $\theta_1 = 34°$, the pressure distribution initially exhibits two ascending stages under the influence of LS and TS, followed by a decline due to the EW generated by reflections. Then, the pressure repeatedly rises and falls because of multiple reflections of SW (CW) and EW within the channel. A total of 11 reflections are observed, significantly more than the case at 33°, reflecting the unique impact of the sudden structural change on the flow field. (b) At $\theta_1 = 35°$, the initial pressure rise shifts leftward, and the strength of LS increases, driving the pressure to a higher level. The reflection intensity of SW and EW weakens, as evident from the changes in the peaks and troughs of the pressure distribution. However, the number of reflections within the channel increases to 13, likely due to the reduced width of the supersonic channel, which shortens the reflection intervals and increases the reflection count. (c) At $\theta_1 = 36°$, the reflection intensity within the channel further weakens, and the number of reflections decreases to 11. Near the compression corner ($x = 41.098\,mm$), the pressure undergoes a final rise, reaching its peak at $x = 40.3739\,mm$. The compression of the flow likely causes this phenomenon as it passes the compression corner. (d) At $\theta_1 = 37°$, the number of reflections significantly decreases. A pressure rise is still near the compression corner ($x = 40.571\,mm$), with the peak pressure occurring at $x = 39.5881\,mm$.

Based on the preceding discussion, this study statistically analyzes the number of reflections within the supersonic channel for both gas models, as shown in Fig. 36. The distribution for both models initially increases with $\theta_1$, which is attributed to the contraction of the supersonic channel, as evidenced by comparing Fig. 34(a.1) with (a.2) and Fig. 34(b.1) with (b.2). Then, the curves decline, reflecting the weakening of reflection intensity. In addition, for IG at $\theta_1 = 36° \rightarrow 36° \rightarrow 37°$, the number of reflections decreases from 13 to 11 and then to 5, exhibiting an accelerated reduction trend.

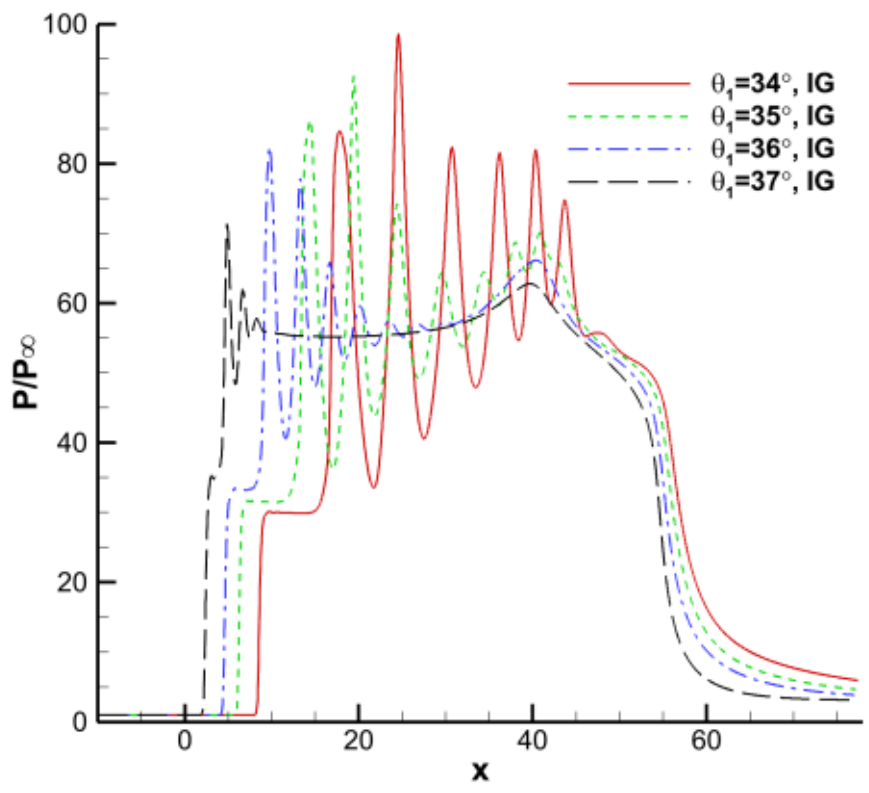

Fig. 35 Pressure distributions along the streamlines passing through TS for Type IV$_r$ interaction flow fields in IG

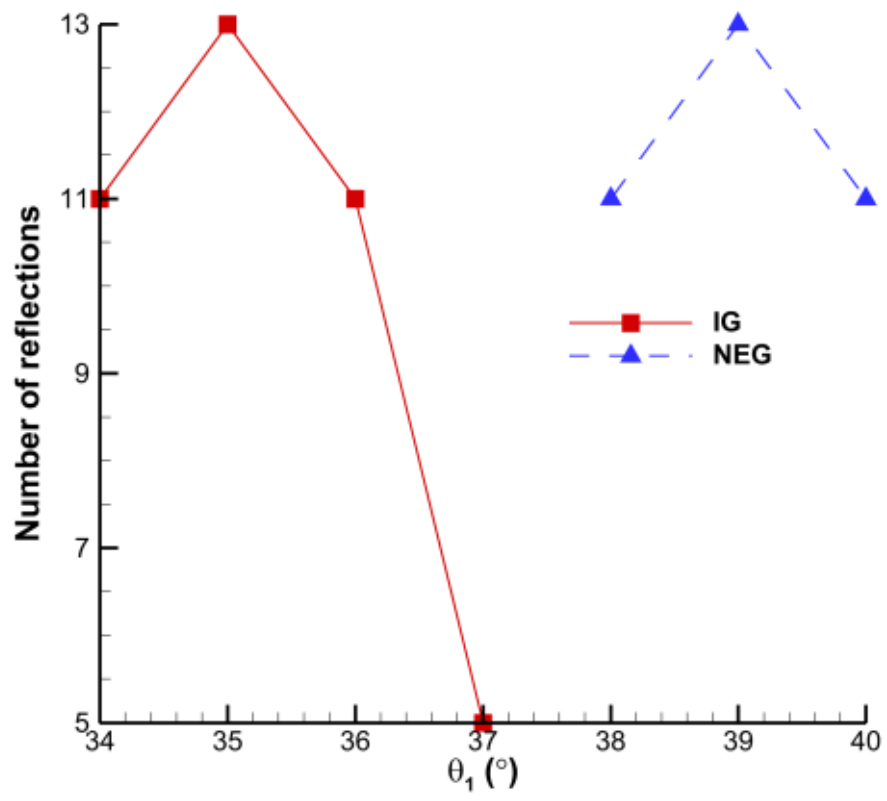

Fig. 36 Variation in the total number of reflections within the supersonic channel in different gas models

Similarly, shock polar is employed to analyze typical states and clarify the mechanisms of the interaction. Specifically, the research selects IG at $\theta_1 = 34°$ as an example. The sketch is illustrated in Fig. 37(a), with typical regions labeled by numbers, where "BL" represents the boundary layer. Fig. 37(b) depicts the corresponding shock polar curves, with clockwise flow deflection angles considered positive. The states at each region are determined as follows: (a) The state at region 1 is defined by the shock polar in black line under freestream conditions at a fore wedge angle of 34°; (b) The state at region 2 is defined by the shock polar of region 1 (red line) at the separation angle, namely 43°, as determined by the numerical simulation; (c) States at regions 3 and 4 are determined by the interaction of the shock polar curve of region 1 (red line) with the freestream polar in black, considering the slip line; (d) States at region 5 is identified by the interaction of the shock polar curve of region 2 (pink line) with that at regions 4 (brown line). In addition, the separation angle in the calculations is defined as the angle between the tangent to the streamlines near the separation point on the small-scale separation vortex and the fore wedge. Due to slight variations in the streamlines, the value of 43° is obtained by averaging multiple measurements.

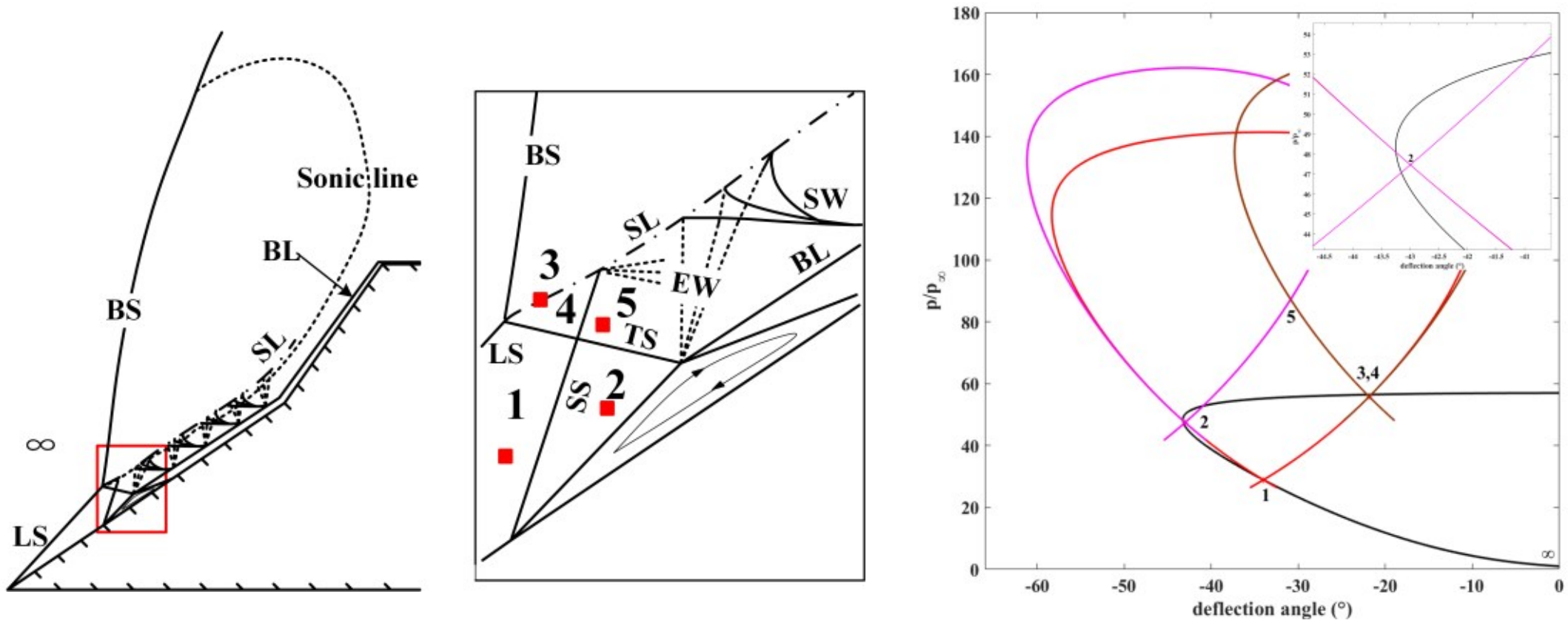

(a) Sketch of interaction (b) Shock polar

Fig. 37 Sketch of interaction and corresponding shock polar of the case of the IG model at $\theta_1 = 34°$

Four representative points are selected from the five regions in Fig. 37(a), and their coordinates are provided in Table 4 to compare theoretical predictions with computational results. In addition, the computed values at these points are compared to the theoretical predictions, as listed in Table 4. Except for Point 2 in Table 4, the measured values at the other locations show minor deviations from the predicted values. The larger discrepancy at Point 2 is likely attributed to the influence of the small-scale separation vortex formed after the sudden structural change.

Table 4. Coordinates and affiliated regions of chosen points and pressure ratios obtained by predictions and measurements of the case of the IG model at $\theta_1 = 34°$

| Point | Coordinates | Associated region | $r_p = p/p_\infty$ | | $\frac{\lvert r_{p,m} - r_{p,sp}\rvert}{r_{p,m}}$ (%) |
|---|---|---|---|---|---|
| | | | Shock polar (sp) | Measured (m) | |
| P1 | (10.0793, 9.4340) | 1 | 28.93 | 29.98 | 3.50% |
| P2 | (16.6316, 13.7600) | 2 | 47.47 | 42.16 | 12.59% |
| P3 | (15.7317, 16.2095) | 3&4 (slip line) | 55.83 | 58.69 | 4.87% |
| P4 | (18.2313, 15.8096) | 5 | 87.17 | 85.60 | 1.83% |

3.3.5 Bow (Detached) Shock — corresponds to $\theta_1 = 38° - 40°$ (IG)

The calculations reveal that the Type $IV_r$ interaction continues to evolve as $\theta_1$ further increases, with $\theta_1 = 37°$ being the critical angle at which the flow field in IG still maintains the Type $IV_r$ interaction. When $\theta_1$ increases by 1°, $\theta_1 = 38°$, the structure in Type $IV_r$ interaction immediately transforms, and only a bow shock wave remains. Accordingly, this research also refers to this discontinuous alteration as another sudden structural change. The pressure contours of the flow field at $\theta_1 = 38°$, 39°, and 40° are depicted in Fig. 38, where only the BS exists, and the subsonic region behind the shock covers the surfaces of the whole wedge. At $\theta_1 = 38°$, the flow field retains only the BS connects to the wedge vertex, which significantly differs from the Type $IV_r$ interaction shown in Fig. 34(a.4), indicating a relatively small-scale abrupt transition in the flow field as $\theta_1$ changes from 37° to 38°. As $\theta_1$ continues to increase, the BS moves upstream and detaches from the body after $\theta_1 \geq 39°$. Under inviscid conditions, the detachment of the shock occurs at $\theta_{1,inviscid} = 42.2546°$, indicating that the current viscous occurrence is rare, close to the inviscid case. Since the $\theta_1$ employed in this study is $0° - 40°$, the current results indicate that the computation of NEG will engender a flow dominated solely by the BS when $\theta_1 > 40°$, which is not investigated in this study.

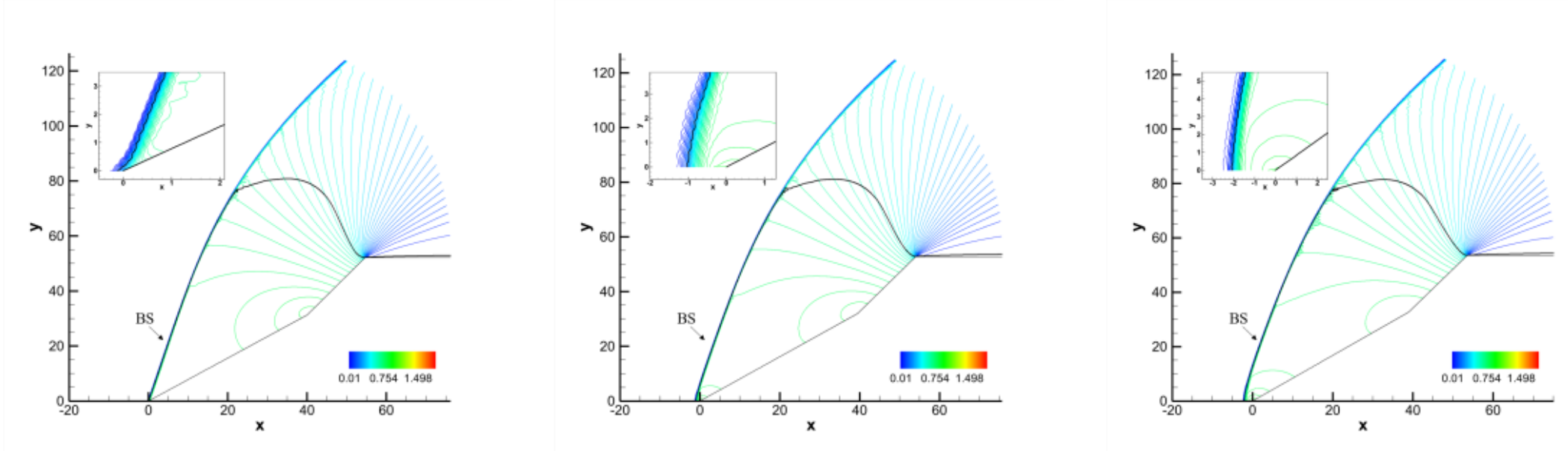

(a) $\theta_1 = 38°$ (b) $\theta_1 = 39°$ (c) $\theta_1 = 40°$

Fig. 38 Pressure contours of the flow field in IG at $\theta_1 = 38°$, 39° and 40° (Black solid lines denote sonic lines, and BS denotes bow shock).

The evolution of interaction patterns and the sudden structural change were investigated in the preceding sections, and the angular ranges for the existence of different interaction patterns were described. The study reveals that, as $\theta_1$ increases, transitions in the interaction structure occur within the same interaction pattern and across different patterns. The former primarily refers to transforming expansion waves into shock waves in Type VI interaction, while the latter regards the transition of interaction patterns. During the latter process, a sudden structural change occurs if the flow structure undergoes discontinuous changes. In the present calculations, the observed sudden structural changes include the transition from Type III to Type $IV_r$ interaction and the transition from Type $IV_r$ interaction to a flow with only the BS remaining. A quantitative study of the wall's aerodynamic properties will be conducted in the following section.

## 3.4 Aerodynamic distribution of wall properties at different $\theta_1$

Section 3.3 mainly concerns the evolution and sudden change of different interaction patterns. The wall pressure and heat flux distributions are analyzed for interactions before sudden structural change (namely, Type VI, Type VI → V, and Type III interaction), Type $IV_r$ interaction and flow fields dominated by the BS to conduct comparative studies on the results with different $\theta_1$.

### 3.4.1 Aerodynamic distributions of interactions before a sudden structural change

Due to significant differences in separation structures before and after the sudden structural change, the aerodynamic performances are separately investigated on the wall. In addition, considering the large amount of computations, only part of them are selected for analysis. First, the distribution of the wall pressure coefficient and its parametric variations are analyzed and discussed, specifically focusing on the results of IG at $\theta_1 = 0°$, 18°, 27°, and 33°, which are chosen as representatives of interactions evolving from Type VI to Type III. The distributions are shown in Fig. 39. For ease of comparison, the horizontal axis chooses the wedge surface distance S from the origin, where $S = 50.8\,mm$ and $S = 76.2\,mm$ correspond to the compression and expansion corner, respectively indicated by vertical dashed lines. In the figure, it can be observed that the main characteristics of different $\theta_1$ are generally consistent with each other, including the rise after SS, the plateau-like distribution within the separation zone, the second rise during the reattachment process, and the drop at the expansion corner. However, at $\theta_1 = 27°$, the continued rise after the expansion corner is observed, caused by the TS impinging on the third wedge surface. When $\theta_1 < \theta_{1,uns}^{min} = 28°$, as $\theta_1$ increases, the SS position is indicated to shift downstream, the separation zone range shortens, and the peak pressure coefficient increases, with the rate of increase gradually slowing down and ceasing eventually. When $\theta_1 > \theta_{1,uns}^{max} = 30°$, based on the changes in the pressure distribution from $\theta_1 = 27°$ to 33°, it can be seen that as $\theta_1$ increases, the position of the first rise shifts forward, the plateau range expands, and the wall pressure coefficient increases.

Fig. 40 illustrates the wall heat flux distributions at different $\theta_1$. The characteristics include a sharp drop after passing the separation point, a plateau-like distribution within the separation zone, reaching the minimum at the compression corner, beginning to rise on the aft wedge surface, and reaching the maximum at the expansion corner. When $\theta_1 < \theta_{1,uns}^{min}$, the minimum value roughly decreases first and then increases as $\theta_1$ increases, while the maximum value gradually decreases. In addition, at $\theta_1 = 0°$, the heat flux distribution experiences a drop and a rise before reaching the minimum value, which is speculated to be caused by the vortex-induced by secondary separation. When $\theta_1 > \theta_{1,uns}^{max}$, the minimum value of the heat flux distribution decreases, and the maximum value rises with the increase of $\theta_1$.

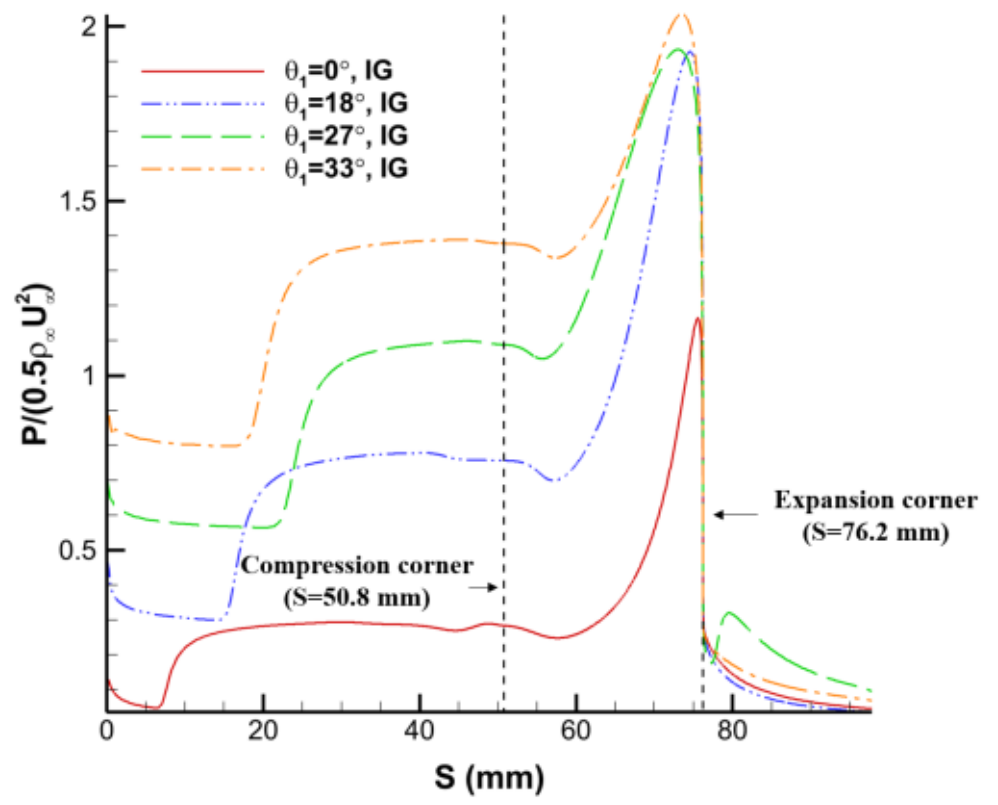

Fig. 39 Distributions of wall pressure coefficient of IG at different $\theta_1$ in interactions before the sudden structural change

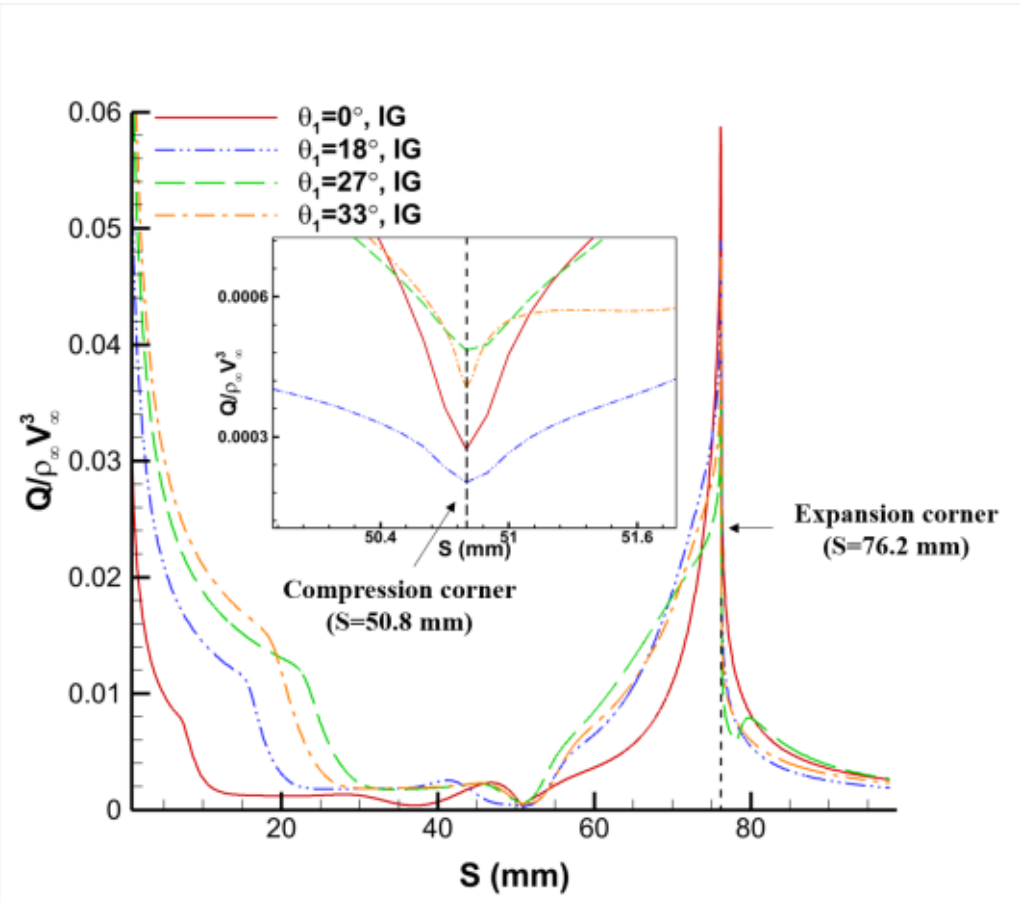

Fig. 40 Distributions of wall heat flux of IG at different $\theta_1$ in interactions before the sudden structural change

The results are compared to the same $\theta_1$ but by different gas models after analyzing the distribution trends of wall parameters with $\theta_1$. Specifically, cases with $\theta_1 = 15°$ and 25° are selected. First, the wall pressure coefficient distributions are presented, as shown in Fig. 41. The results in the wall pressure coefficient distributions for the two gas models essentially coincide before the separation zone. However, the first rise in pressure occurs earlier relatively for NEG, and within the plateau region, the distributions of IG have higher values, whereas the pressure peak of NEG is slightly higher. Fig. 42 shows the wall heat flux distributions for $\theta_1 = 15°$ and 25°. The heat flux distribution of IG is generally higher than that of NEG before the second sharp drop at the expansion corner. In addition, the minimum heat fluxes at the compression corner of IG and their peak heat fluxes at the expansion corner are larger than those of NEG.

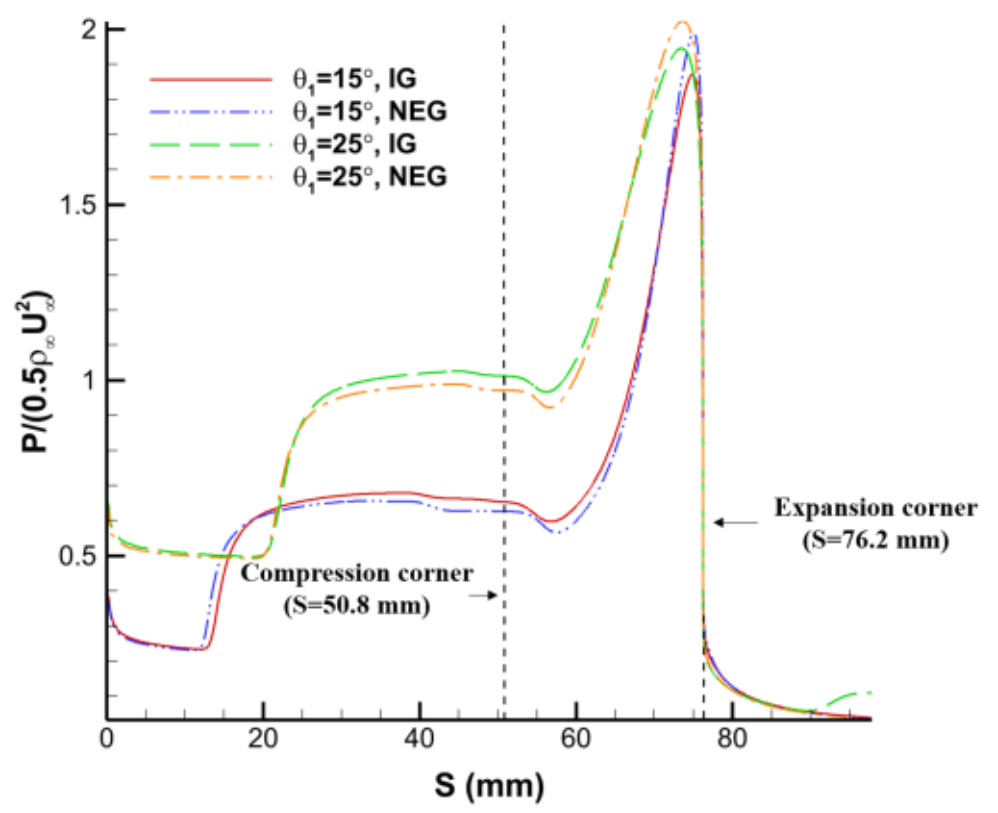

Fig. 41 Distributions of wall pressure coefficient of different gas models in interactions at $\theta_1 = 15°$ and 25° respectively

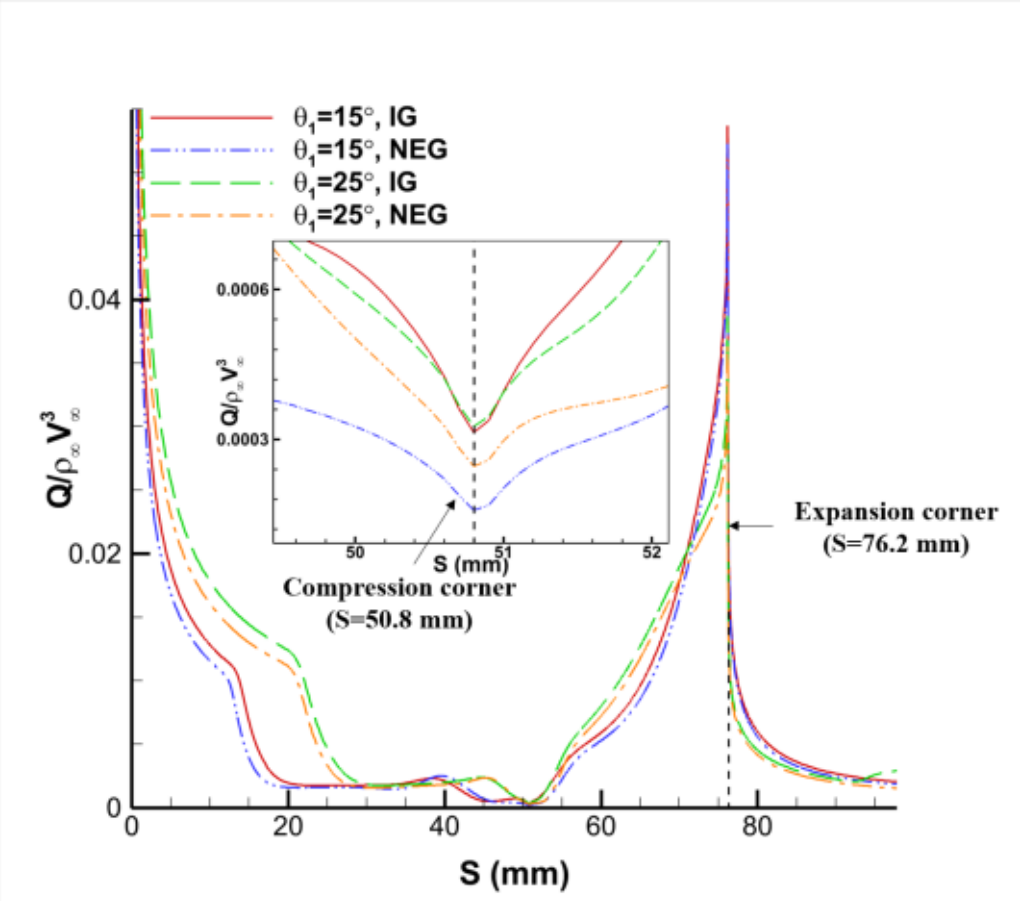

Fig. 42 Distributions of wall heat flux of different gas models in interactions at $\theta_1 = 15°$ and 25° respectively

3.4.2 Aerodynamic distributions in Type $IV_r$ interaction

When the interaction pattern evolves into Type $IV_r$, the emergence of supersonic jets and the disappearance of large-scale separation vortices significantly affect the wall distributions of aerodynamic properties. Cases with $\theta_1 = 34°$, 36° (IG) and $\theta_1 = 38°$ and 40° (NEG) are selected for analysis, and they are so chosen that the structures of the two models resemble each other sequentially. Fig. 43 shows the distributions of wall pressure coefficient by different gas models. Due to the disappearance of large-scale vortices, the plateau structure, such as those in Fig. 41, vanishes. The distributions for $\theta_1 = 34°$ (IG) and $\theta_1 = 38°$ (NEG) exhibit similar trends: the pressure first rises to a peak after passing SS, then decreases and is followed by an immediate second rise due to reflections of SW and EW within the supersonic channel. Then, the pressure undergoes a third rise due to compression near the compression corner. For $\theta_1 = 36°$ (IG) and $\theta_1 = 40°$ (NEG), as the strength of SS weakens with increasing $\theta_1$, the initial pressure rise is reduced. In addition, due to fewer reflections, the flow only experiences a second and moderate compression at the compression corner, resulting in a lower peak pressure. The fluctuations in wall pressure differ significantly from those shown on the streamlines within the supersonic channel (Fig. 35), indicating that the boundary layer can mitigate the impact of wave reflections within the channel on the wall distribution.

Fig. 44 shows the heat flux distributions by different gas models. Considering the results by IG at $\theta_1 = 34°$ as an example, a first fast drop occurs at S $\approx 18\ mm$, indicating the onset of separation. A rise follows whose location corresponds to the impingement of the transmitted shock on the boundary layer (Fig. 34(a.1)). The distribution exhibits a "wavy" pattern similar to the pressure distribution in Fig. 35, indicating that the reflections within the channel affect the heat flux, which is transmitted to the wall surface through the boundary layer of temperature. Then, the heat flux drops to a minimum at the compression corner. On the aft wedge surface, the heat flux increases for a period before decreasing, and a rise in heat flux occurs as the flow passes the corner due to the thinning of the boundary layer. For other angles, the overall trends in the heat flux distribution are similar to those at $\theta_1 = 34°$. However, the regions of rise or fall in the curves vary due to differences in the number of wave reflections and the position of LS. In addition, as $\theta_1$ increases, the minimum heat flux at the compression corner increases, while the peak heat flux at the expansion corner decreases.

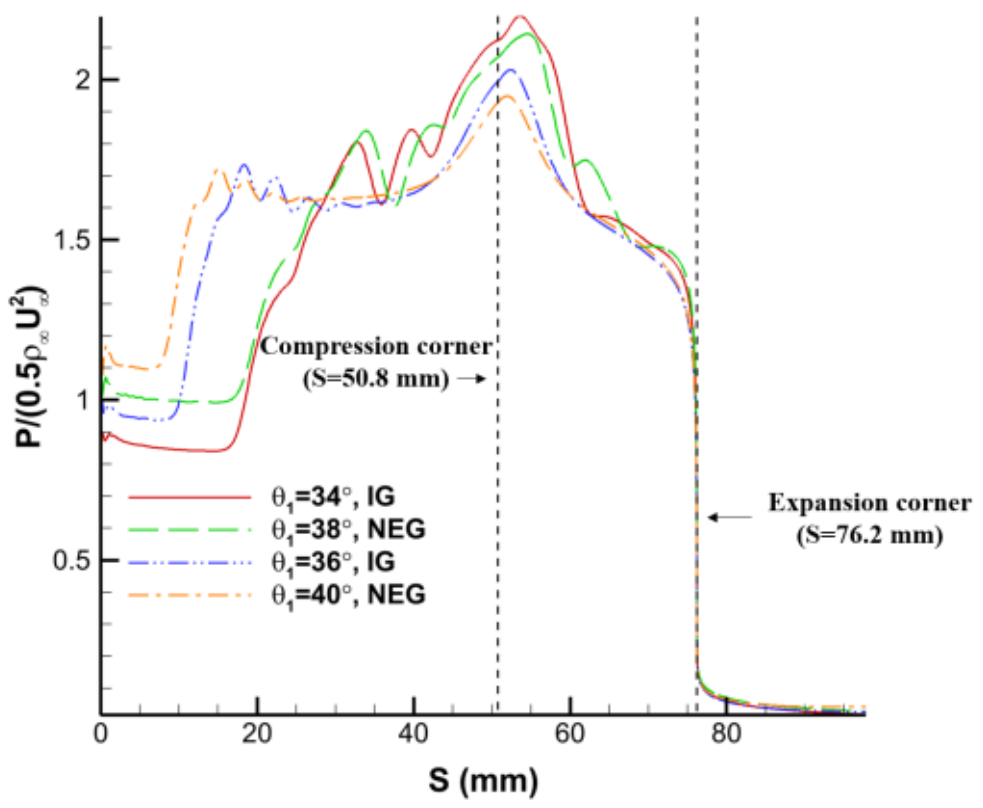

Fig. 43 Distributions of wall pressure coefficient of Type $IV_r$ interactions with $\theta_1$ by different gas models

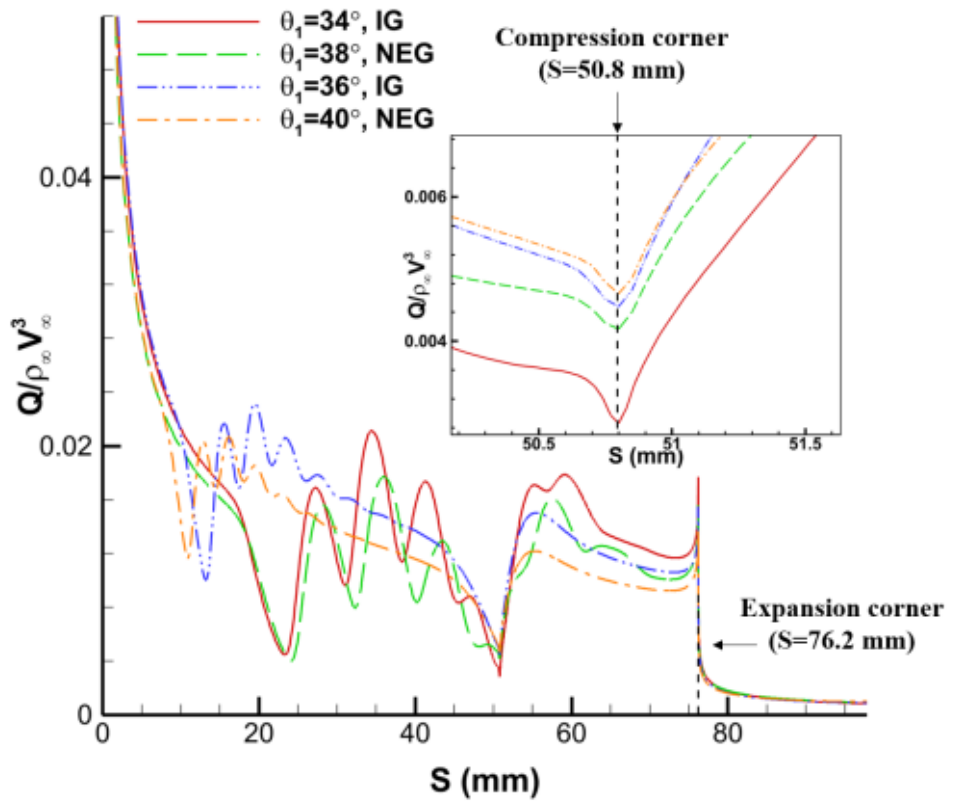

Fig. 44 Distributions of wall heat flux of Type $IV_r$ interactions with $\theta_1$ by different gas models

3.4.3 Aerodynamic distributions in cases dominated by the BS

As $\theta_1$ continues to increase, the forward movement of the BS leads to a sudden structural change, resulting in a flow field dominated by the BS. The BS also determines the aerodynamic properties of the wall. Fig. 45 shows the wall pressure coefficient distributions of IG at different $\theta_1$. At $\theta_1 = 39°$ and 40°, the shock detaches, causing an initial decreasing distribution other than the increase counterpart at $\theta_1 = 38°$. In the latter, the BS remains attached to the vertex of the double wedge, and the pressure rises rapidly before gradually leveling off. On the fore wedge surface, the pressure coefficient generally increases as it approaches the compression corner, with an overall small change. The peak pressure at the compression corner decreases as $\theta_1$ increases. On the aft wedge, the wall pressure coefficient gradually decreases and drops rapidly toward a stable low value at the expansion corner.

Fig. 46 shows the wall heat flux distributions at different $\theta_1$. The heat flux initially decreases with S, reaching a minimum at the compression corner, where the minimum value increases with $\theta_1$ (as shown in the zoomed window). On the aft wedge, the wall heat flux increases with S until reaching a peak at the expansion corner. The peak value increases with $\theta_1$, and after passing the corner, the heat flux rapidly decreases and stabilizes at a constant value.

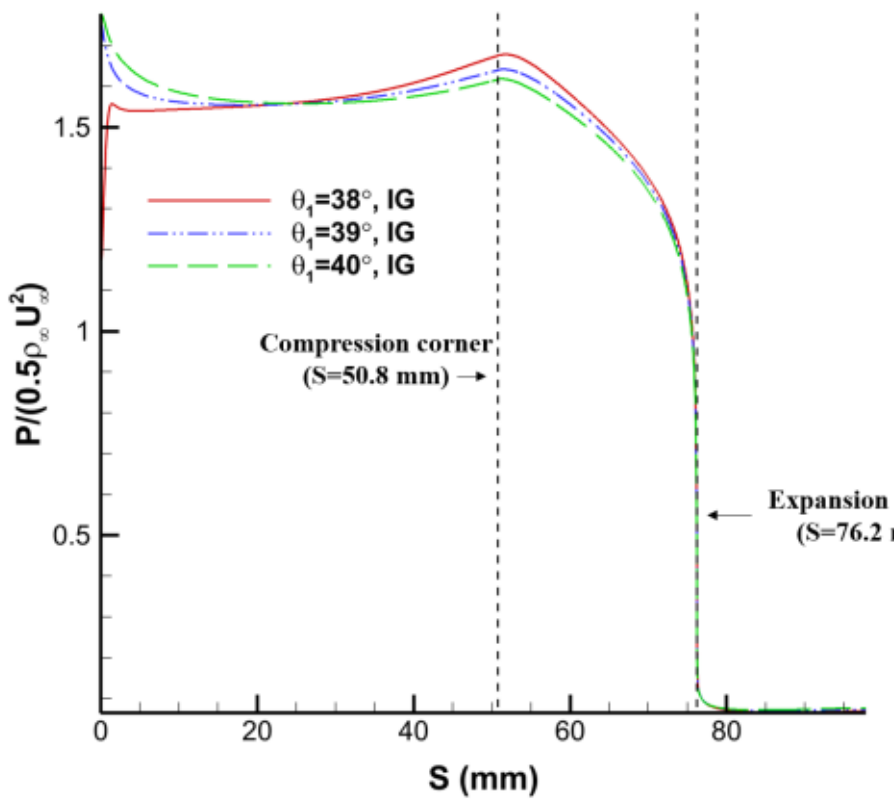

Fig. 45 Distributions of wall pressure coefficient of BS-dominating flows at different $\theta_1$

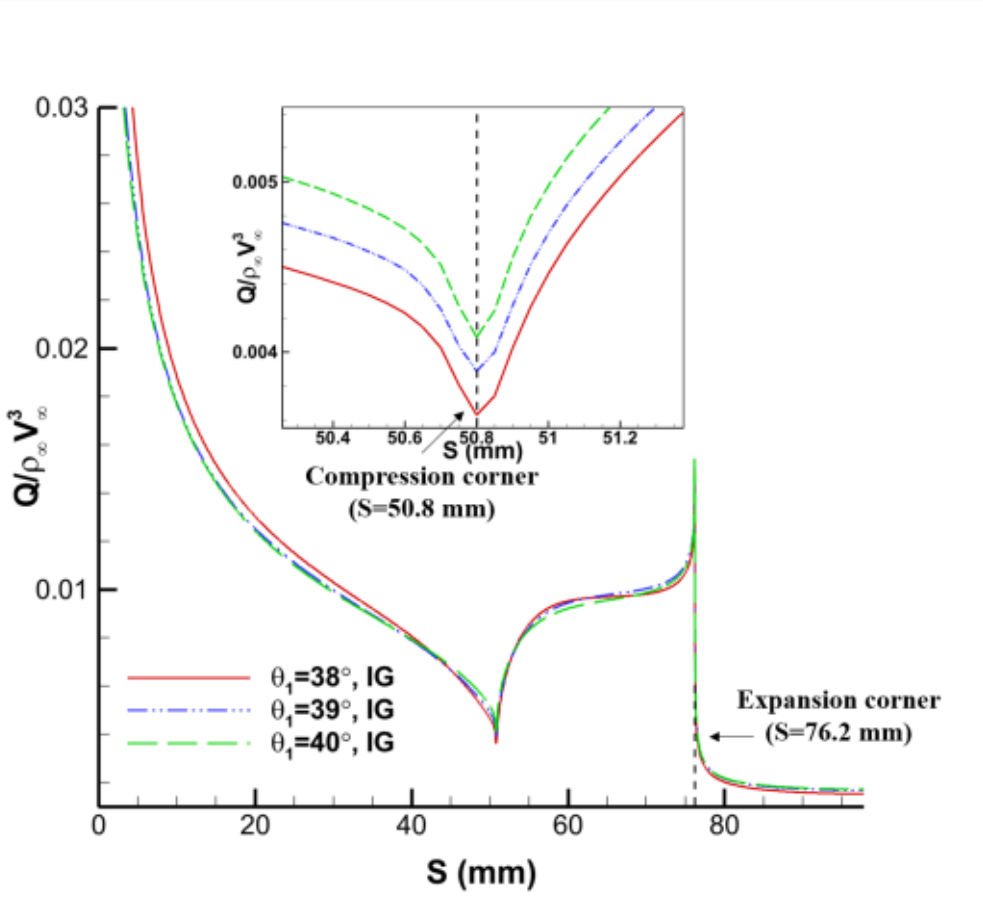

Fig. 46 Distributions of wall heat flux of BS-dominating flows at various $\theta_1$

## 4 Conclusions

This study conducts numerical simulations and analyses to examine the evolution of complex interaction patterns in the double wedge configuration under conditions of $Re = 1 \times 10^5\ /m$ and $\theta_1 = 0° - 40°$. The simulations reference the experimental conditions and geometry specified by Swantek and Austin [1, 2], employing the laminar Navier-Stokes equations and the WENO3-PRM$^2_{1,1}$ scheme, using the IG and NEG models, respectively. The evolution of the flow field structure and flow parameters with $\theta_1$ is investigated utilizing different gas models. Through analysis and discussion, the following results are obtained:

(1) A grid convergence study is conducted to determine the appropriate grid resolution for subsequent investigations by selecting the results from the IG model at $\theta_1 = 32°$ and the NEG model at $\theta_1 = 35°$.

(2) Regarding the steady and unsteady characteristics of the flows at various $\theta_1$, computations confirm that unsteady behavior occurs for IG at $\theta_1 = 28°$ to 30° and for NEG at $\theta_1 = 30°$ to 33°. The study shows that the cause of unsteadiness is related to the position of the TS intersection point. When the TS approaches the expansion corner from both sides, the flow structure loses stability, causing a transition from steady to unsteady flow fields.

(3) Through computational studies, evolution and the characteristics of steady interactions with $\theta_1$ are derived, indicating the sudden structural change in the flow field structure. The quantitative characteristics of the typical features are obtained by conducting quantitative analyses of their shocks and separation.

(4) The corresponding ranges of $\theta_1$ for different patterns are determined by analyzing the

evolution of interaction patterns. These patterns include Type VI, Type VI → V, Type III, Type $IV_r$, and the type where the flow is dominated only by the BS. The study exhibits that during the increase of $\theta_1$, a transition from expansion to shock waves occurs in Type VI interaction. An unconventional reflection of the TS on the separation vortex is observed in Type III interaction. A sudden structural change occurs during the transition from Type III to Type $IV_r$ interaction, and a small-scale one-Type $IV_r$ interaction to flow fields dominated by the BS.

(5) The evolution patterns of pressure and heat flux distributions are obtained using different gas models in various interaction patterns by analyzing the aerodynamic properties of the wall. These aerodynamic properties exhibit distinct characteristics in three scenarios: interaction patterns before a sudden structural change (Type VI, Type VI → V, and Type III interaction), Type $IV_r$ interaction, and flow fields dominated by the BS.

(6) The numerical results obtained from the two gas models indicate that the interaction patterns, their characteristics, and evolution trends are qualitatively consistent but with quantitative differences. The evolution rates of patterns vary between the models. Specifically, the evolution of IG results is faster than that of NEG.